\documentclass[sigconf,screen]{acmart}

\AtBeginDocument{%
  }

\usepackage{balance}

\usepackage[normalem]{ulem}
\usepackage{algorithmic}
\usepackage{graphicx}
\usepackage{textcomp}
\usepackage{xcolor}
\usepackage{microtype}
\usepackage{braket}
\usepackage{soul}
\usepackage{booktabs}
\usepackage{adjustbox}

\usepackage{bm}
\usepackage{url}

\usepackage[linewidth=1pt]{mdframed}

\usepackage{tikz}
\newcommand*\circled[1]{\tikz[baseline=(char.base)]{
            \node[shape=circle,draw,inner sep=0.5pt] (char) {#1};}}
\usepackage{mathtools}

\usepackage{caption}
\usepackage{subcaption}

\usepackage{tabularx}

\setcopyright{acmcopyright}
\acmPrice{15.00}
\acmDOI{10.1145/3567955.3567958}
\acmYear{2023}
\copyrightyear{2023}
\acmSubmissionID{asplosa23main-p56-p}
\acmISBN{978-1-4503-9915-9/23/03}
\acmConference[ASPLOS '23]{Proceedings of the 28th ACM International Conference on Architectural Support for Programming Languages and Operating Systems, Volume 1}{March 25--29, 2023}{Vancouver, BC, Canada}
\acmBooktitle{Proceedings of the 28th ACM International Conference on Architectural Support for Programming Languages and Operating Systems, Volume 1 (ASPLOS '23), March 25--29, 2023, Vancouver, BC, Canada}
\received{2022-03-31}
\received[accepted]{2022-06-16}

\begin{document}

\title{CAFQA: A Classical Simulation Bootstrap for Variational Quantum Algorithms}


\author{Gokul Subramanian Ravi}
\authornote{Correspondence: gravi@uchicago.edu}
\affiliation{%
  \institution{University of Chicago, USA}
}

\author{Pranav Gokhale}
\affiliation{%
  \institution{Super.tech, USA}
}

\author{Yi Ding}
\affiliation{%
  \institution{Massachusetts Institute of Technology, USA}
}

\author{William Kirby}
\affiliation{%
  \institution{Tufts University, USA}
}

\author{Kaitlin Smith}
\affiliation{%
  \institution{University of Chicago, USA}
}

\author{Jonathan M. Baker}
\affiliation{%
  \institution{University of Chicago, USA}
}

\author{Peter J. Love}
\affiliation{%
  \institution{Tufts University, USA}
}

\author{Henry Hoffmann}
\affiliation{%
  \institution{University of Chicago, USA}
}

\author{Kenneth R. Brown}
\affiliation{%
  \institution{Duke University, USA}
}

\author{Frederic T. Chong}
\affiliation{%
  \institution{University of Chicago, USA}
}

\renewcommand{\shortauthors}{Ravi, et al.}

\begin{abstract}
Classical computing plays a critical role in the advancement of quantum frontiers in the NISQ era. In this spirit, this work uses classical simulation to bootstrap  Variational Quantum Algorithms (VQAs). VQAs rely upon the iterative optimization of a parameterized unitary circuit (ansatz) with respect to an objective function.
Since quantum machines are noisy and expensive resources, it is imperative to classically choose the VQA ansatz initial parameters to be as close to optimal as possible to improve VQA accuracy and accelerate their convergence on today's devices.

This work tackles the problem of finding a good ansatz initialization, by proposing CAFQA, a \underline{C}lifford \underline{A}nsatz \underline{F}or \underline{Q}uantum \underline{A}ccuracy. The CAFQA ansatz is a hardware-efficient circuit built with only Clifford gates. In this ansatz, the parameters for the tunable gates are chosen by searching efficiently through the Clifford parameter space via classical simulation. The resulting initial states always equal or outperform  traditional classical initialization (e.g., Hartree-Fock), and enable high-accuracy VQA estimations.
CAFQA is well-suited to classical computation because: a) Clifford-only quantum circuits can be exactly simulated classically in polynomial time, and b) the discrete Clifford space is searched efficiently via Bayesian Optimization.

For the Variational Quantum Eigensolver (VQE) task of molecular ground state energy estimation (up to 18 qubits), CAFQA's Clifford Ansatz achieves a mean accuracy of nearly 99\% and recovers as much as 99.99\% of the molecular correlation energy that is lost in Hartree-Fock initialization. 
CAFQA achieves mean accuracy improvements of 6.4x and 56.8x, over the state-of-the-art, on different metrics.
The scalability of the approach allows for preliminary ground state energy estimation of the challenging chromium dimer (Cr$_2$) molecule.
With CAFQA's high-accuracy initialization, the convergence of VQAs is shown to accelerate by 2.5x, even for small molecules.

Furthermore, preliminary exploration of allowing a limited number of non-Clifford (T) gates in the CAFQA framework, shows that as much as 99.9\% of the correlation energy can be recovered at bond lengths for which Clifford-only CAFQA accuracy is relatively limited, while remaining classically simulable.


\end{abstract}



\maketitle


\begin{figure}
     \centering
         \includegraphics[width=0.97\columnwidth,trim={0.35cm 0.3cm 0.3cm 0.3cm},clip]{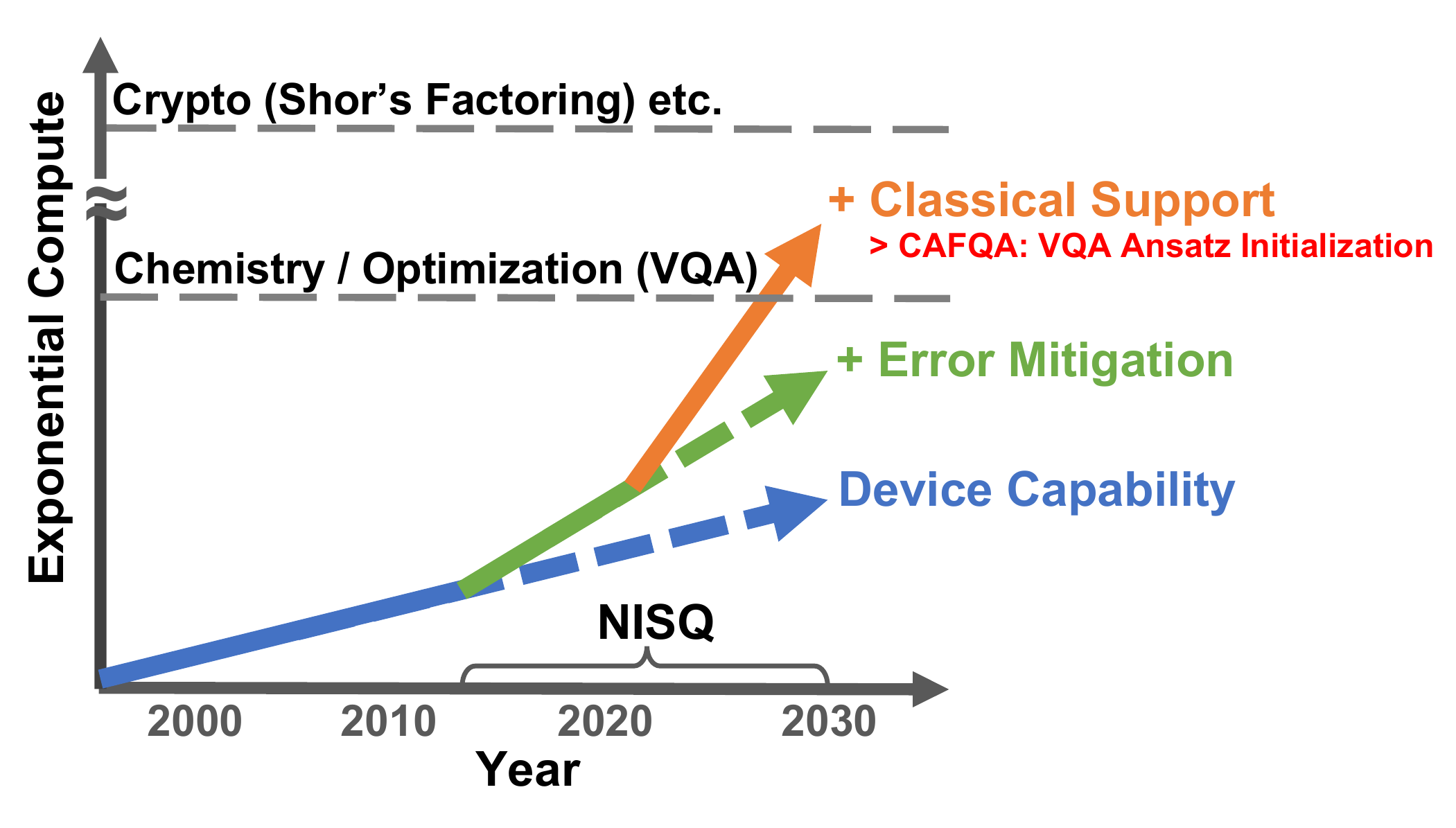}
        \caption{Advancing the NISQ frontiers with error mitigation, and classical support (eg., CAFQA).}
        \label{fig:frontier}
\end{figure}

\section{Introduction}

\textbf{\emph{Quantum Computing in the NISQ era:}} Quantum computing (QC) is a revolutionary computational model to solve certain classically intractable problems and is projected to give QCs a significant advantage in cryptography~\cite{Shor_1997}, chemistry~\cite{kandala2017hardware}, optimization~\cite{moll2018quantum} and machine learning~\cite{biamonte2017quantum}.
In the ongoing Noisy Intermediate-Scale Quantum (NISQ) era, we expect to work  with quantum machines which comprise hundreds to thousands of imperfect qubits ~\cite{preskill2018quantum}. 
On the one hand, NISQ era machines will be unable to execute large-scale quantum algorithms like Shor Factoring~\cite{Shor_1997} and Grover Search~\cite{Grover96afast}, which would require error correction comprised of millions of qubits to create fault-tolerant quantum systems~\cite{O_Gorman_2017}.
On the other hand, a variety of error mitigation techniques~\cite{czarnik2020error,Rosenberg2021,barron2020measurement,botelho2021error,wang2021error,takagi2021fundamental,temme2017error,li2017efficient,giurgica2020digital,ding2020systematic,smith2021error} have been proposed that have improved execution fidelity on today's quantum devices.
However, the resulting fidelity is still insufficient for most real-world use cases.

\textbf{\emph{Advancing NISQ with classical support:}}
There has been a recent impetus toward classical computing support to boost NISQ applications / devices to the realm of real-world applicability.
These include compiler level optimizations~\cite{murali2019noise,tannu2019not,murali2020software,ravi2021vaqem}, improved classical optimizers~\cite{9259985}, circuit cutting with classical compensation~\cite{CutQC,du2021accelerating,zhang2021variational} etc.
We are still in the early days of exploring this synergistic quantum-classical paradigm.
There is tremendous potential for sophisticated application-specific classical bootstrapping to advance the NISQ frontiers, and CAFQA is one such approach.
An illustration of advancing the NISQ frontiers towards real-world applicability is shown in Fig.\ref{fig:frontier}.

\textbf{\emph{Variational Quantum Algorithms:}} Variational quantum algorithms (VQAs) are expected to be a good match for NISQ machines.
This class of algorithms has a wide range of applications, such as the estimation of electronic energy of molecules~\cite{peruzzo2014variational}, MAXCUT approximation~\cite{moll2018quantum}. 
The quantum circuit for a VQA is parameterized by a list of angles which are optimized by a classical optimizer over many iterations towards a specific target objective which is representative of the VQA problem.
VQAs are more suitable for today's quantum devices because these algorithms adapt to the characteristics and noise profile of the quantum machine on~\cite{peruzzo2014variational, mcclean2016theory}.
Unfortunately, VQA accuracy obtained on today's NISQ machines, even with error mitigation, is often considerably far from the stringent accuracy requirements in fields such as molecular chemistry, especially as we scale to larger problem sizes~\cite{wang2021error, kandala2017hardware, ravi2021vaqem}.

\textbf{\emph{Aiding VQAs in the NISQ Era:}}
For NISQ VQAs to progress towards real-world applicability, it is imperative to classically choose a VQA's parameterized circuit (ansatz) wisely and its initial parameters to be as close to optimal as possible, prior to quantum exploration.
This would improve accuracy and accelerate convergence of the algorithm on the noisy quantum device~\cite{mcclean2018,wang2020noise}.
Suitable ansatz circuits for today’s devices, referred to as ``hardware efficient ansatz”~\cite{kandala2017hardware}, are often application-agnostic and can especially benefit from a wise choice of initial parameters, but these can be difficult to estimate classically.

\textbf{\emph{Classical simulation support for VQAs with Cliffords:}}
This work helps to initialize the VQA ansatz using classical simulation. 
In general, classical simulation of quantum tasks is not a scalable solution, primarily only suited to trivial quantum problems, and is, in fact, the motivation for quantum machines.
An exception to the above is the classical simulation of the Clifford space.
Circuits made up of only Clifford operations can be exactly simulated in polynomial time~\cite{gottesman1998heisenberg}. 
Clifford operations  do not provide a universal set of quantum gates - hence, the stabilizer states produced by Clifford-only circuits are limited in how effectively they can explore the quantum space of a given problem such as those targeted by VQAs. 
However, exploring the Clifford space of the VQA problem through ideal classical simulation can potentially find good noise-free initial states, which is particularly beneficial in the NISQ era.
Note: in addition to Clifford gates, it is possible for a small number of T gates to also be efficiently classically simulated~\cite{Bravyi2016}.
While we primarily focus on the Clifford space in this work, we show preliminary results for beyond-Clifford exploration in Section \ref{FW}.

\begin{figure}
     \centering
         \includegraphics[width=0.85\columnwidth]{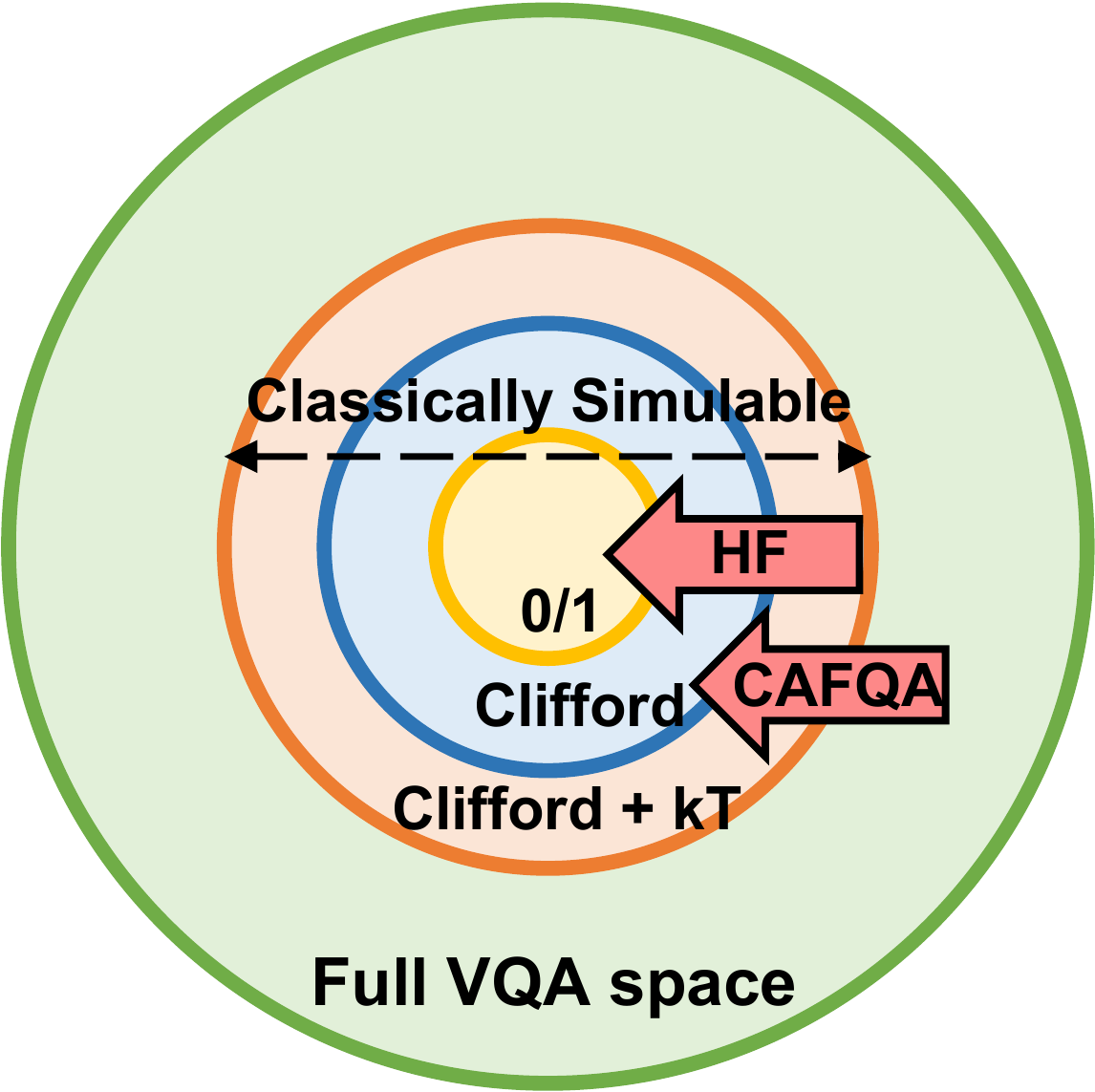}
        \caption{Figure illustrates the VQA(E) state space explored by different techniques. Hartree-Fock (HF), only explores the yellow computational basis (i.e., classical bits) space. CAFQA goes further by exploring the Clifford space (blue) for higher accuracy. Both of these are classically simulable. Classical exploration can be extended to incorporate a few T gates (orange). The rest of the state space is shown in green - this space cannot be  efficiently simulated classically and requires quantum exploration. By searching efficiently through the classically simulable Clifford space, CAFQA provides a good VQA ansatz initialization for quantum exploration.}
        \label{fig:space}
\end{figure}

\textbf{\emph{CAFQA:}}
This work tackles the challenge of finding initial ansatz parameters by proposing CAFQA, a \underline{C}lifford \underline{A}nsatz \underline{F}or \underline{Q}uantum \underline{A}ccuracy. 
The CAFQA Clifford Ansatz is a hardware-efficient parameterizable circuit that is parameterized with only Clifford gates.
In this ansatz, the initial parameters for the tunable gates are chosen by searching efficiently through the Clifford parameter space via classical simulation, thereby producing a suitable stabilizer state. 
The proposed approach is attractive for multiple reasons: \circled{a} the Clifford-initialized ansatz produces stabilizer states that perform equal to or better than the traditional classical approach of finding a suitable computational basis state (e.g., Hartree-Fock~\cite{hartree1935self}) because (i) it can explore a larger state space and (ii) the stabilizer can have direct chemical relevance in some molecular systems; \circled{b} the Clifford-only quantum circuits can be perfectly simulated in polynomial (quadratic or even linear) time on classical computers~\cite{gottesman1998heisenberg}; \circled{c} the produced initial states are obtained ideally, since classical simulation is noise-free; \circled{d} the discrete Clifford space is searched efficiently via Bayesian Optimization using a random forest surrogate model and a greedy acquisition function; \circled{e} while the Clifford space is significantly smaller compared to the entire quantum space, the stabilizer states produced from the optimal Clifford parameters are able to achieve solutions of high accuracy even prior to execution/exploration on a quantum device; and finally \circled{f} the selected ansatz can then be tuned over the entire parameter space on a quantum device, allowing for accelerated accurate convergence on NISQ devices (and beyond).
Fig.\ref{fig:space} provides a break down of the VQA parameter space and CAFQA's scope.

\vspace{0.2in}
\textbf{\emph{Key CAFQA results:}}

\circled{1}\ For the VQE task of ground state energy estimations of molecular systems up to 18 qubits, CAFQA's Clifford Ansatz is observed to achieve a mean accuracy of near 99\% and is able to recover up to 99.99\% of the molecular correlation energy lost in state-of-the-art Hartree-Fock initialization. 
CAFQA achieves mean accuracy improvements over the state-of-the-art of 6.4x when averaged over all bond lengths and 56.8x at highest bond lengths (maximum of 3.4*10$^5$x).

\circled{2}\ Quantum exploration post CAFQA initialization can lead to faster and highly accurate VQA convergence, even on reasonably noisy quantum machines --- we show 2.5x faster convergence compared to HF for a small molecule. Greater benefits can be expected for larger problem sizes, which can be usefully evaluated when NISQ machines improve.

\circled{3}\ The scalability of the approach allows for accurate ansatz initialization for ground state energy estimation of the challenging Chromium dimer (often considered a benchmark for variational quantum advantage) with greater than Hartree-Fock accuracy.

\circled{4}\ Preliminary exploration of allowing a very limited number of non-Clifford (T) gates in the CAFQA framework shows that as much as 99.9\% of the correlation energy can be recovered at bond lengths for which Clifford-only CAFQA accuracy is relatively limited, while remaining classically simulable.

\vspace{0.1in}
\textbf{\emph{Key CAFQA insights:}}

\circled{1}\ CAFQA uses classical simulation to explore the Clifford space of a VQA problem and produces high accuracy VQA ansatz initialization, considerably outperforming the state-of-the-art.


\circled{2}\ CAFQA's benefits are especially significant because it is classically simulable, it searches the search space efficiently, and its evaluations are ideal.



\circled{3}\ CAFQA highlights the potential for quantum inspired classical techniques as well as a synergistic quantum-classical paradigm, to boost NISQ-era quantum computing (with focus on VQA) towards real world applicability. 

\section{Background and Motivation}

\subsection{{VQAs in the NISQ Era}}
\label{Real-VQA}

\textbf{\emph{VQE:}} While CAFQA is suited widely across variational algorithms (eg., QAOA~\cite{farhi2014quantum}), in this paper we primarily focus on the Variational Quantum Eigensolver (VQE)~\cite{peruzzo2014variational}.
VQE is used to estimate an upper bound on the ground state energy of a Hamiltonian.
Here, a Hamiltonian is a mathematical representation of some problem from, say, optimization or molecular chemistry, and is a linear combination of multiple Pauli terms.
For example, a 4-qubit Hamiltonian could be $[H = 0.1*XYXY + 0.5*IZZI]$.
VQE tries to find suitable parameters for an appropriately chosen parameterized circuit (ansatz) such that the expectation value of the target Hamiltonian is minimized. 
At a high level, VQE can be conceptualized as a repetitive ``classical guess'' + ``quantum check'' algorithm~\cite{gokhale2019minimizing}. 
The check stage involves the preparation of a quantum state corresponding to the guess.
This preparation stage is done in polynomial time on a quantum computer, but would incur exponential cost  in general on a classical computer. 
This contrast gives rise to a potential quantum speedup for VQE~\cite{Gokhale:2019}.
In chemistry, VQE is a critical step in computing the energy properties of molecules and materials.
While conventional computational chemistry provides methods to approximate such properties, they can lack sufficient accuracy in molecular systems due to an inadequate treatment of the correlations between constituent electrons. 
These interactions require computation that scales exponentially in the size of the system~\cite{tilly2021variational,PhysRevX.10.041038}.

\textbf{\emph{NISQ era accuracy:}} Estimating the VQE global optimum with high accuracy has proven challenging in the NISQ era even with sophisticated optimizers, a well-chosen ansatz, and error mitigation~\cite{ravi2021vaqem,czarnik2020error,Rosenberg2021,barron2020measurement,botelho2021error,wang2021error,tilly2021variational,takagi2021fundamental}. 
As an example, ground state energy estimation of molecules (the energy required to break a molecule into its sub-atomic components), a key use case for VQE, requires energy estimates with an estimation error of less than $1.6 \times 10^{-3}$ Hartree, or what is known as ``chemical accuracy"~\cite{peterson2012chemical}, for applicability in understanding chemical reactions and their rates.
Unfortunately, for instance, previous work on the estimation of ground state energy of BeH$_2$ on a superconducting transmon machine resulted in an error greater than 10$^{-1}$ Hartree, which is roughly 100x worse than the required accuracy~\cite{kandala2017hardware}. 
Considering the significant disparity between NISQ VQA accuracy and real world requirements, it is imperative to aid VQA to the best extent possible.


\begin{figure}[t]
\centering
\includegraphics[width=0.95\columnwidth,trim={0cm 0cm 0cm 0cm}]{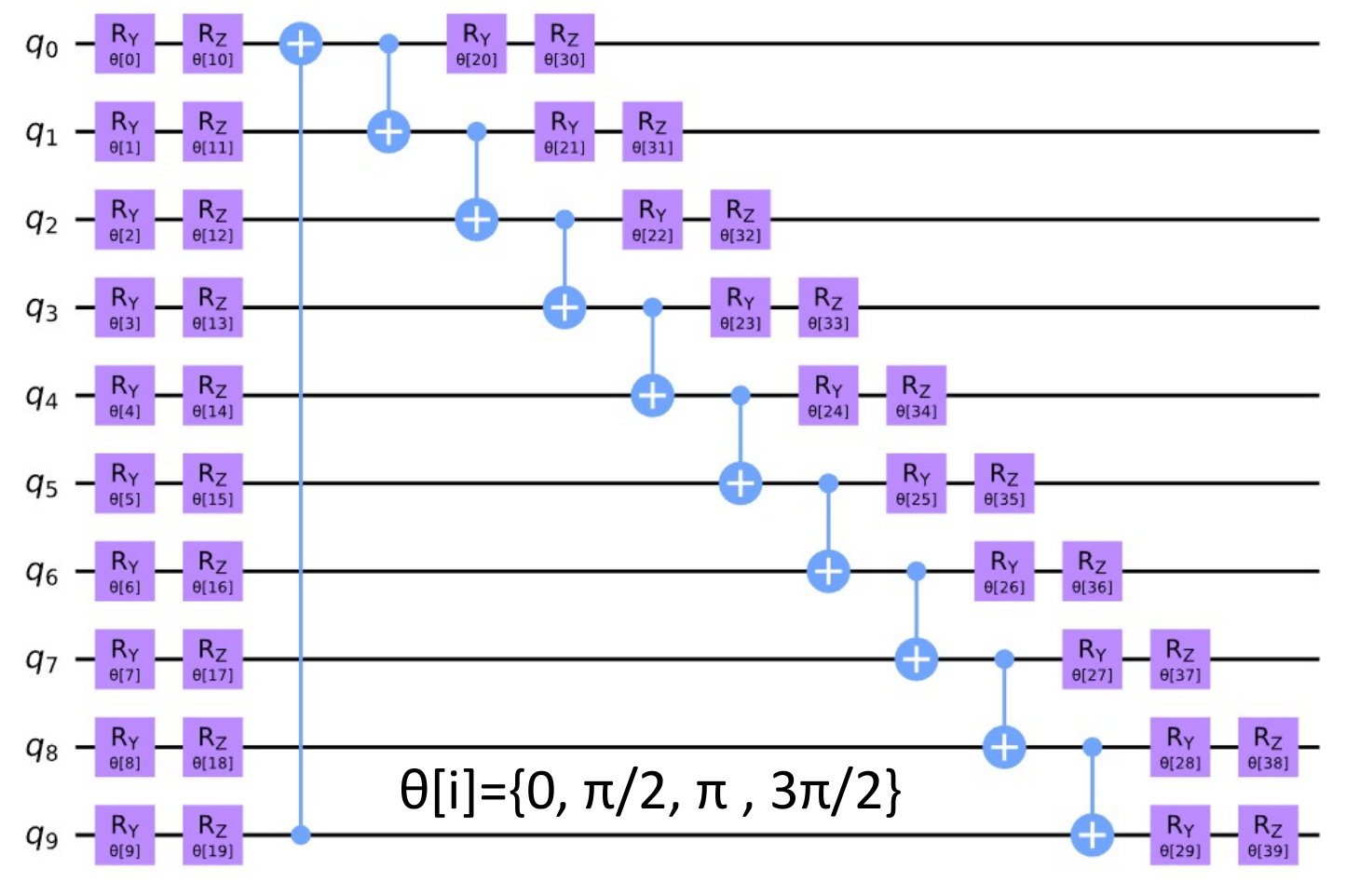}
\caption{A Clifford Ansatz, a circuit with only Clifford gates. In this example, all fixed components are CXs and the tunable rotational gate angles are multiples of $\pi/2$. }
\label{fig:ansatz}
\end{figure}

\subsection{{VQA Ansatz and its Initialization}}
\label{bm_ansatz}
\label{bm_init}
\textbf{\emph{Ansatz:}} An ansatz is a parameterized circuit which is used to explore the quantum Hilbert space of the target VQA Hamiltonian, to find its ground state energy. 
An ansatz with parameterized gate rotation angles is shown in Fig.~\ref{fig:ansatz}.
Many ansatz structures are suitable for VQAs.
In the context of VQE for molecular chemistry, the Unitary Coupled Cluster Single-Double (UCCSD) ansatz is considered the gold standard~\cite{romero2018strategies, Gokhale:2019}. 
Unfortunately, the UCCSD ansatz is generally of considerable circuit depth, making it less suitable for today's NISQ machines, except for very small molecules such as H$_2$.
More suitable to the NISQ-era are hardware-efficient ansatz like the SU2~\cite{IBM-SU2} which are low depth parameterized circuits (but have Hilbert space coverage limitations~\cite{tilly2021variational, holmes2021connecting}).
Fundamentally, this ansatz is constructed by repeating blocks of parameterized single-qubit rotation gates and ladders of entangling gates~\cite{tilly2021variational} (Fig.\ref{fig:ansatz}).
CAFQA builds atop a traditional hardware efficient ansatz~\cite{kandala2017hardware}.
Suitability to other ansatz structures is discussed in Section \ref{FW}.

\textbf{\emph{Optimization surface:}} While a good choice of classical optimizer improves VQA convergence~\cite{9259985}, VQAs can: a) have complex optimization surfaces and b) suffer from a barren plateau problem. 
The optimization contour worsens as the noise and complexity of the problem increases in relation to the increase in the depth of the circuit, the number of parameters and the spread of the entanglement~\cite{9259985}.
The barren plateau is the phenomenon in which the gradients of the VQE parameters vanish exponentially.
While barren plateaus can become a critical issue for a variety of reasons~\cite{mcclean2018, PhysRevResearch.3.033090,marrero2021entanglement,Cerezo2021,Uvarov2021}, in the context of this work, they can become significant in the presence of noise~\cite{wang2020noise} and with poor (random) ansatz initialization~\cite{mcclean2018}.
Thus, well-chosen initialization of the ansatz can help avoid barren plateaus and effects of noise, and therefore enable fast accurate convergence on the VQA problem.


\textbf{\emph{Hartree-Fock initialization:}} A popular and simple approach to construct a fair initial state for quantum systems is derived from Hartree-Fock (HF) theory~\cite{hartree1935self}.
Although such approximation / optimization problems are generically hard, HF usually rapidly converges to good solutions, especially for closed-shell molecules at equilibrium geometries~\cite{Microsoft-HF}.
HF yields an initial state that has no entanglement between the electrons (i.e., simply a bitstring of 0s and 1s on the circuit's qubits).
HF assumes that each electron’s motion can be described as a stand-alone particle function, independent of the instantaneous motion of other electrons.
In doing so, HF neglects the correlation between electrons, which is where classical computing is limited in solving such molecular chemistry problems.
Therefore, although HF has reasonable accuracy for many molecules, it is generally insufficient to make highly accurate quantitative predictions~\cite{Ornl-HF}. 
Thus, its usefulness as a suitable initialization on today's very noisy quantum devices is limited - there is too much ground left for the quantum device to cover, which is challenging considering the complex noisy optimization surface and barren plateaus as  described earlier.
Therefore, initialization with greater accuracy, especially for strongly correlated systems and/or away from equilibrium geometry, necessitates quantum states that go beyond HF.


\subsection{{Clifford Circuits}}
\label{bm_cliff}
Classical simulation of quantum problems usually requires exponential resources (otherwise, the need for quantum computers is obviated).
Even with high-performance supercomputers, simulation is restricted to under 100 qubits~\cite{tilly2021variational,Haner2017,Raedt2019,Boixo2018,45919}.

However, not all simulations are non-scalable. 
The Gottesman-Knill theorem states that \emph{``Any quantum computer performing only: a) Clifford group gates, b) measurements of Pauli group operators, and c) Clifford group operations conditioned on classical bits, which may be the results of earlier measurements, can be perfectly simulated in polynomial time on a probabilistic classical computer"}~\cite{gottesman1998heisenberg}. 

While the Clifford group operations and Pauli group measurements do not provide a universal set of quantum gates, there are quantum domains that have applications focused on the Clifford-space including quantum networks~\cite{Veitch2014}, error correction codes~\cite{QECIntro}, teleportation~\cite{gottesman1999demonstrating} and error mitigation~\cite{czarnik2020error,strikis2021learningbased}.

CAFQA explores the benefits of Clifford-only circuits as an ansatz for variational algorithms.
An example is shown in Fig.\ref{fig:ansatz}.
Extending beyond Cliffords is discussed in Section \ref{FW}.



\begin{figure*}[t]
\centering

\includegraphics[width=\textwidth,trim={0cm 0cm 0cm 0cm}]{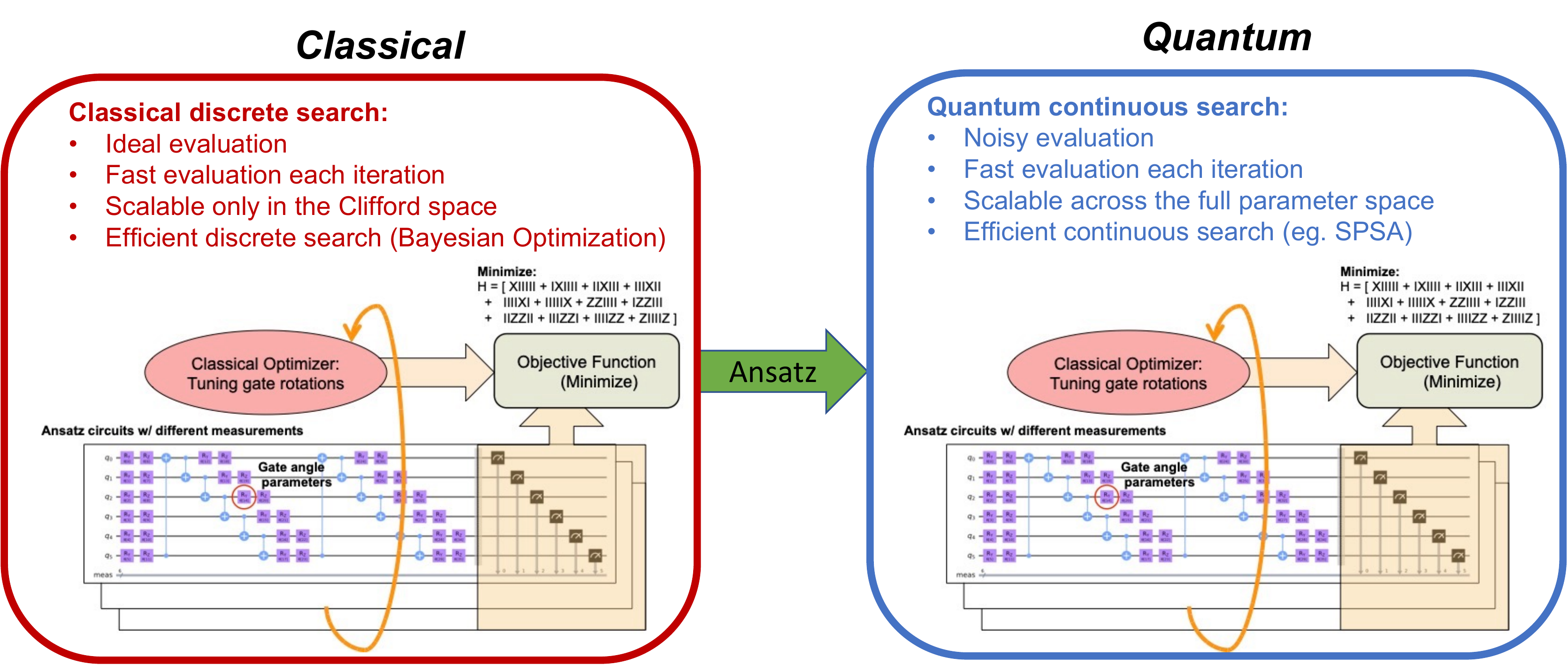}

\caption{The red box on the left shows the CAFQA framework. CAFQA performs ansatz parameter tuning inspired by traditional VQA. But CAFQA's tuning is suitable for classical compute since it restricts the search space to the Clifford space alone (in an ansatz that has its fixed components to also be Clifford, often the case with a hardware-efficient ansatz). The search space is discrete and is searched with Bayesian Optimization. Although the search space is limited, its evaluation is classically efficient and noise-free. On the other hand, traditional quantum variational tuning, while scalable and suited to the entire quantum space, is extremely noisy in the NISQ era. Once CAFQA finds a suitable Clifford initialization, traditional VQA tuning is performed (blue box / right), leading to faster and more accurate convergence. }
\label{fig:qafca-intro}
\end{figure*}

\section{CAFQA Proposal}

\label{proposal_overview}

Fig.\ref{fig:qafca-intro} provides an illustrative overview of how CAFQA complements traditional VQA tuning.
CAFQA is illustrated in the red box and is discussed below.

\circled{1}\ CAFQA begins with a parameterized circuit in which all fixed gates are Clifford. 
This is usually the case with hardware-efficient ansatz, as described in Section \ref{bm_ansatz}.
Focusing on a hardware-efficient ansatz is justified considering that other ansatz options are generally less suitable to noisy execution on today's NISQ devices.
However, extensions are discussed in Section \ref{FW}.

\circled{2}\ Given this parameterized circuit, CAFQA performs a discrete search over the tunable circuit parameters. The tunable search space is limited to angles which make the tunable gates Clifford.
Extensions discussed in Section \ref{FW}.

\circled{3}\ Since both the fixed gates as well as the tunable gates are Clifford, the resulting circuit in each iteration of the tuning process can be simulated classically, even as the size of the circuits grow (as discussed in Section \ref{bm_cliff}). 

\circled{4}\ Simulating the ansatz circuits corresponding to the Hamiltonian and measuring the expectation produces the objective function value for the iterative tuning process. 

\circled{5}\ For molecular chemistry, if any electron and spin preservation constraints have to be imposed on the problem they can can be added to the Hamiltonian~\cite{ryabinkin2018constrained} or directly to the objective function - CAFQA uses the latter. More in Section \ref{eval_mol}. 

\circled{6}\ Since simulations are performed classically, they are free of noise, thus having the potential to eliminate a considerable portion of the noise impact that variational tuning on the real quantum device could suffer, i.e., noise-induced barren plateaus~\cite{wang2020noise} etc. 

\circled{7}\ Also worth noting is that only one-shot simulation is required for each Pauli term, since the expectation value produced by each term is strictly +1, -1, or 0 for stabilizer states (i.e., for Clifford circuits)~\cite{nielsen2002quantum}.

\circled{8}\ The search is continued until the convergence of the minimum value obtained or for a specific tuning budget.

\circled{9}\ The resulting circuit, with Clifford parameters corresponding to the minimum objective function value observed, is the Clifford ansatz and is then ready for traditional VQA optimization. 

\circled{10}\ Subsequent VQA tuning on the quantum device is noisy but is able to explore the entire quantum space allowed by the ansatz. The initial state produced by CAFQA can enable faster and more accurate convergence of traditional VQA.

\section{Qualitative Analysis}

\begin{figure}[t]
\centering
\includegraphics[width=0.95\columnwidth,trim={0cm 0cm 0cm 0cm}]{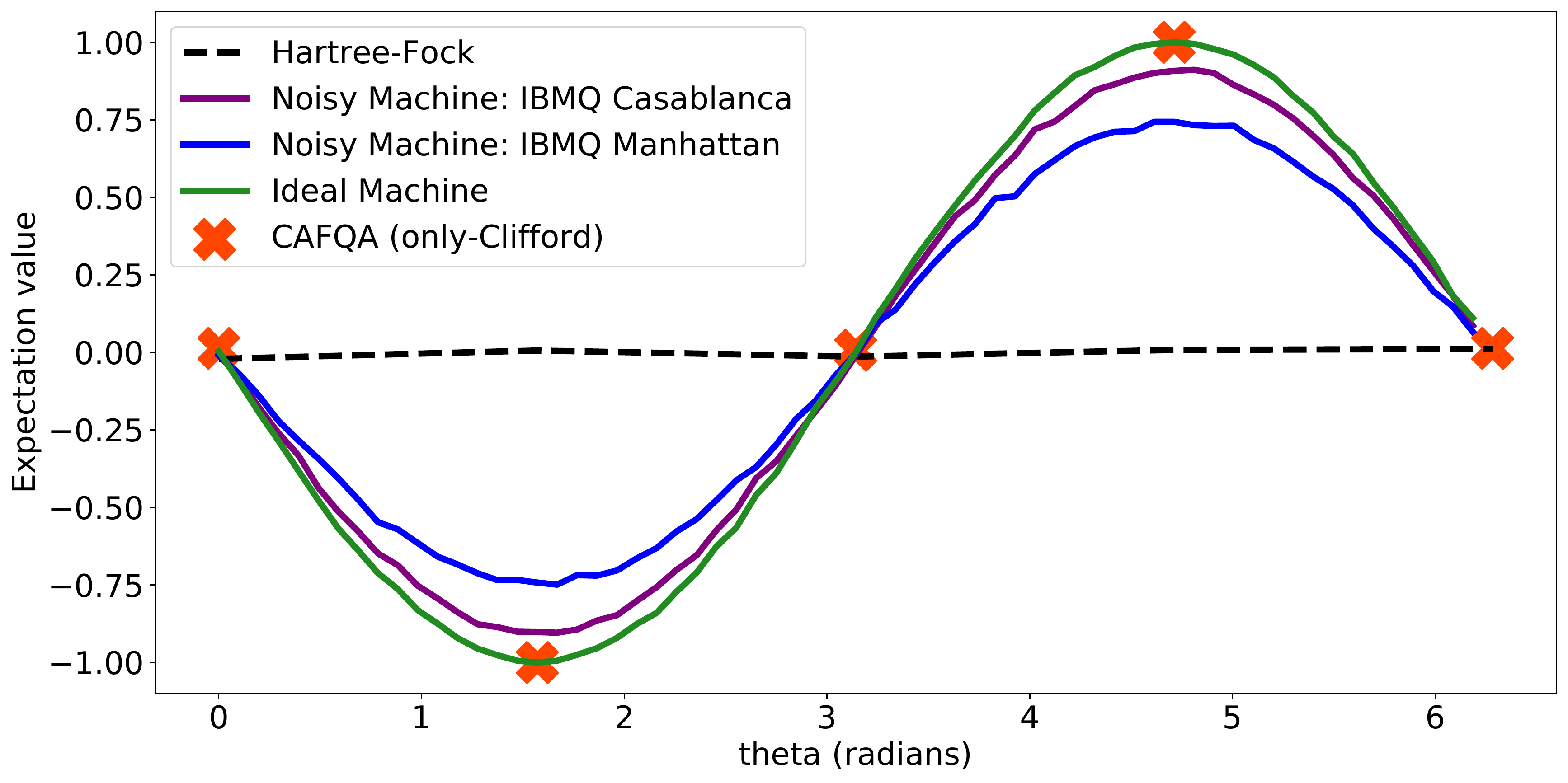}
\caption{Comparison of different methods of ansatz tuning for a 2qubit XX Hamiltonian employing an ansatz with only one tunable rotation angle parameter. Although the search space for CAFQA is limited, it is able to achieve the global minimum (equaling the ideal machine). Furthermore, CAFQA outperforms the noisy machines, which are limited by noise although they can explore the entire tuning space. On the other hand, the HF method of initialization is unable to recover the expectation value at all since the XX Hamiltonian does not have an uncorrelated component.}
\label{fig:cliff}
\end{figure}

\subsection{ CAFQA Benefits on a Microbenchmark}
\label{CAFAXX}
In Fig.\ref{fig:cliff} we use a 2-qubit `XX' Hamiltonian system and a 2-qubit hardware-efficient ansatz with only one tuning parameter to show the benefits of CAFQA. 
The Y-axis shows the estimated expectation values of the Hamiltonian while the X-axis sweeps the tuning parameter: 

\circled{1}\ The green line represents tuning the one ansatz parameter on an ideal noise-free quantum device.
Sweeping through all rotations produces an expectation value mimima = -1.0.

\circled{2}\ Next, the same tuning is performed on two noisy quantum devices, IBMQ Casablanca and Manhattan (simulated with noise models). These are shown with the purple and blue lines.
Clearly, the noisy devices are able to sweep through the entire parameter space, but the effect of noise limits the minimum obtained, achieving only -0.7 / -0.85. Note that this microbenchmark is too simplistic to suffer from barren plateaus, etc., but it is expected that the deviation from the ideal / exact minimum will increase with more complex problems and increasing noise.

\circled{3}\ Next, the dashed line shows the expectation value produced by HF initialization. In this example, HF is unable to produce any useful result since the chosen Hamiltonian does not have any diagonal Pauli terms suitable for HF. 
This can be thought to represent pure electron correlation energy in the context of molecular chemistry (as described in Section \ref{bm_init}). 

\circled{4}\ Finally, CAFQA is shown with orange `X's. Note that there are only 4 unique discrete points in the Clifford space for the one tunable parameter. Even so, CAFQA is able to produce the expectation value global minimum (= -1.0) via one of those 4 Clifford points. Not only does the Clifford minimum match the minimum of the entire tuning space, but also the ability to simulate the Clifford space ideally without noise produces more accurate estimations than the noisy devices.

\begin{figure*}[t]
\centering
\includegraphics[width=\textwidth,trim={0.6cm 0.5cm .6cm 0.5cm}]{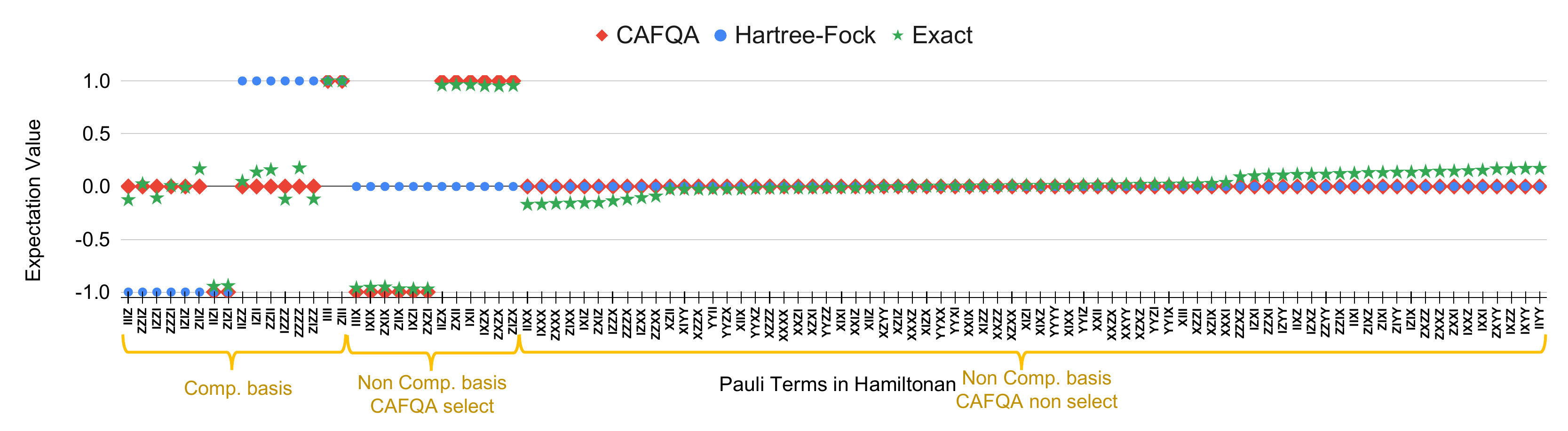}
\caption{LiH ground state energy at bond length of 4.8{\AA} (3x equilibrium ). Expectation value of each Pauli term, as obtained from different methods, is shown. The Pauli terms along the X-axis are arranged as: i) Computational basis terms, ii) Non computational basis terms selected by CAFQA, and iii) remaining terms which are beyond the Clifford reach (sorted by Exact expectation value). While HF is only able to obtain non zero expectations (of +/- 1) for diagonal Pauli (computational basis) terms, the Clifford Ansatz enables expectations of +/- 1 for non-diagonal Pauli terms as well. Further, the similarity in expectation value between the Clifford Ansatz and the exact (i.e., ideal) LiH tuning is evident. }
\label{fig:LiH-Terms}
\end{figure*}

\subsection{High Accuracy CAFQA Stabilizer States}
\label{Qual}
In Fig.\ref{fig:LiH-Terms} we break down the expectation value returned by the Clifford Ansatz in comparison to Hartree-Fock and ideal / exact minimum from noise free simulation.
This is shown for the ground-state energy estimation of the LiH molecule, represented by a 4 qubit Hamiltonian system, at a bond length of 4.8{\AA}.
The Y-axis shows the expectation value of each Pauli term and the X-axis lists the Pauli terms in the Hamiltonian.

\circled{1}\ In the figure, the resulting expectation value of each Pauli term for the Hartree-Fock (HF) initialization is shown in blue. 
Since HF is a `classical' computational basis state (i.e., a bitstring), and therefore Clifford, as noted in Section \ref{proposal_overview}, all the HF Pauli term expectation values are +1 / -1 / 0.
Further, since HF is `classical', all non-diagonal Pauli terms (i.e., any terms apart from the tensor products of I and Z) have an expectation value of zero.  
Only calculating the expectation values for the diagonal terms leads to HF ignoring the correlation energy, which is known to cause serious errors for some larger molecules (described in Section \ref{bm_init}). 

\circled{2}\ The expectation value of each Pauli term for the Clifford ansatz produced by CAFQA is shown in red. 
Again, note that all expectation values are +1 / -1 / 0.
Moreover, note that for the Clifford ansatz, there are multiple non-diagonal Pauli terms which produce an expectation of +1 / -1. 
The non-zero expectation on non-diagonal terms is indicative of CAFQA producing a non-computational basis state, albeit a Clifford one. 
By doing so, it is able to capture some of the correlation energy that is contributed by the non-diagonal Pauli terms. 
This is important because it is qualitatively indicative of the potential for high(er) estimation accuracy through CAFQA as the complexity of the problem scales. 

\circled{3}\ The expectation value of each Pauli term for the exact minimum from  ideal noise-free simulation is shown in green.
Ideal noise-free simulation is possible since LiH is a very small molecule.
While expectation values range from -1 to 1, it is evident that the expectation values are close to those produced by the Clifford Ansatz, both for diagonal terms and the non-diagonal terms. 
This is a clear indicator of the effectiveness of CAFQA for ground-state energy estimation for the LiH molecule (at the chosen bond length).
This is confirmed later in Section \ref{eval_lih} which shows that CAFQA is able to achieve high accuracy in the range of 10$^{-2}$ Hartree for LiH.
This trend is also observed across other molecules and bond lengths, leading to high accuracy overall.

\begin{figure}[t]
\centering
\includegraphics[width=0.98\columnwidth,trim={0cm 0cm 0cm 0cm}]{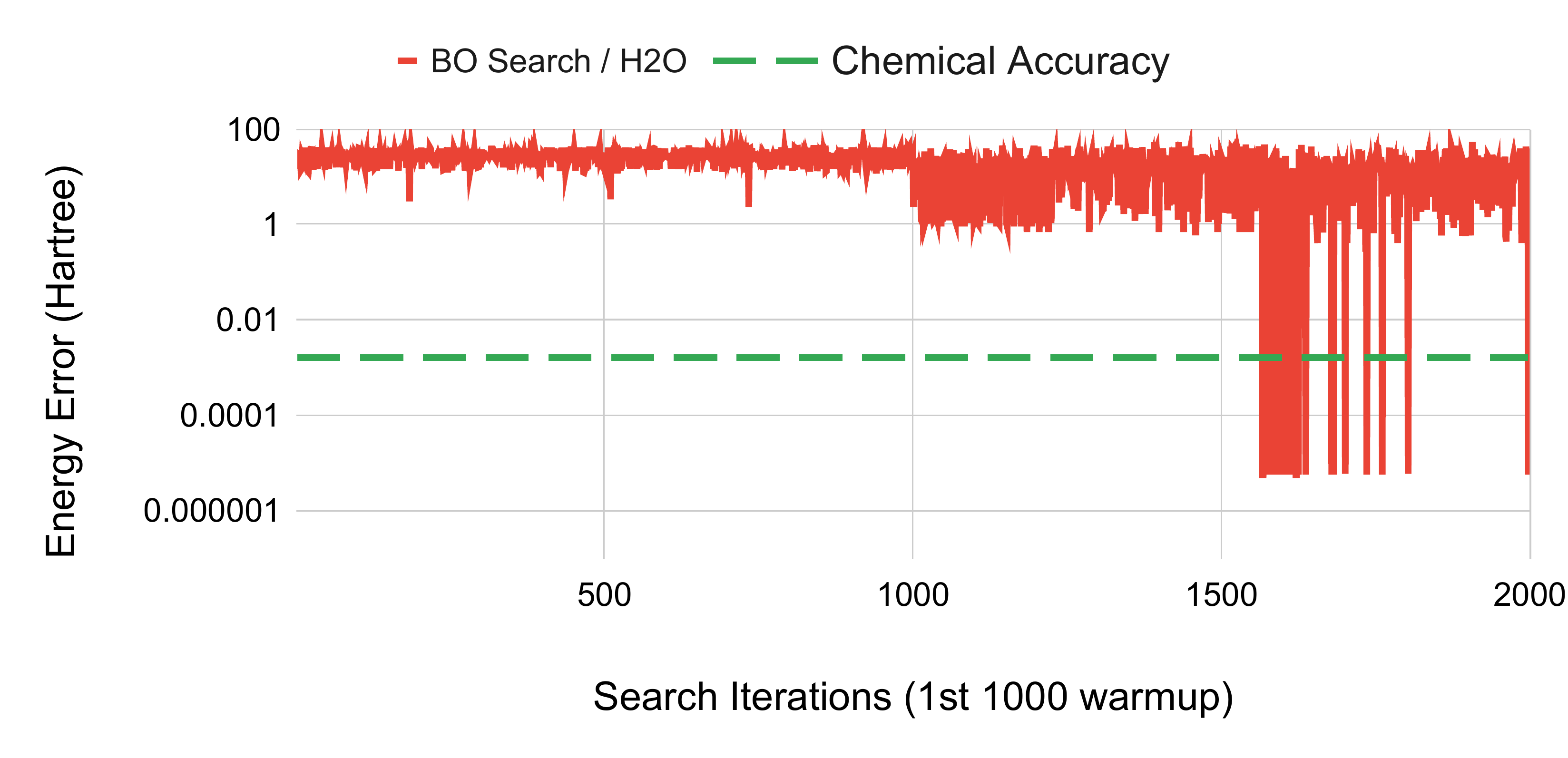}
\caption{H$_2$O ground state energy estimation via CAFQA's Clifford ansatz discrete search at bond length of 4{\AA} (4x equilibrium). The first 1000 iterations are a warm up period. Beyond this, the search can achieve estimations within the chemical accuracy target in an additional 600 iterations. In this instance, post-CAFQA variational tuning on a quantum machine is not required.}
\label{fig:BO}
\end{figure}

\section{Discrete Search over the Clifford Space}
\label{DS-proposal}
\label{bm_ds}

To efficiently search through the discrete Clifford parameter space, CAFQA requires a sample-efficient search technique to find the performant ansatz parameters as quickly as possible. 
Bayesian optimization is one such technique that actively and intelligently queries the most informative samples at each round to reduce the number of samples required~\cite{frazier2018tutorial}. 
Due to this merit, Bayesian optimization has been successfully applied to different domains in computer systems optimization such as compiler tuning~\cite{nardi2019practical}, resource allocation~\cite{Clite2020}, and configuration optimization~\cite{dingbayesian,ding2021,cherrypick2017,Bliss2021}.
Bayesian optimization iteratively alternates between intelligent sampling and model updates. 
As such, it includes two components: a surrogate model and an acquisition function. 
The surrogate model tries to learn the unknown underlying function that maps the search parameters to the problem objective (e.g., ground-state energy).
The acquisition function is the search strategy that selects the next sample to query to update the surrogate model. 

CAFQA searches through the Clifford space with Bayesian optimization to identify optimal Clifford gates for the tunable circuit parameters. 
Each tunable parameter is able to take one of four rotational angles as was shown in Fig.\ref{fig:ansatz}.
This creates a discrete search space complexity of $O(4^{\# params})$ that scales exponentially in the number of parameters, although it is considerably smaller than the entire quantum tuning space. 
While Bayesian Optimization efficiency can degrade with increased number of search parameters (or dimensionality)~\cite{li2017hyperband}, it is still observed to be effective in searching through the Clifford space since each parameter chooses only from four different rotational angles.

Since the Clifford parameter space is discrete, CAFQA chooses the random forest as the surrogate model as it is flexible enough to model the discrete space and scales well~\cite{nardi2019practical}. 
CAFQA uses a greedy acquisition function~\cite{nardi2019practical,ding2021} to select samples with the lowest energy estimates predicted from the surrogate model. 
Empirically, the combination of the random forest surrogate model and the greedy acquisition function gives highly accurate results, as illustrated in Section \ref{eval}.
Details on the implementation of the search algorithm can be found in \cite{nardi2019practical}.
Here, we limit ourselves to an illustrative example.

Fig.\ref{fig:BO} shows the discrete search employed by CAFQA to produce a Clifford Ansatz for H$_2$O ground state energy estimation at a bond length of 4{\AA}.
The first 1,000 iterations are a warm-up period, which involves randomly sampling and mapping the search space, a key component to BO.
The search algorithm then uses these random samples to efficiently search the parameter space.
In the figure, note that as soon as the random sampling is complete, the search algorithm begins to find better expectation values compared to random.
A potentially global minimum is found after an additional 600 search iterations. 
Notably, in this use case, the identified minimum is well within the chemical accuracy requirements.
Although 2000 iterations are shown here, the search can be constrained by a tuning budget or by the saturation of the obtained minimum.

\section{Methodology}
\label{6-method}

\begin{table}[t!]
\centering
\caption{VQA applications and their characteristics. }
\resizebox{\columnwidth}{!}{%
\centering
\begin{tabular}{|l|l|l|l|l|}
\hline
\textbf{App} &
  \textbf{\# Qu.} &
  \textbf{\begin{tabular}[c]{@{}l@{}}Bond Len.\\ (Eqbm.)\end{tabular}} &
  \textbf{\begin{tabular}[c]{@{}l@{}}Bond Len.\\ (Range)\end{tabular}} &
  \textbf{\begin{tabular}[c]{@{}l@{}}Mol Orbitals\\ Total / Used\end{tabular}} \\ \hline
\textbf{H$_2$}         & 2  & 0.74 {\AA} & 0.37 - 2.96 {\AA} & 2 / 2    \\ \hline
\textbf{LiH}        & 4  & 1.6 {\AA}  & 0.8 - 4.8 {\AA}   & 4 / 3   \\ \hline
\textbf{H$_2$O}        & 12 & 1 {\AA}    & 0.5 - 4.0 {\AA}     & 7 / 7   \\ \hline
\textbf{H$_6$}         & 10 & 0.9 {\AA}  & 0.45 - 3.6 {\AA}  & 6 / 6    \\ \hline
\textbf{N$_2$}         & 12 & 1.09 {\AA} & 0.55 - 4.36 {\AA} & 10 / 7   \\ \hline
\textbf{Cr$_2$}        & 34 &    1.68 {\AA}    & 1.25 - 3.5 {\AA}              & 36 / 18 \\ \hline
\textbf{NaH}        & 12 & 1.9 {\AA}  & 0.95 - 7.6 {\AA}  & 10 / 7  \\ \hline
\textbf{H$_2$-S1}      & 18 & -      & -             & -       \\ \hline
\textbf{BeH$_2$}       & 12 & 1.32 {\AA} & 0.66 - 5.28 {\AA} & 7 / 7    \\ \hline
\end{tabular}%
}
\label{Table-1}
\end{table}

\emph{\textbf{Ground state energy estimation of molecules:}}
We use VQE to estimate the ground state energy of the following molecules: H$_2$, LiH, H$_2$O, H$_6$, Cr$_2$, N$_2$, NaH, H$_2$-S1 and BeH$_2$.
Hamiltonians are constructed in the STO-3G basis with parity mapping and Z2 symmetry / two qubit reduction.
Hamiltonians are constructed for spin corresponding to the singlet (0 unpaired electrons in the orbitals) electronic state, which usually has the lowest energy near equilibrium geometries (more on this in Section \ref{H6_eval}). 
We provide detailed evaluations for the first five and only mean accuracy results for the other three.
H$_2$-S1\_STO-3G\_singlet (H$_2$-S1) is obtained from Contextual Subspace VQE~\cite{kirby2021}. 
Details about these molecules and their representative Hamiltonians are provided in Table \ref{Table-1}. 

H$_6$ is known to be a prototypical strongly correlated molecule, thus widening the gap between ideal results and classical methods.
Also notable is the Chromium Dimer (Cr$_2$), which has long been a benchmark molecule for evaluating the performance of different computational methods due to its unusual bonding properties in its ground and excited states~\cite{vancoillie2016potential,elfving2020quantum}.
Cr$_2$ is especially challenging to simulate, requiring a system of as many as 72 qubits; therefore, we are unable to compare against its exact estimates.
Furthermore, we freeze the lower 18 (out of Cr$_2$'s 36 orbitals) to reduce the system to 34 qubits, to ease the burden of iterative tuning given our reasonable yet limited computational resources - but this is not a strict limitation.
Freezing lower orbitals is least detrimental to bond dissociation energy estimations -  electrons closer to the nucleus are tightly attached and have high ionization energies~\cite{nist}.

For all Hamiltonians above, we use a hardware-efficient SU2 parameterized circuit~\cite{IBM-SU2} with one layer of linear entanglement as ansatz. 
An example of this for 10 qubits is shown in Fig.\ref{fig:ansatz}.
Different initialization comparisons are performed on this circuit.

\emph{\textbf{Evaluation Comparisons:}}
We compare the following - 

\circled{1} \emph{CAFQA:} Our proposed approach, which uses a Clifford-only ansatz, and potentially produces the best possible stabilizer initial state for the target Hamiltonian. 

\circled{2} \emph{Exact:} The exact energy estimations computed classically (but possible only for small problem sizes).

\circled{3} \emph{Hartree-Fock (HF):}  HF is the best computational basis state for the target Hamiltonian under specified electron and spin preservation constraints. 


\emph{\textbf{Evaluation Metrics:}}
We evaluate CAFQA across four metrics detailed below:

\circled{1} \emph{Ground State Energy:} Potential energy as a function of nuclear coordinates, as estimated by different techniques, expressed in Hartree units.

\circled{2} \emph{Energy estimation accuracy:} Absolute energy difference between energy estimates from different techniques and exact estimates, expressed in Hartree units. Chemical accuracy region is shown in orange.

\circled{3} \emph{Recovered correlation energy:}  Percentage of the difference between the Exact energy and the Hartree-Fock limit that is recovered by CAFQA.

\circled{4}\ \emph{Relative accuracy:} Relative energy estimation accuracy between CAFQA and state-of-the-art HF (only Fig.\ref{fig:ovl}).

\emph{\textbf{Infrastructure:}}
The CAFQA framework is implemented in Python.
Hartree-Fock estimations are performed via Psi4~\cite{turney2012psi4} while CAFQA evaluations are performed using Qiskit~\cite{Qiskit}.
Qiskit interfaces with the PySCF library~\cite{sun2017pythonbased} in the process of constructing Hamiltonians from molecular specifications.
The discrete search to find the optimal Clifford gates is performed through Bayesian Optimization via the HyperMapper~\cite{nardi2019practical} framework.
Classical computations are predominantly carried out on the Google Compute Cloud.

\section{Evaluation}
\label{eval}
\label{eval_vqe}

\subsection{Detailed Molecular Analysis}
\label{eval_mol}
\begin{figure}


\subfloat[H$_2$ Energy\label{Energy.H2}]{\includegraphics[width=\columnwidth]{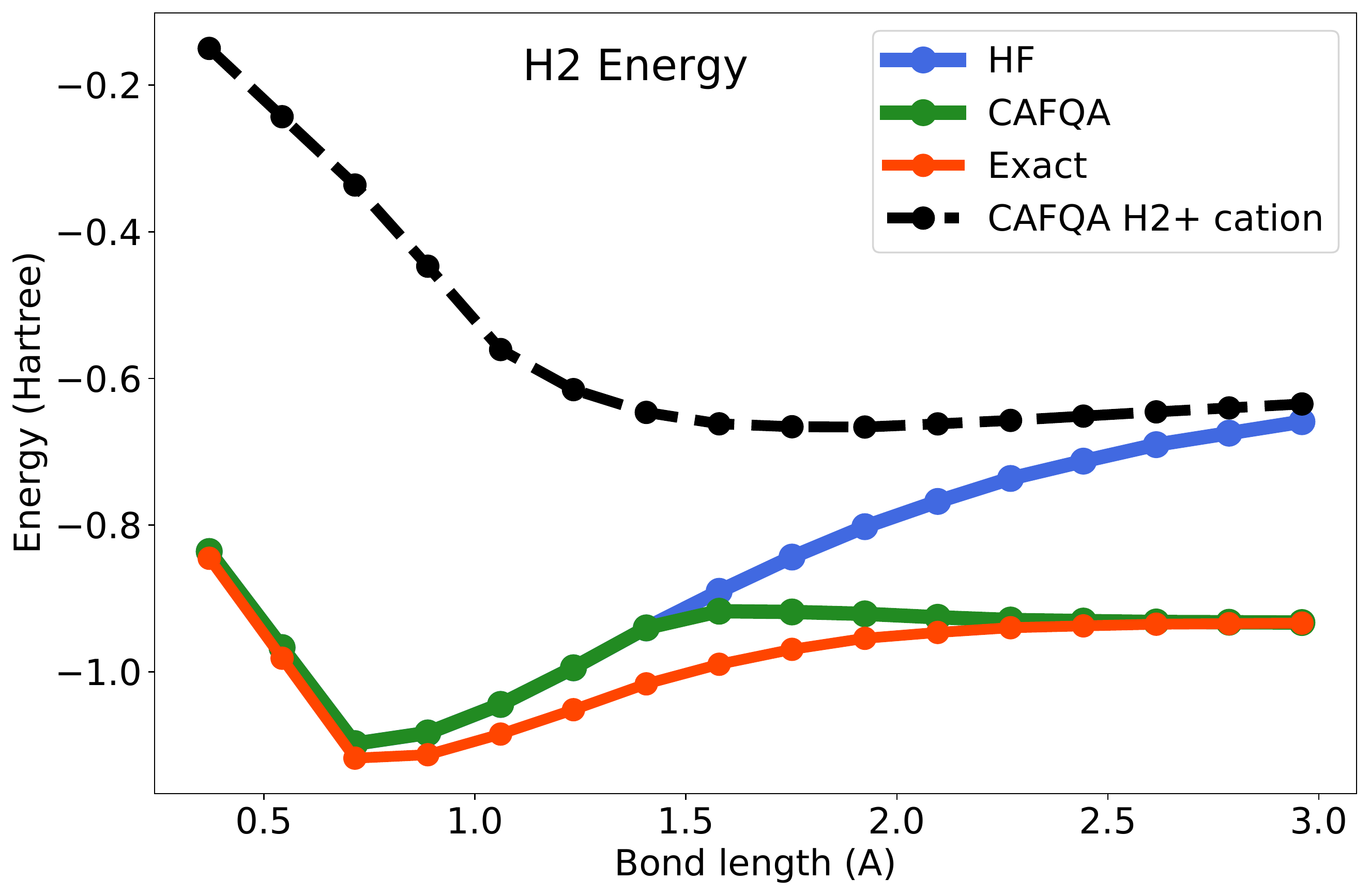}}

\subfloat[H$_2$ Accuracy\label{Acc.H2}]{\includegraphics[width=\columnwidth]{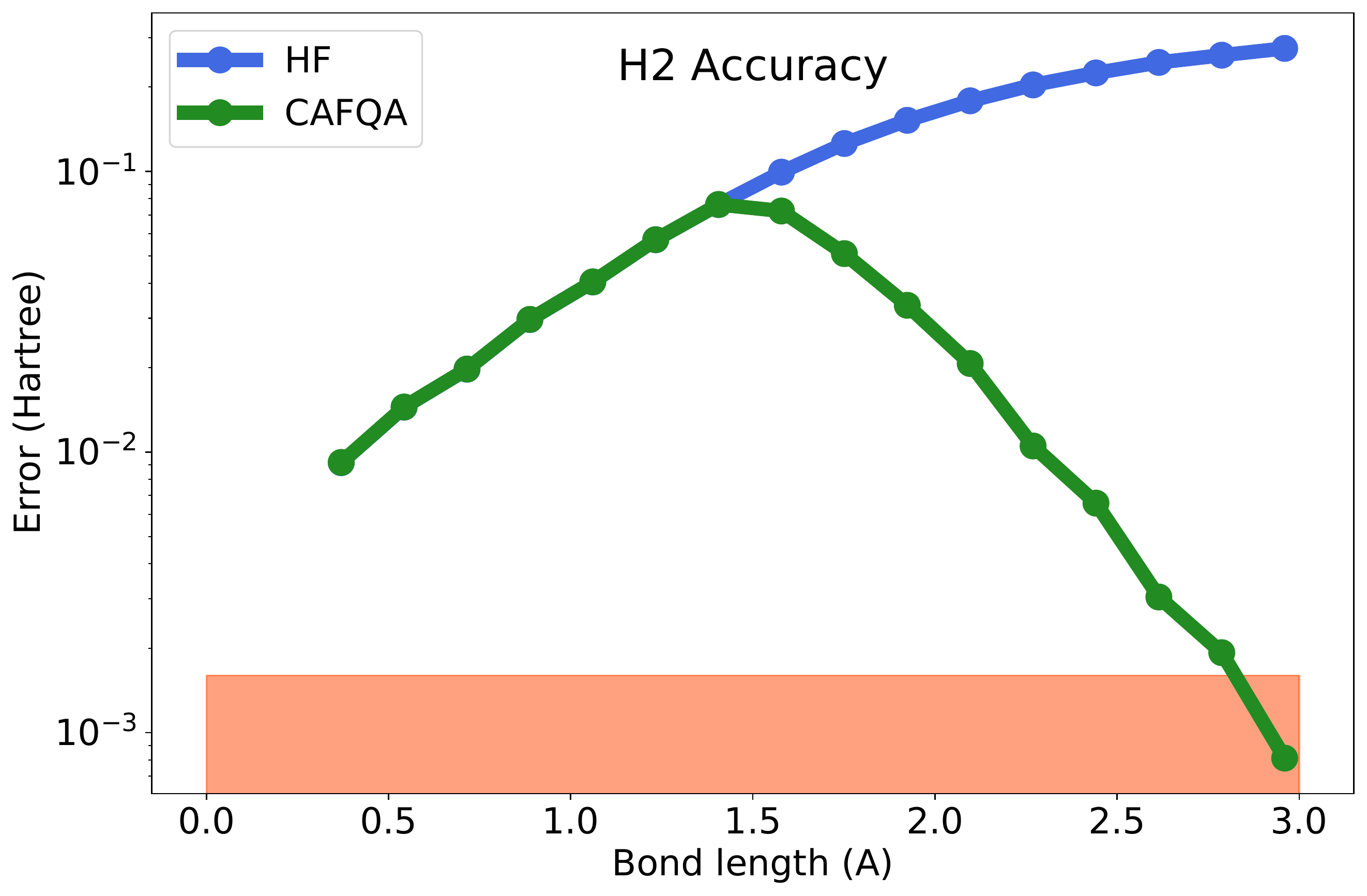}}

\subfloat[H$_2$ Correlation Energy\label{Corr.H2}]{\includegraphics[width=\columnwidth]{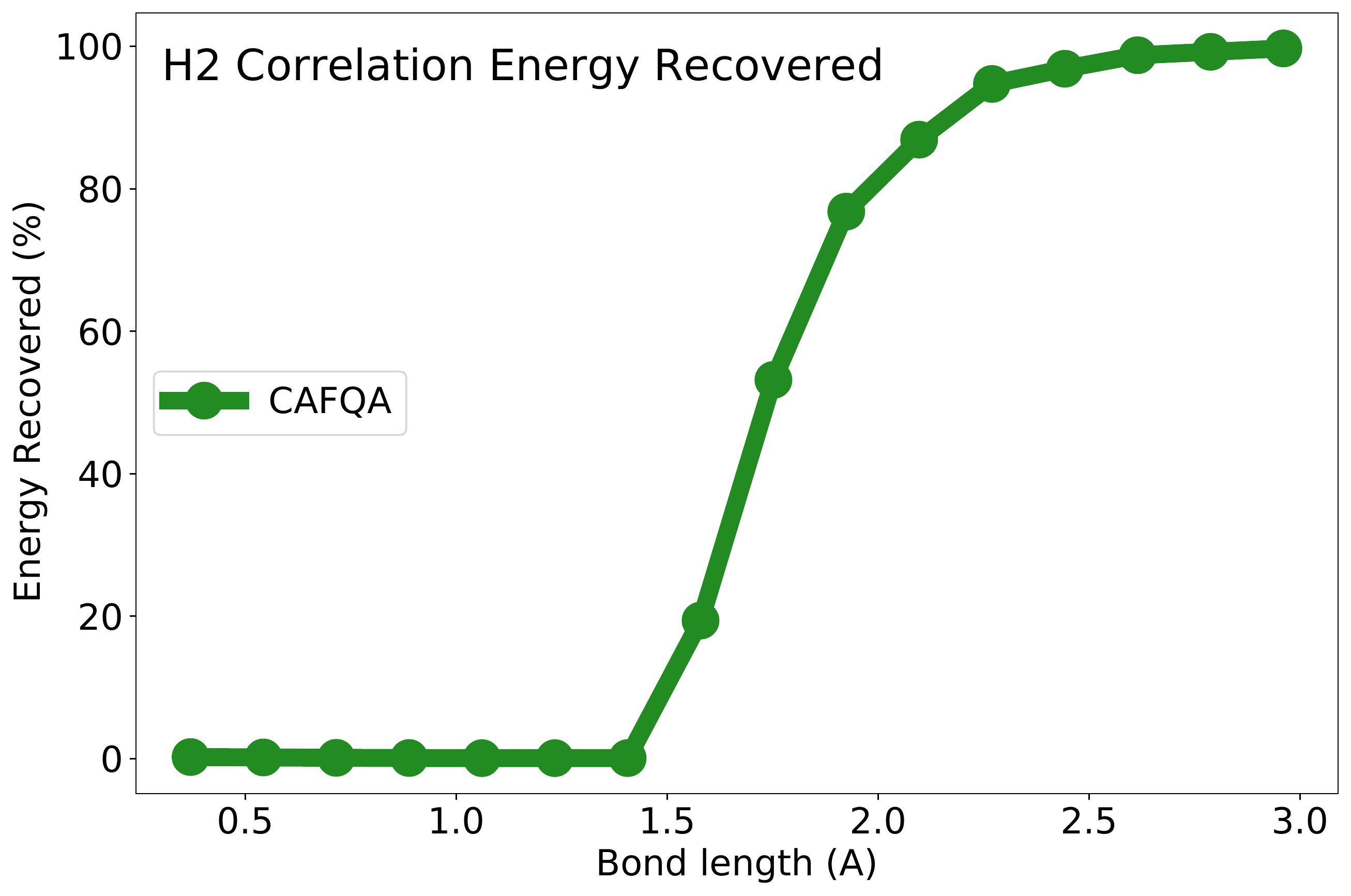}}

\caption{Dissociation curves for $H_2$. 
Evaluation of CAFQA in terms of ground state energy, energy estimation error and correlation energy recovered. Comparisons to Exact / Chemical Accuracy and Hartree-Fock are shown. Energy for $H_{2}^{+}$ cation is also shown.
}
\label{fig:eval1}

\end{figure}

\begin{figure}

\subfloat[LiH Energy\label{Energy.LiH}]{\includegraphics[width=\columnwidth]{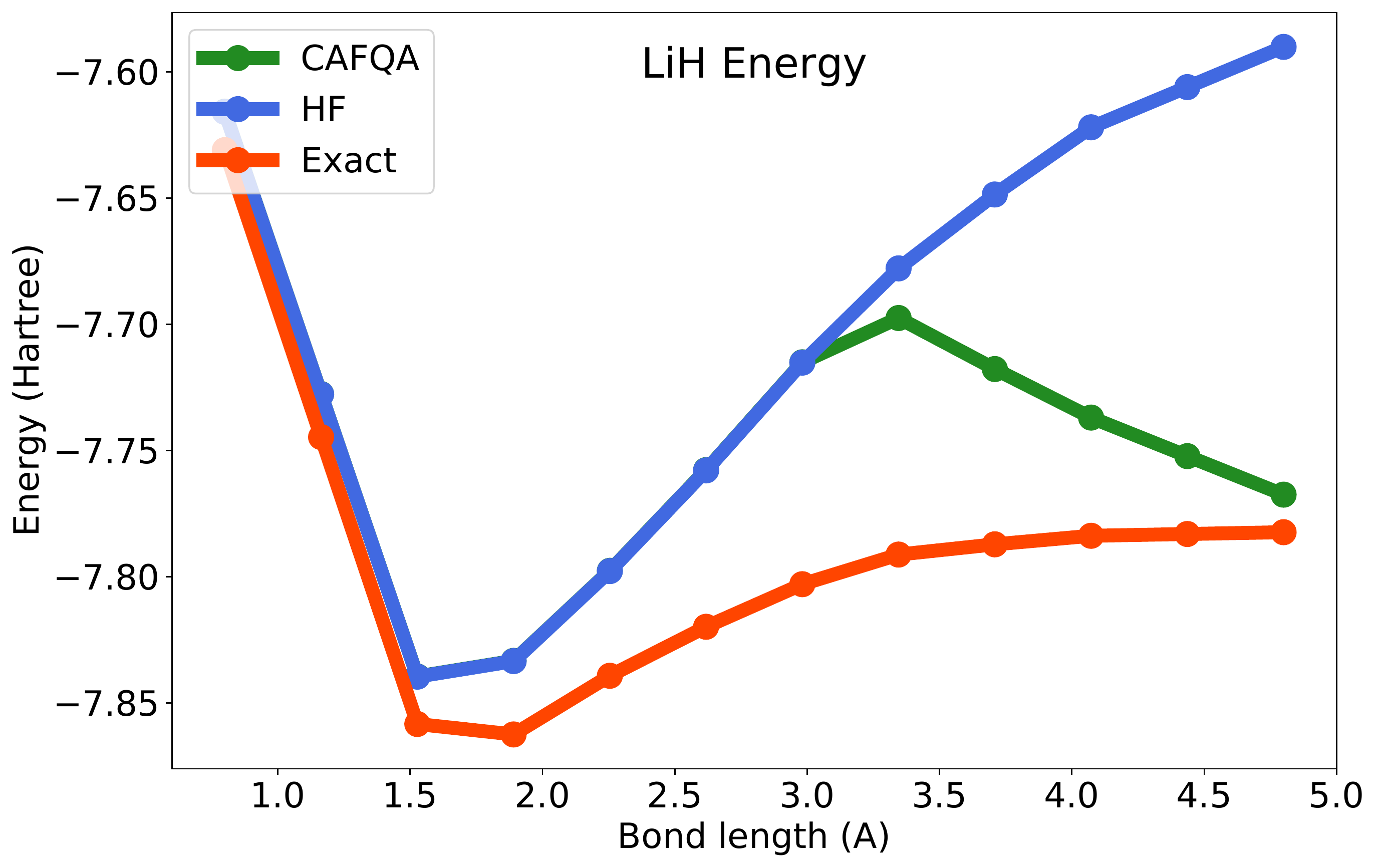}}

\subfloat[LiH Accuracy\label{Acc.LiH}]{\includegraphics[width=\columnwidth]{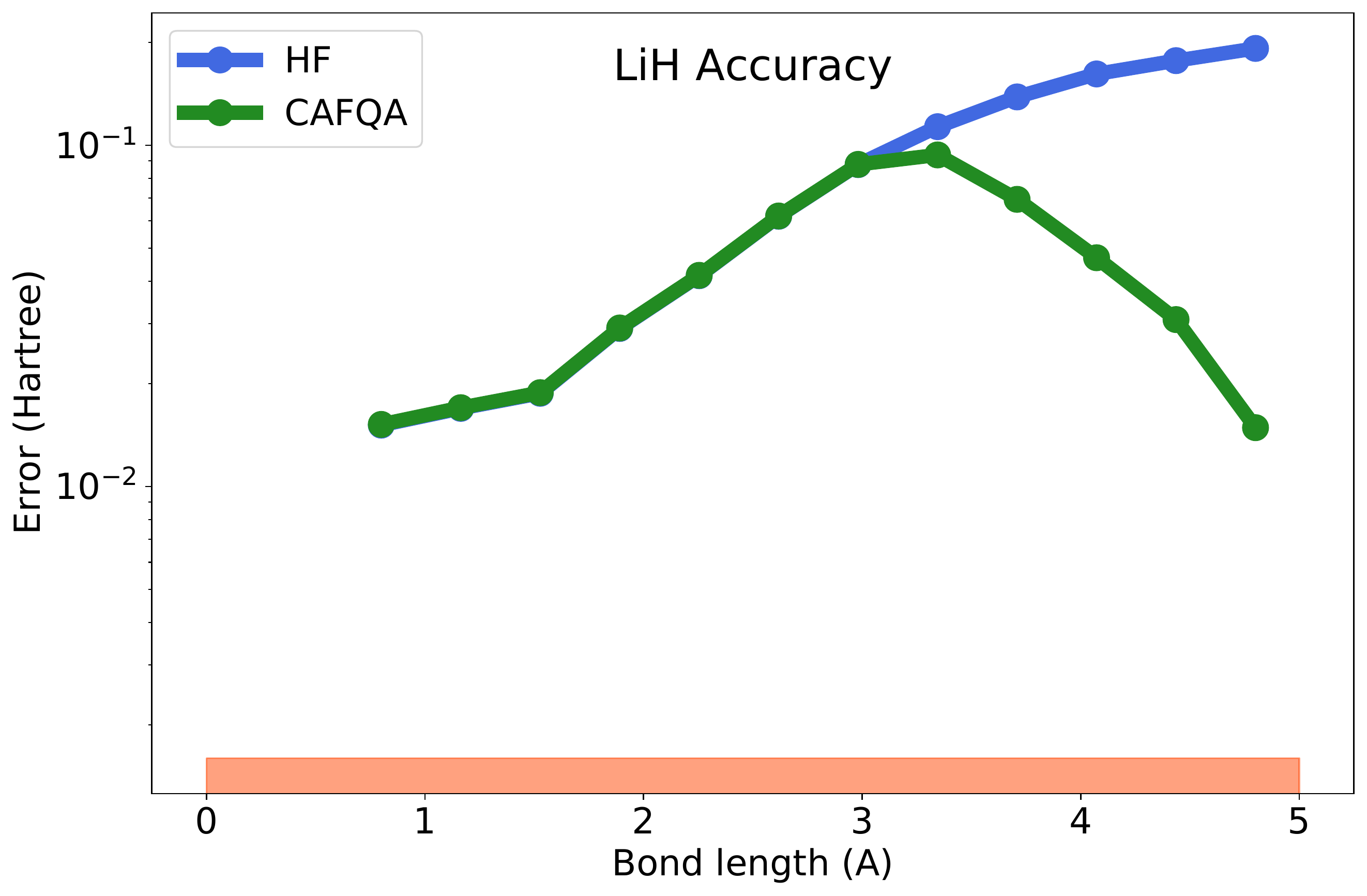}}

\subfloat[LiH Correlation Energy\label{Corr.LiH}]{\includegraphics[width=\columnwidth]{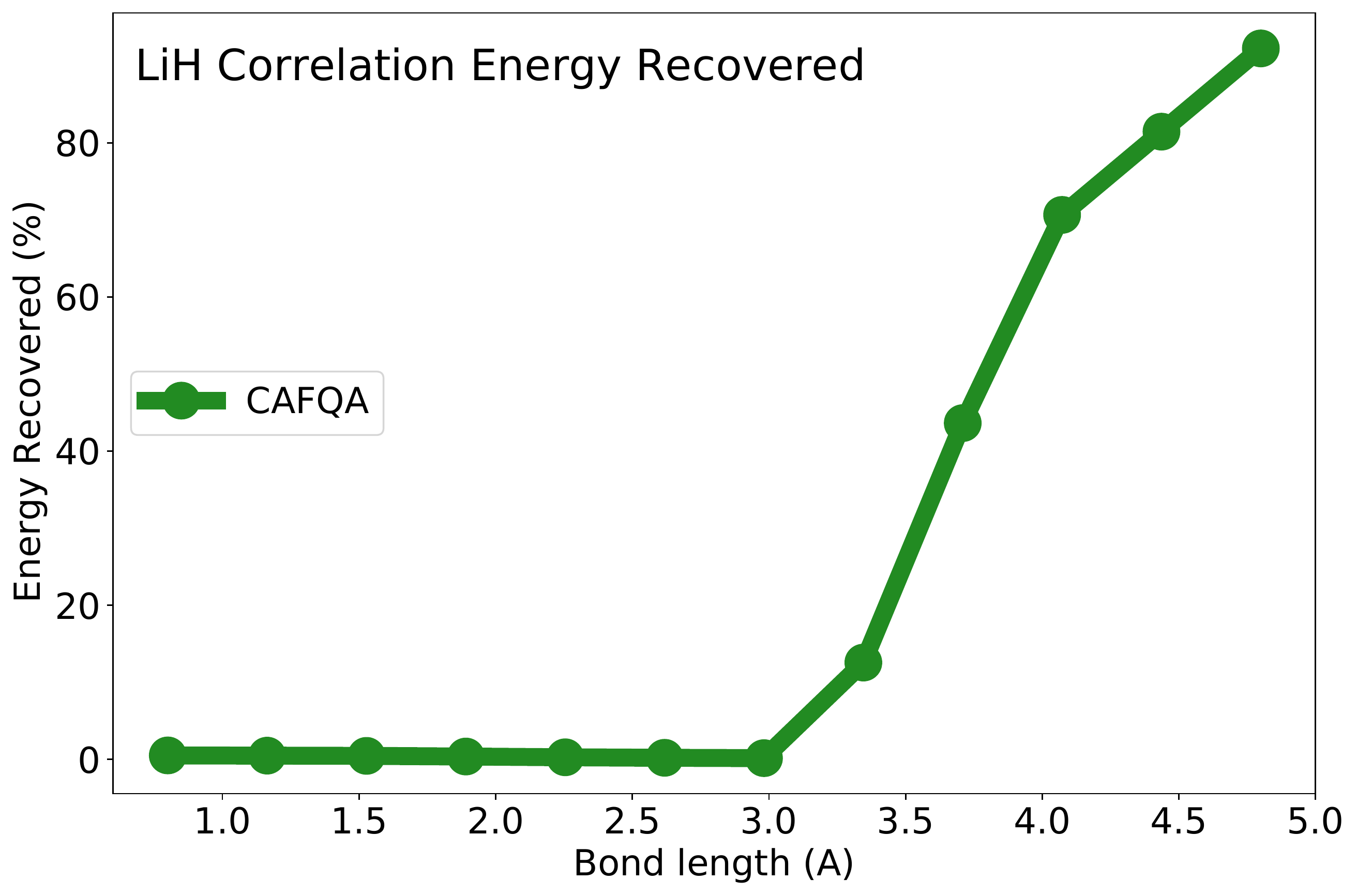}}

\caption{Dissociation curves for $LiH$. 
Evaluation of CAFQA in terms of ground state energy, energy estimation error and correlation energy recovered. Comparisons to Exact / Chemical Accuracy and Hartree-Fock are shown.
}
\label{fig:eval2}

\end{figure}

In Figures \ref{fig:eval1}-\ref{fig:eval4} we show ground state energy estimation through VQE for 4 molecules, over different bond lengths. 
Each figure shows different evaluation metrics for the target molecule molecule: the top subfigure shows the absolute ground state energy (in Hartree), the middle subfigure shows the error in energy estimation, and the bottom subfigure shows the correlation energy recovered by CAFQA over HF. 
We compare CAFQA in green against exact evaluations in orange; and  Hartree-Fock initialization (HF) in blue.
More details on these molecules, metrics and comparisons are discussed in Section \ref{6-method}.

\subsubsection{H$_2$}
First, we look at H$_2$ shown in Fig.\ref{fig:eval1} (a)-(c).
In Fig.\ref{fig:eval1} (a) and (b), we see that the HF steadily deviates away from exact energies as bond lengths increase.
This is not surprising, as HF is known to work best at / near equilibrium geometry, as discussed in Section \ref{bm_init}.
CAFQA matches HF at low bond lengths but achieves lower energy estimates at higher bond lengths, thus being closer to exact estimates.

Fig.\ref{fig:eval1} (a) also shows CAFQA energy estimates for the H$_{2}^{+}$ cation.
The cation is in a higher energy state than its neutral counterpart; this is intuitive, as H$_2$ does not naturally ionize. 
For a given molecular system, the Fock (i.e., energy) space represented by the problem Hamiltonian combines the energy spaces of the molecular forms with all possible numbers of electrons and all electron-spin combinations~\cite{ryabinkin2018constrained}.
Thus, when solving for the ground-state energy of higher-energy cations and anions, some explicit enforcement of electron preservation constraints on VQE is often required (whereas this is usually not required for the lowest-energy neutral molecule).
For H$_{2}^{+}$, this means that only electronic energies corresponding to a one electron system should be considered for VQE.
With CAFQA, we impose electron count constraints through the search objective function.
Thus, CAFQA ensures that any required constraints are maintained, along the lines of prior work~\cite{ryabinkin2018constrained}.

Fig.\ref{fig:eval1} (b) shows that CAFQA's error is always less than 10$^{-1}$ Hartree, and is able to achieve estimates near chemical accuracy.
On the other hand, as the bond lengths increase, the HF error is 10$^{-1}$ Hartree and greater.
Thus, the benefits of CAFQA are clearly evident.
Finally, Fig.\ref{fig:eval1} (c) shows that CAFQA is able to recover up to 99.7\% of the correlation energy as bond lengths increase.

Overall, CAFQA achieves more accurate energy estimates than HF's best computational basis state and is able to incorporate the expectation of non-diagonal Pauli terms (as discussed in Section \ref{Qual}). 
It produces a non-computational basis state which is uncommon in other classical approaches - an example of this was illustrated earlier for LiH in Fig.\ref{fig:LiH-Terms}.
Furthermore, the stabilizer state produced appears intuitively suited to a molecule such as H$_2$.
The opposing attractive and repulsive forces of similar strengths acting on the electrons (especially at higher bond lengths) can result in optimal configurations bearing resemblance to stabilizers.
Further examination at a molecular level is beyond our current scope.

\subsubsection{LiH}
\label{eval_lih}
Next, we examine LiH shown in Fig.\ref{fig:eval2} (a)-(c).
In Fig.\ref{fig:eval2} (a) and (b), we see that HF deviates considerably from the exact value at medium-high bond lengths, but is closer to the exact value at low bond lengths.
As before, CAFQA is closest to exact, especially accurate at low and high bond lengths, but always achieves equal or more accurate energy estimates compared to HF. 
Fig.\ref{fig:eval2} (b) shows that CAFQA's estimation error is usually in the 10$^{-1}$ to 10$^{-2}$ Hartree range, with higher accuracy at high bond lengths.
Finally, Fig.\ref{fig:eval2} (c) shows that CAFQA is able to recover up to 93\% of the correlation energy at medium-high bond lengths.
As before, improving beyond HF, CAFQA is able to produce a non-computational basis state as its ansatz initialization state.

\begin{figure}


\subfloat[H$_2$O Energy\label{Energy.H2O}]{\includegraphics[width=\columnwidth]{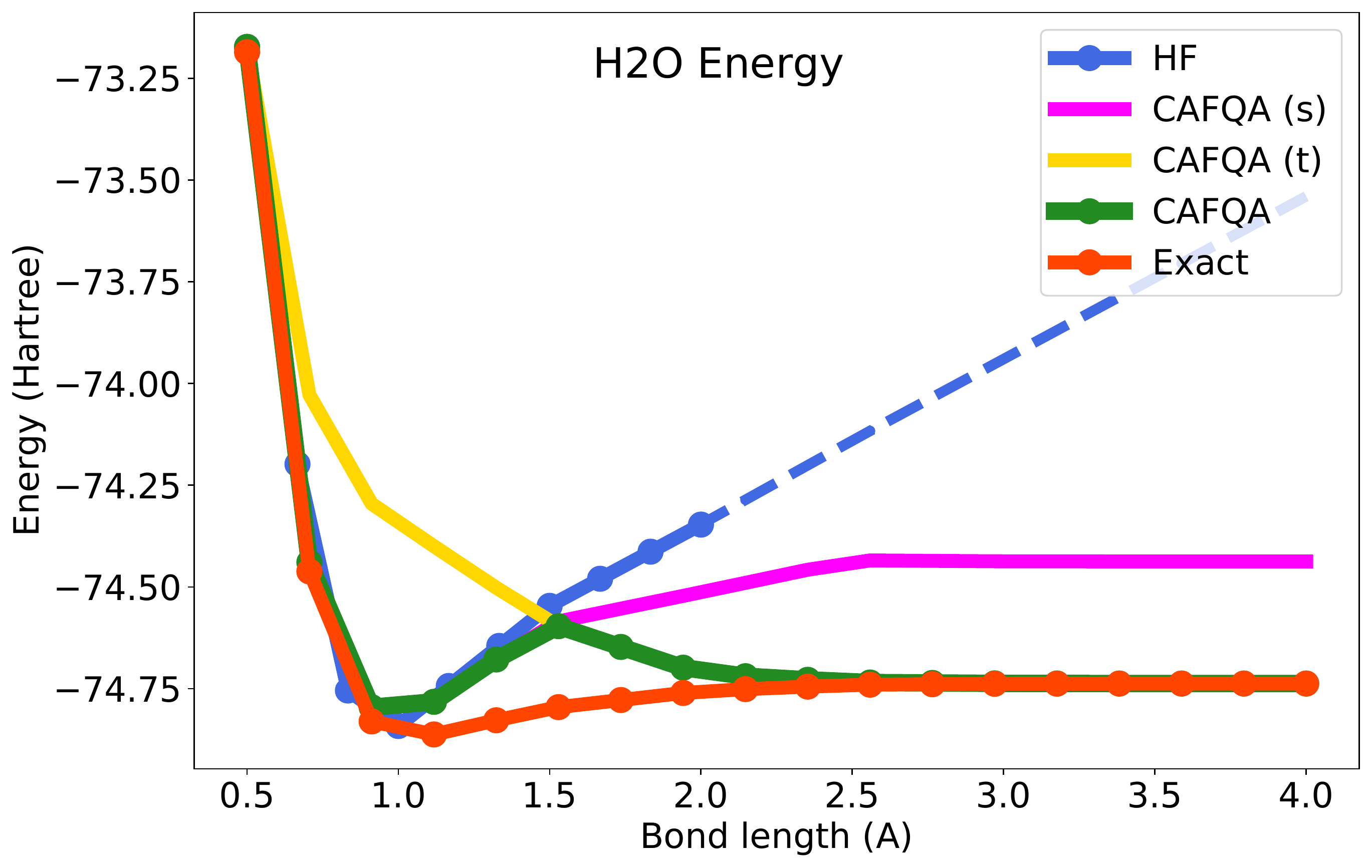}}

\subfloat[H$_2$O Accuracy\label{Acc.H2O}]{\includegraphics[width=\columnwidth]{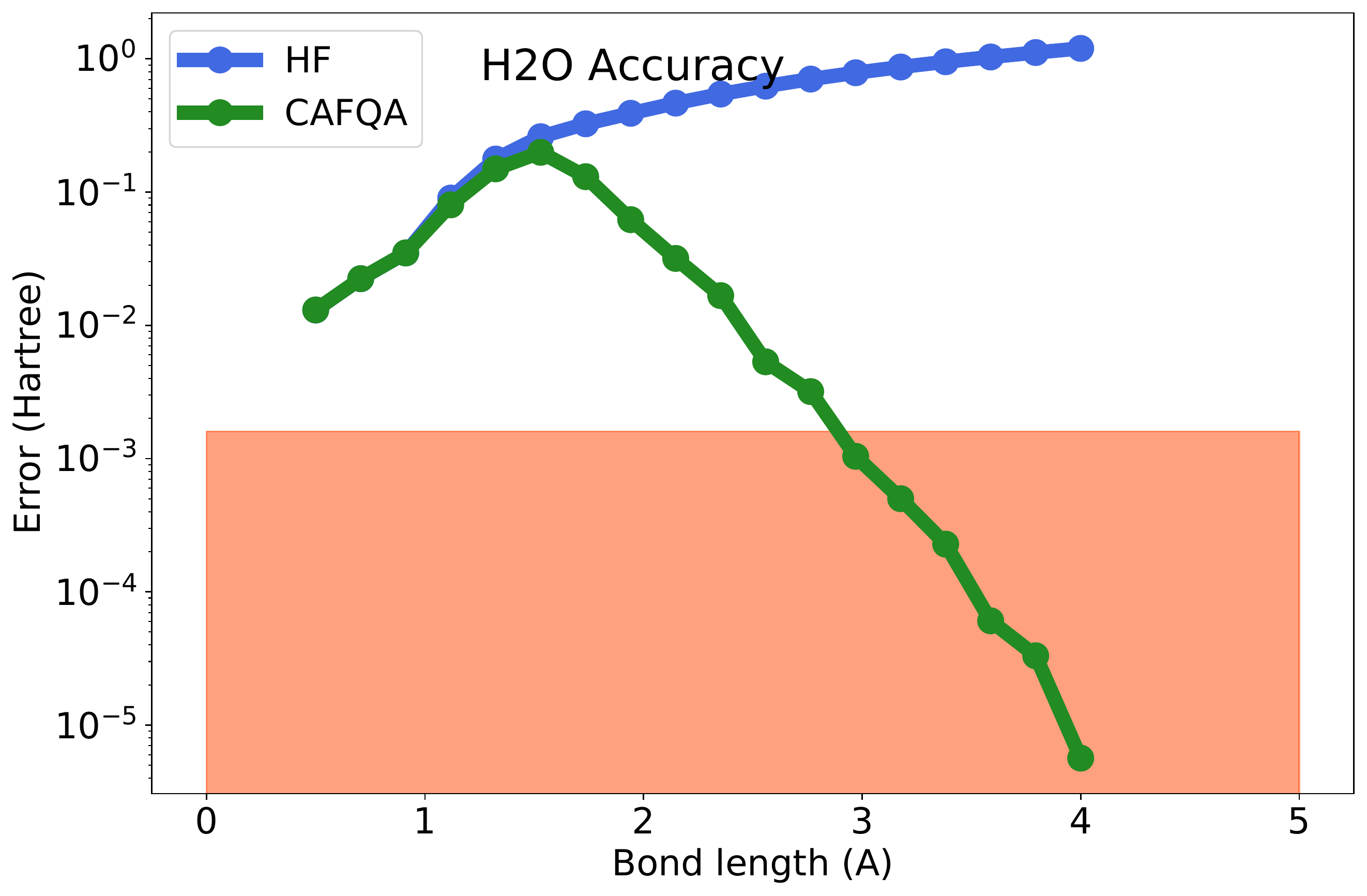}}

\subfloat[H$_2$O Correlation Energy\label{Corr.H2O}]{\includegraphics[width=\columnwidth]{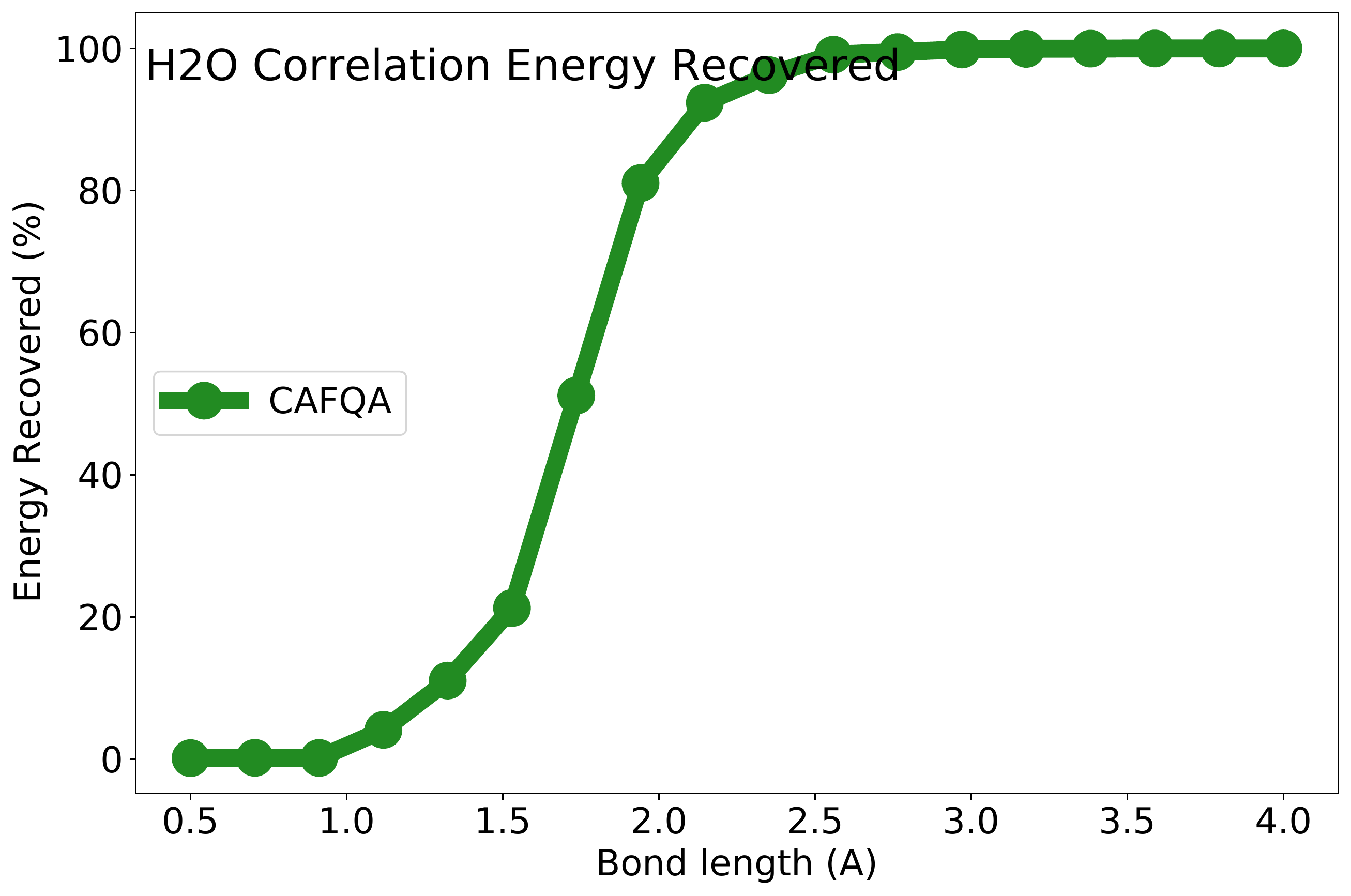}}

\caption{Dissociation curves for $H_{2}O$. 
Evaluation of CAFQA in terms of ground state energy, energy estimation error and correlation energy recovered. Comparisons to Exact / Chemical Accuracy and Hartree-Fock are shown. CAFQA (s) and (t) refer to energies corresponding to singlet and triplet states. 
}
\label{fig:eval3}

\end{figure}

\subsubsection{H$_2$O}
Next, we look at H$_2$O shown in Fig.\ref{fig:eval3} (a)-(c).
The first aspect to be noted is that the HF Psi4 estimations do not converge at high bond lengths, so we extrapolate the expected trend as shown in Fig.\ref{fig:eval3} (a).
In Fig.\ref{fig:eval3} (a) and (b), we see that HF steadily deviates away from the exact at higher bond lengths.
CAFQA matches HF at lower bond lengths but achieves considerably better energy estimates compared to HF at medium / higher bond lengths.

It is interesting to observe the kink in the energy estimation near a bond length of 1.5 {\AA}. 
This appears to match the prior observation that this is caused by the energy crossing of the lowest singlet (0 unpaired electrons) and triplet (2 unpaired electrons) electronic states for the H$_2$O molecule~\cite{ryabinkin2018constrained}.
Accordingly, the singlet and triplet states from CAFQA are plotted in pink and yellow, respectively.

Next, Fig.\ref{fig:eval3} (b) shows that CAFQA is able to impressively achieve chemical accuracy at higher bond lengths while HF has high error in the range of 10$^{-1}$ Hartree.
At lower bond lengths, CAFQA achieves error rates of around 10$^{-2}$ Hartree.
Finally, Fig.\ref{fig:eval3} (c) shows that CAFQA is able to recover up to 99.998\% of the correlation energy over HF as the bond lengths increase.

\begin{figure}

\centering

\subfloat[H$_6$ Energy\label{Energy.H6}]{\includegraphics[width=\columnwidth]{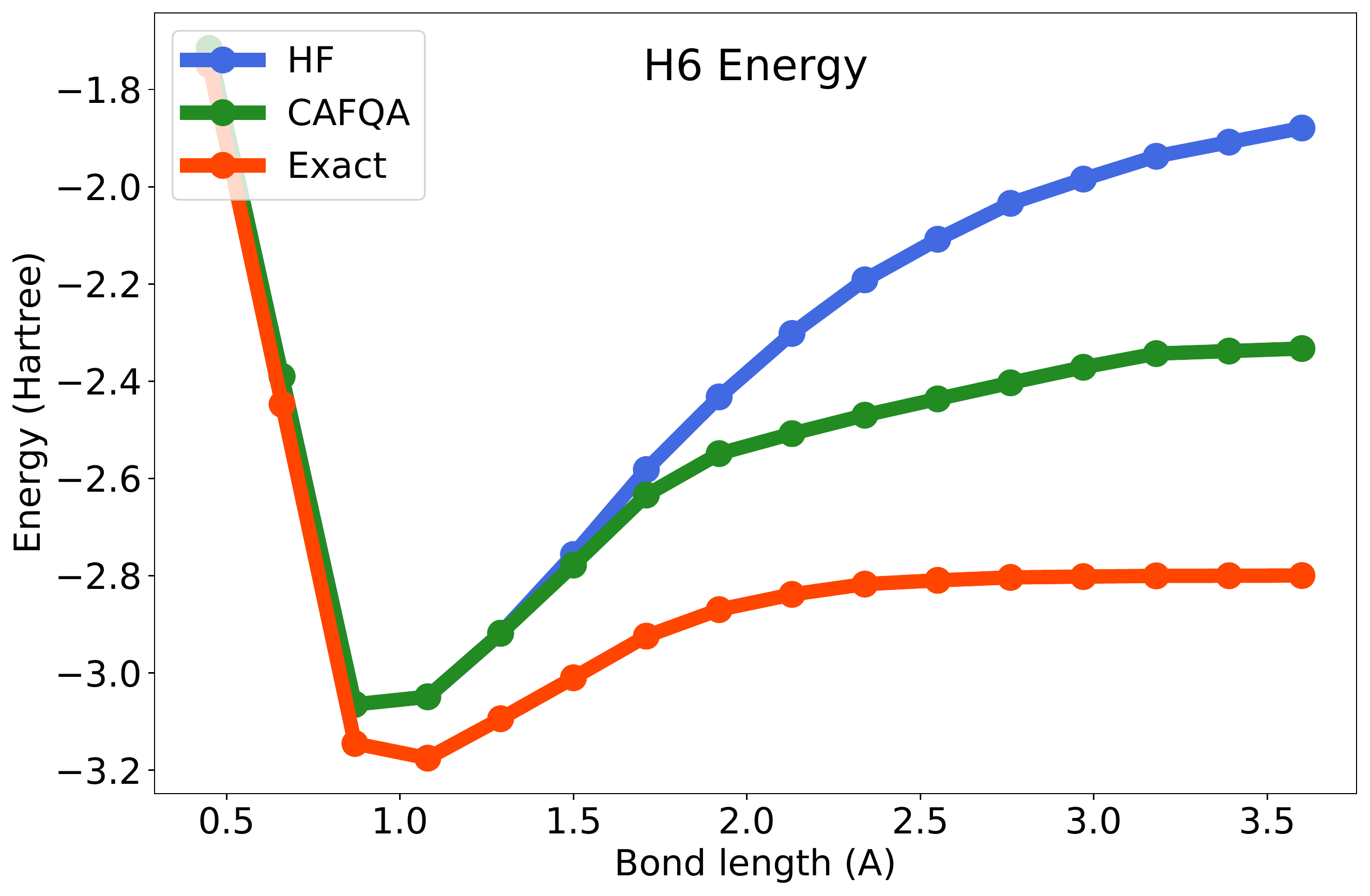}}

\subfloat[H$_6$ Accuracy\label{Acc.H6}]{\includegraphics[width=\columnwidth]{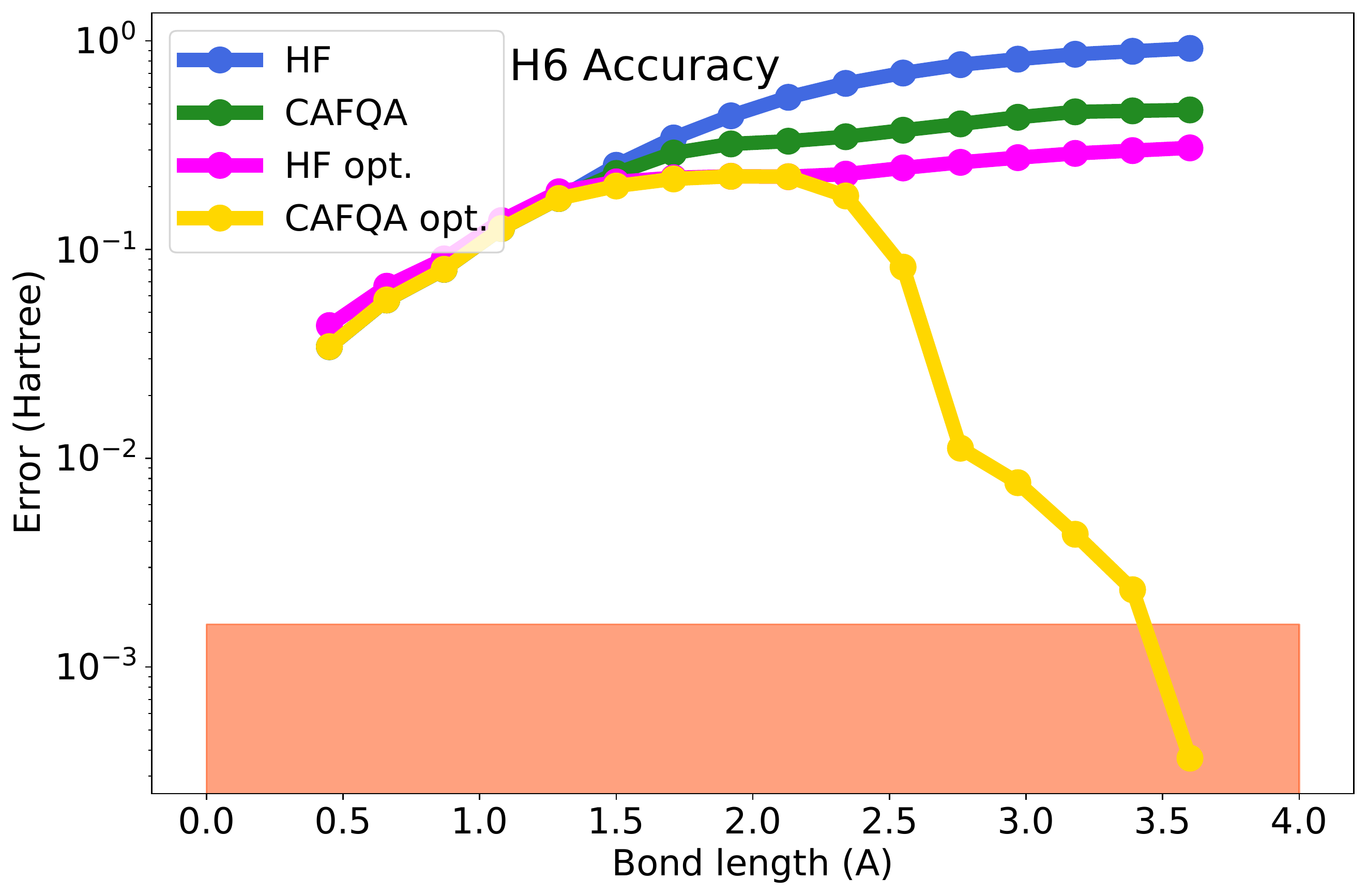}}

\subfloat[H$_6$ Correlation Energy\label{Corr.H6}]{\includegraphics[width=\columnwidth]{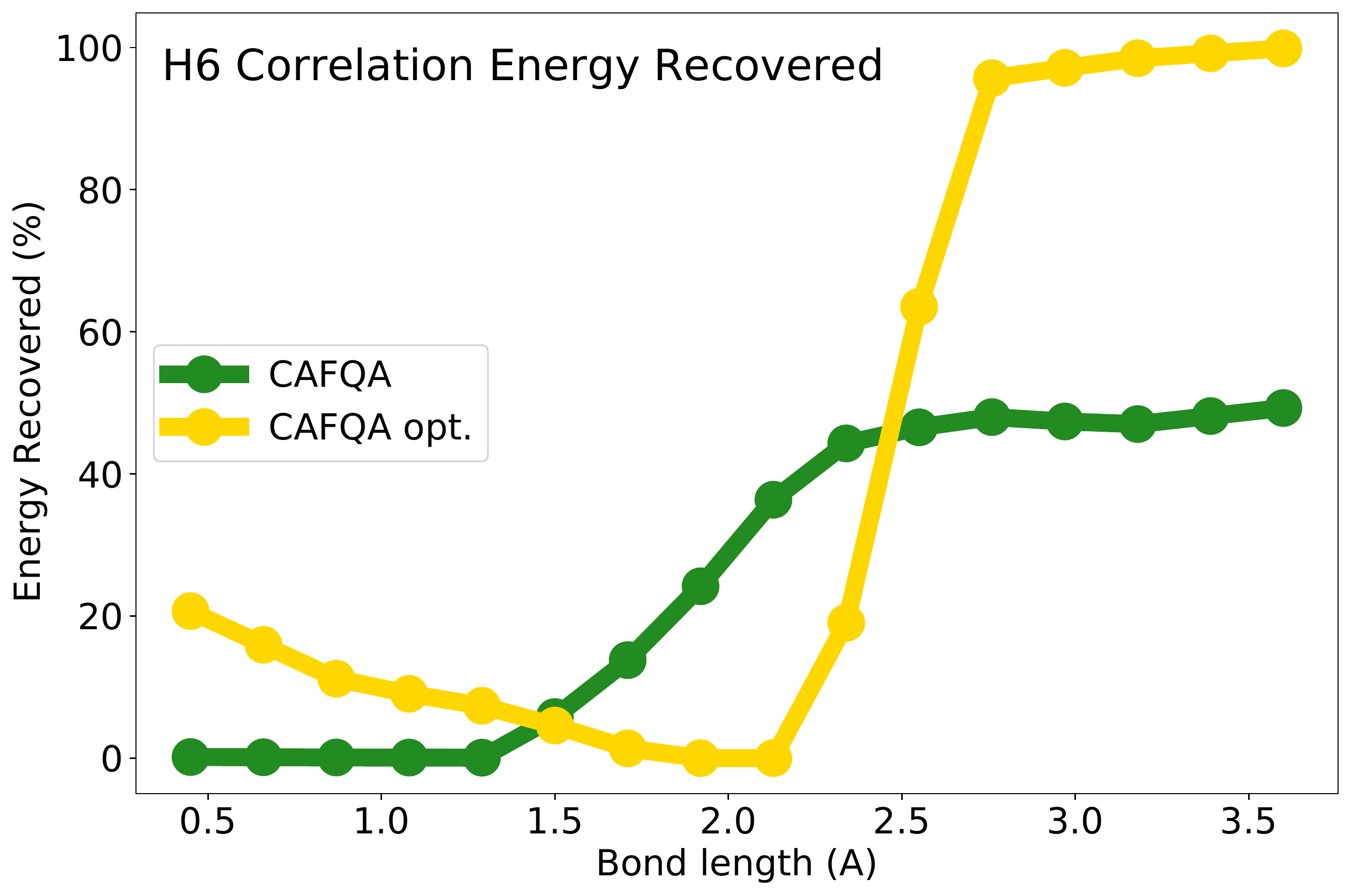}}



\caption{Dissociation curves for $H_6$. 
Evaluation of CAFQA in terms of ground state energy, energy estimation error and correlation energy recovered. Comparisons to Exact / Chemical Accuracy and Hartree-Fock are shown. CAFQA opt. refers to best estimates over multiple spin-optimized Hamiltonians.
}
\label{fig:eval4}

\end{figure}

\subsubsection{H$_6$}
\label{H6_eval}
Next, we look at H$_6$ shown in Fig.\ref{fig:eval4} (a)-(c).
In Fig.\ref{fig:eval4} (a) and (b) we show two versions of HF and CAFQA.
The traditional HF and CAFQA results correspond to Hamiltonians generated for orbitals optimized for the singlet electronic state as mentioned in Section \ref{6-method}.
On the other hand, the optimized results (labeled `opt.') are produced by generating unique Hamiltonians for different spins and with orbitals  optimized accordingly. 
Then, the HF and CAFQA results corresponding to the lowest estimates across all Hamiltonians are selected for every bond length.
It is evident that `opt.' results produce better energy estimates compared to those obtained from the singlet-optimized Hamiltonian at higher bond lengths.
This shows that optimizing the Hamiltonian to the best extent / as widely as possible can considerably improve VQE estimation, at the cost of increased compute. 
The HF techniques and CAFQA are far from exact while CAFQA opt. is near exact at high bond lengths.
Deviations from the exact value are not surprising, as H$_6$ has high correlation energy.

Fig.\ref{fig:eval4} (b) shows that CAFQA errors are in the 10$^{-1}$ Hartree range (except at high bond lengths when CAFQA opt. achieves the chemical accuracy range), thus clearly requiring quantum exploration on a quantum device post-ansatz selection.
Finally, Fig.\ref{fig:eval4} (c) shows that CAFQA is able to recover up to 50\% of the correlation energy over HF as the bond lengths increase, while CAFQA opt. can achieve near 100\% at high bond lengths.
CAFQA is again able to produce a non-computational basis state as its ansatz initialization state, although it is evident that exploration of only the Clifford space limits accuracy.

\begin{figure}[t]
\centering
\includegraphics[width=\columnwidth,trim={0cm 0cm 0cm 0cm}]{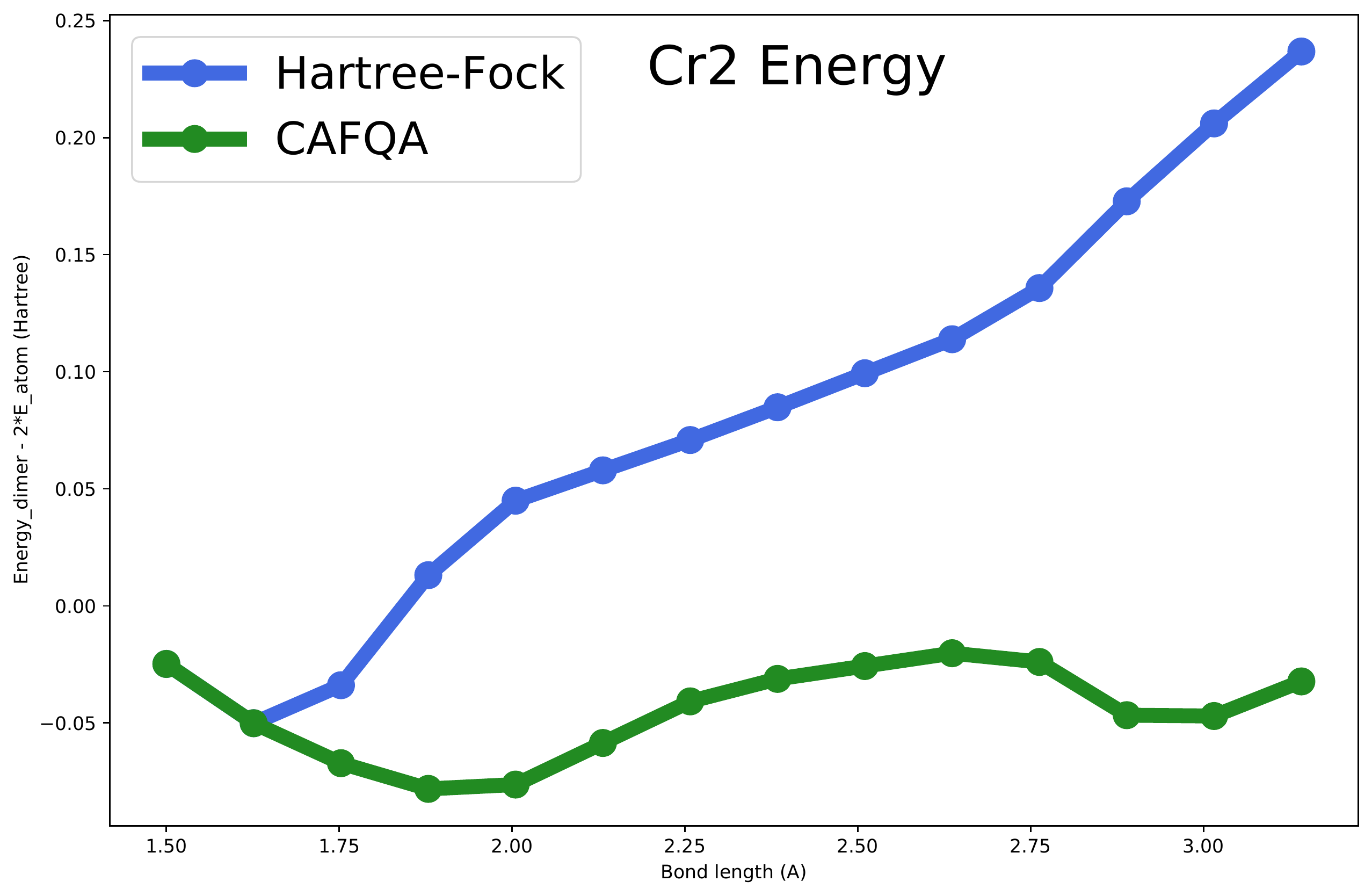}

\caption{Ground state energy for Cr$_2$ with lower 18 out of 36 orbitals frozen. CAFQA's estimates are limited by computational time / resources, thus Clifford estimates can be suboptimal. Comparison to HF is shown.}
\label{fig:cr2}
\end{figure}

\subsubsection{Cr$_2$}
\label{dimer}
In Fig.\ref{fig:cr2} we evaluate CAFQA on Cr$_2$ for ground state energy estimation.
For Cr$_2$, we are unable to generate exact evaluations, since the size of the system is too large (discussed in Section \ref{6-method}).
We compare CAFQA to HF.
It is evident from the figure that CAFQA consistently achieves better initial energy estimates compared to HF across all bond lengths.
In addition, CAFQA has resemblance to experimental estimates~\cite{vancoillie2016potential}, although there are some limitations to the comparison due to the orbital freezing that we utilize (discussed in Section \ref{6-method}), as well as a lack of sufficient molecular specifications.
We note that very recent work~\cite{Cr2_2022} (subsequent to CAFQA) showed high accuracy computational prediction of the $Cr_2$ potential energy curve, consistent with experimental data.
Comparisons to this work are worth pursuing.

Due to the size of the problem, we are limited by resources in running the Bayesian Optimization search extensively at each bond length. 
Limited search means that CAFQA at some bond lengths can produce sub-optimal estimates. 
Our current estimates are obtained over a 1-week period but these estimations can improve with more memory / compute, more execution time, better search strategies, efficient parallelization, limited  exploration of non-Cliffords --- more in Section \ref{FW}.
We reemphasize that CAFQA is only a first step in VQA tasks, with the primary goal of producing an ansatz initial state well suited to further quantum exploration on a quantum device. 

\label{eval2}

\begin{figure}[t]
\centering
\includegraphics[width=\columnwidth,trim={0cm 0cm 0cm 0cm}]{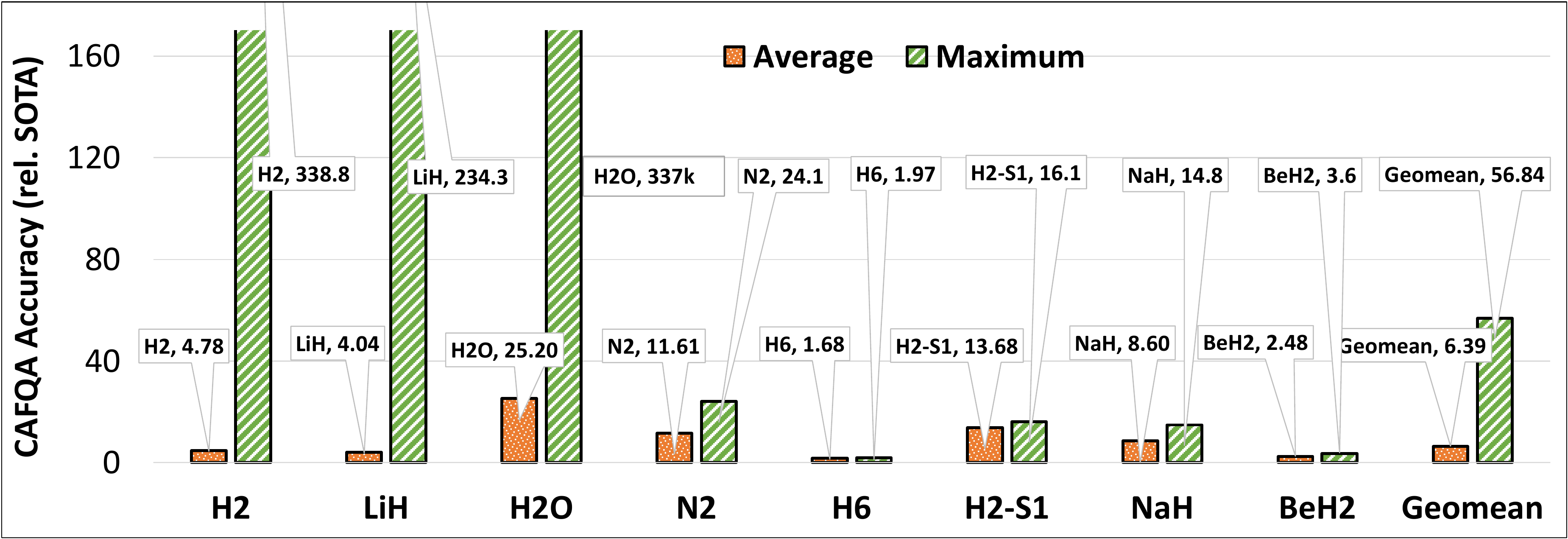}

\caption{CAFQA accuracy compared to state-of-the-art Hartree-Fock. `Average': Relative error reduction averaged over all bond lengths for each molecule. `Maximum': Maximum error reduction for each molecule, usually at the greatest bond length.}
\label{fig:ovl}
\end{figure}

\subsection{Relative Accuracy Compared to SOTA}
\label{eval_ovl}
Fig.\ref{fig:ovl} shows the accuracy achieved by CAFQA in all applications (except Cr$_2$) - 8 VQE molecular chemistry ground state estimation tasks, relative to the state-of-the-art Hartree-Fock approach.
Two sets of results are shown: `Average' and `Maximum'.
For `Average', the relative error reduction of CAFQA compared to HF is averaged across all the evaluated bond lengths (for each molecule).
For `Maximum', the highest error reduction of CAFQA compared to HF is presented, which is usually at the greatest bond length, since Hartree-Fock steadily deteriorates away from equilibrium.

It is evident that the CAFQA is able to achieve significant average relative accuracy improvements over all applications, with a mean of 6.4x (highest of 25x).
Furthermore, the maximum improvements are very substantial, with a mean of 56.8x (highest of 3.4*10$^5$x).
The lowest benefits are obtained for H$_6$, which, as explained in Section \ref{6-method}, has a significant correlation energy component which cannot be entirely recovered by only exploring the Clifford space. 
It is clearly evident that high VQA initialization accuracy can be achieved by CAFQA compared to state-of-the art.

\begin{figure}[t]
\centering
\includegraphics[width=\columnwidth,trim={0cm 0cm 0cm 0cm}]{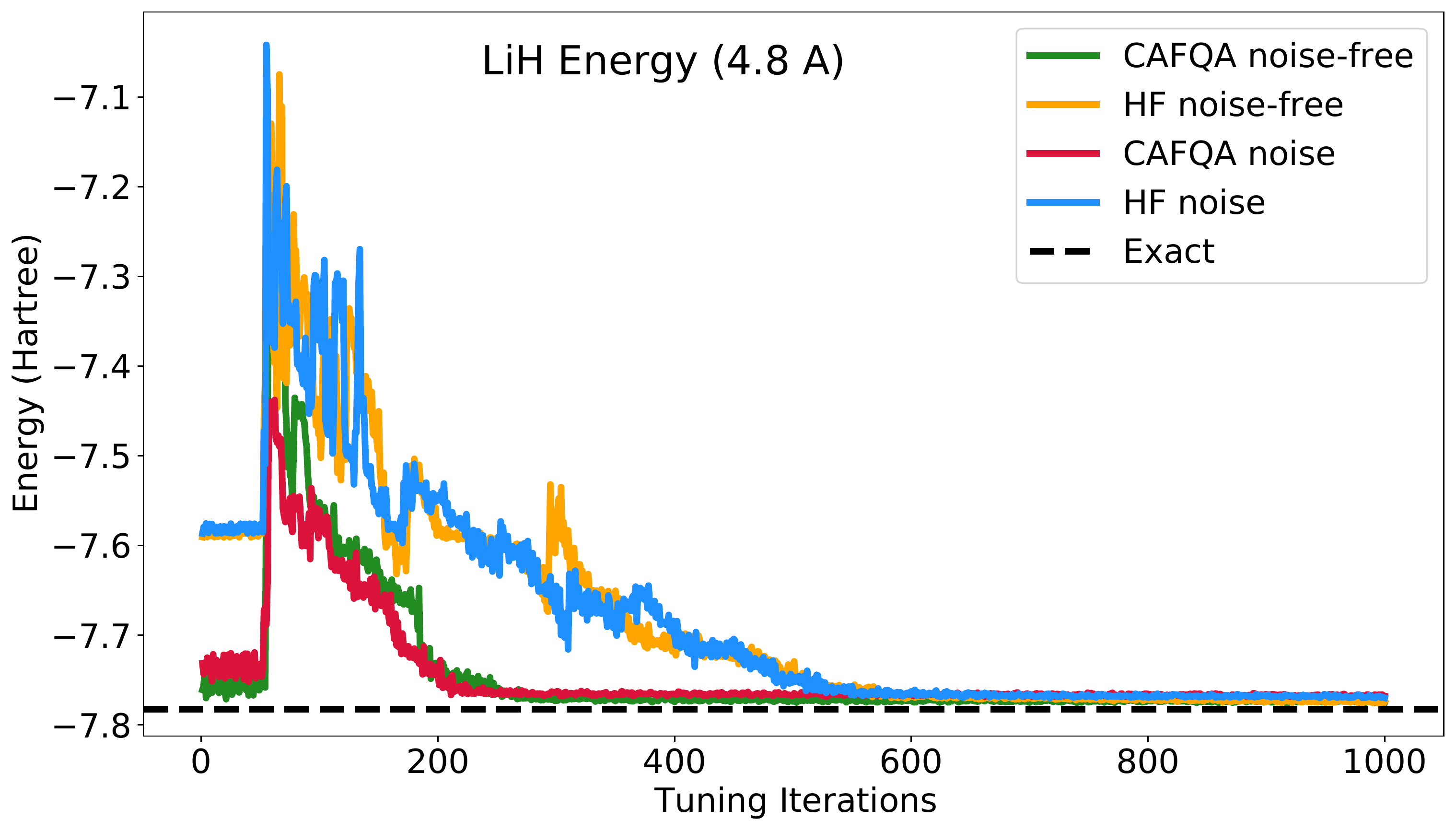}

\caption{Post-CAFQA VQA tuning for LiH. CAFQA initialization leads to 2.5x faster VQA convergence compared to HF, on both ideal and noisy  machines.}
\label{fig:post}
\end{figure}

\subsection{Post-CAFQA VQA Exploration}
\label{postcafqa}


Although today's NISQ machines are often too noisy to improve on CAFQA's estimates, it is expected that NISQ machines in the near future will be able to do so.
In this case, the CAFQA initialization will allow for more focused tuning on the machine, resulting in lower potential for detrimental impact from noise and barren plateaus, thereby leading to faster and more accurate convergence.

This is illustrated in Fig.\ref{fig:post} which shows post-CAFQA VQE tuning for LiH ground state energy estimation.
Evaluation is shown for tuning beginning from HF initialization and CAFQA initialization, respectively. 
Furthermore, two sets of results are shown, one on ideal noise-free simulation and the other on noisy simulation modeled on real machine characteristics. 
In both sets of results it is evident that CAFQA-initialized exploration converges roughly 2.5x faster than HF-initialization, clearly indicative of the benefits from better initialization.

It can be observed that the ideal simulation produces near exact results, improving over the initialization.
Furthermore, the energy estimate produced by noisy simulation (error roughly = 10$^{-2}$ Hartree) is on par with the estimate obtained directly from CAFQA initialization itself. 
While we do not expect the latter trend to hold for more complex Hamiltonians and as machine noise reduces, CAFQA initialization will continue to be useful for fast and accurate convergence.
Greater benefits can be expected for larger problem sizes, which can be realistically evaluated as NISQ machines improve.
Reduced execution on the actual quantum device is also beneficial from the monetary standpoint.
Prior work discusses high execution costs of variational algorithms on the quantum cloud, consuming thousands of dollars to execute problems of reasonably small sizes~\cite{gu2021adaptive}.

\begin{figure}[t]
\centering
\includegraphics[width=\columnwidth,trim={0cm 0cm 0cm 0cm}]{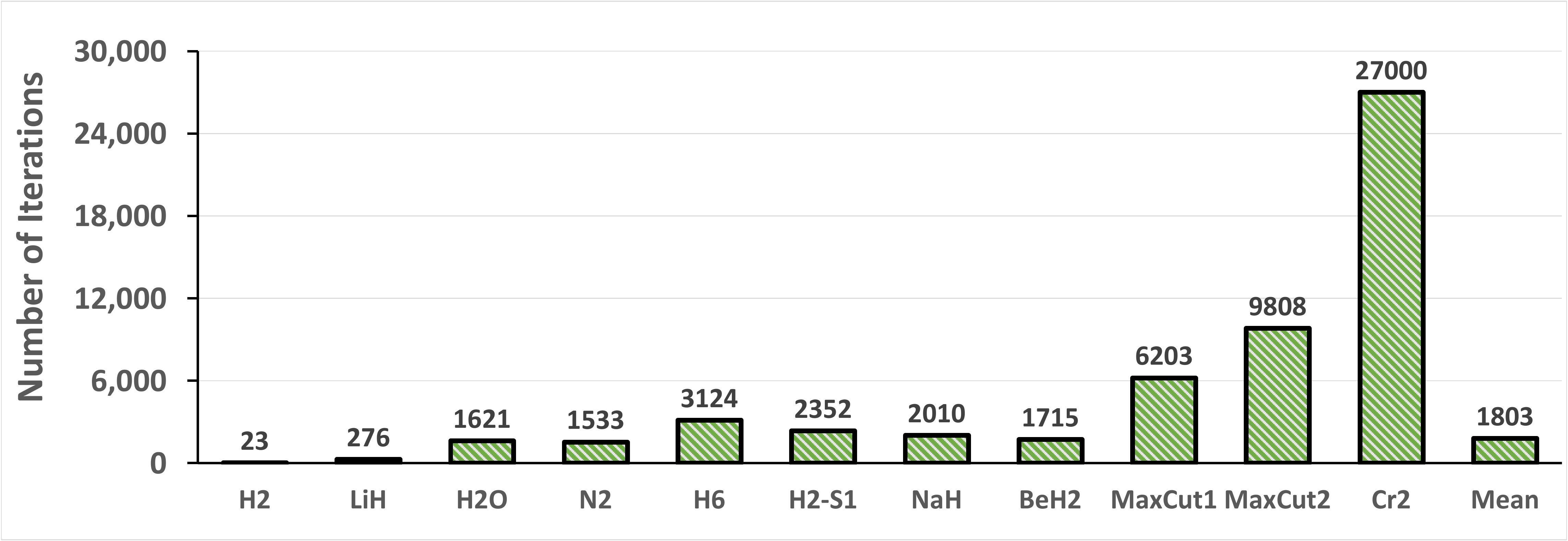}

\caption{BO search iterations for CAFQA to converge to the lowest expectation estimates for each VQA problem. Search overheads are reasonable, especially considering the benefits of reduced variational tuning on NISQ devices.}
\label{fig:itn}
\end{figure}

\subsection{Discrete Search}
\label{eval_search}

Fig.\ref{fig:itn} shows the number of iterations consumed by CAFQA's discrete search to converge to a minimum energy estimate.
It is evident that the number of iterations increases with the size of the problem because the number of tuning parameters increases.
The number of iterations across all applications is very reasonable considering the benefits of reduced variational tuning on noisy quantum devices. 
The current run time of CAFQA varies roughly from a few minutes (H$_2$) to a week (Cr$_2$). 
The execution time can be reduced via increased compute / memory resources, improved search algorithm, parallel search, etc. 

\section{Discussion}

\label{FW}

\begin{figure}[t]

\centering

\subfloat[H$_2$ Energy (<=1T)\label{H2+kT}]{\includegraphics[width=\columnwidth]{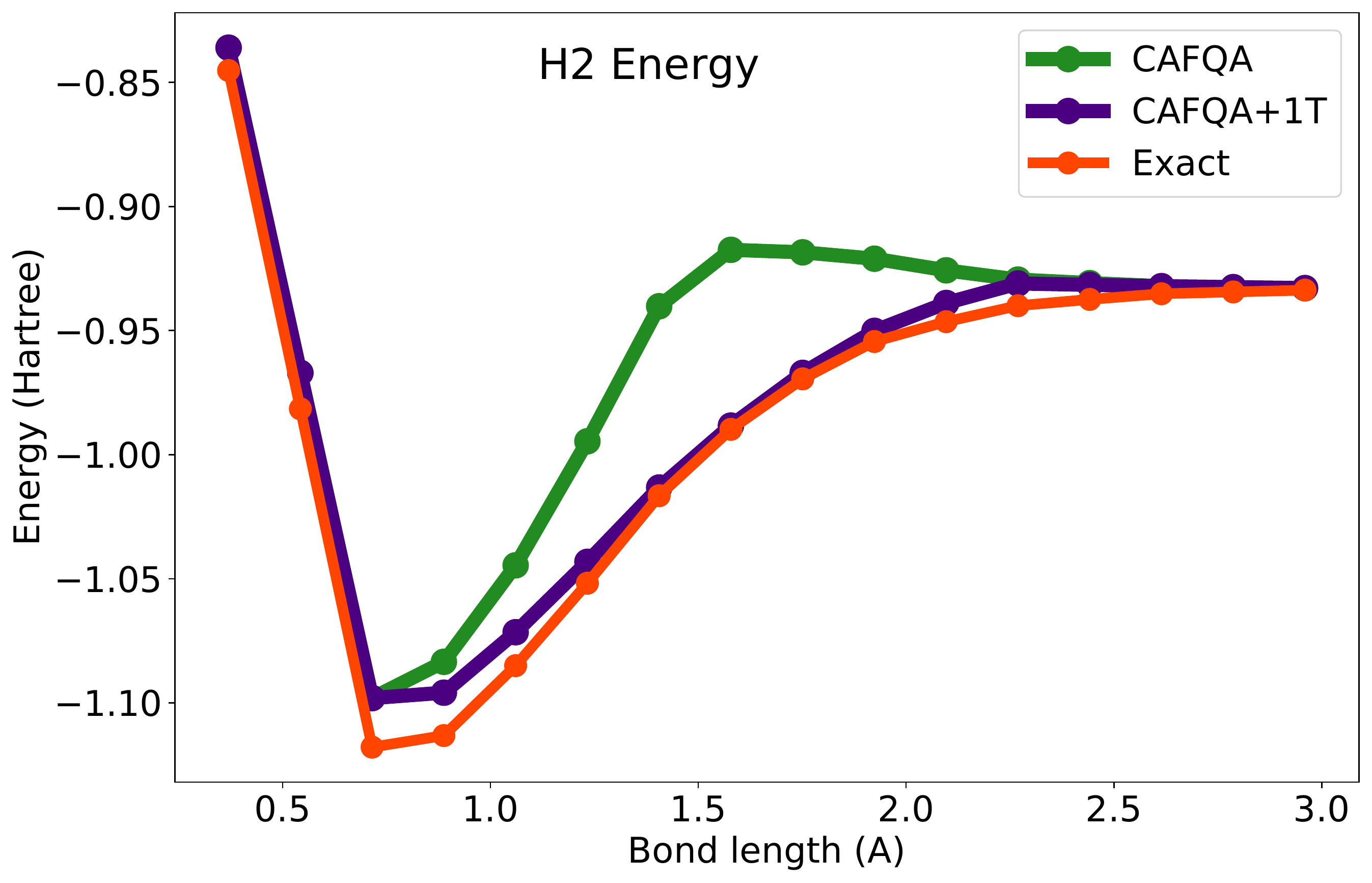}}

\subfloat[$LiH$ Energy (<=4T)\label{LiH+kT}]{\includegraphics[width=\columnwidth]{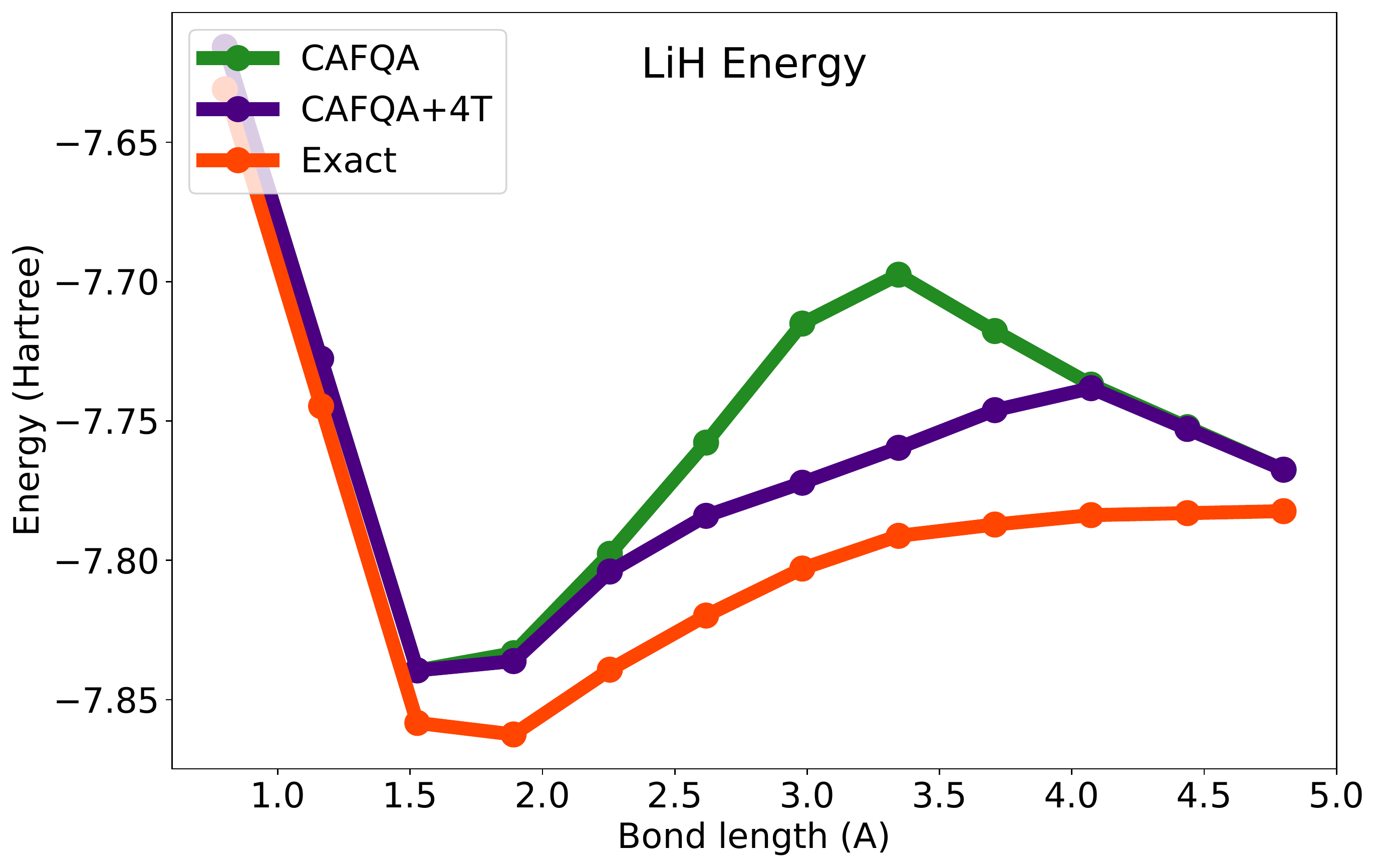}}

\caption{CAFQA + kT dissociation curves, for $H_2$ and $LiH$. 
The addition of just up to 1 T gate for $H_2$ and up to 4 T gates for $LiH$ is seen to significantly improve initialization, while remaining classically simulable. 
}
\label{fig:eval_kT}

\end{figure}

\textbf{\emph{Simulation beyond Cliffords:}}
Prior work has shown that efficient classical simulation can be extended beyond  Clifford-only circuits to constrained Clifford+T circuits wherein T refers to the single-qubit 45-degree phase shift~\cite{Bravyi2016,Bravyi2019}.
Optimally designing a CAFQA ansatz with a mix of Clifford gates and minimal T gates is worth exploring.
We perform preliminary exploration of allowing a few T gates within the CAFQA framework.
Note that the simulation complexity grows exponentially with the number of T gates, so the number and location of the T gates require careful analysis.
Our current exploration only studies the insertion of T gates at prior Clifford gate positions in the ansatz and only attempts the addition of under 10 T gates.
But it is already evident that this direction is promising --- Fig.\ref{fig:eval_kT} shows that the addition of just up to 1 T gate for $H_2$ and up to 4 T gates for $LiH$ significantly improves initialization accuracy, while remaining classically simulable.
\emph{CAFQA+kT (k<=1 / k<=4)} is able to recover as much as 99.9\% of the correlation energy at bond lengths for which Clifford-only CAFQA accuracy is relatively limited.

\textbf{\emph{Beyond a hardware-efficient ansatz:}}
A hardware-efficient ansatz can be limited in its capabilities because it is application-agnostic.
Thus, expanding beyond this ansatz can be beneficial if efficiently suited to the CAFQA approach.
Expanding the ansatz search to Clifford plus limited non-Clifford gates (discussed above) is a first step in this direction - potentially allowing for a dynamically evolving ansatz structure similar to ADAPT-VQE~\cite{adaptvqe}.

\textbf{\emph{Optimization:}}
The discrete search via Bayesian Optimization employed by CAFQA is able to produce high accuracy results and in a reasonable number of iterations and runtime.
Although this is clearly efficient for the target problem space (our largest is a 34-qubit system), the search could face scalability challenges on significantly larger problems.
This is especially important since the Clifford search space (stabilizer state space) scales exponentially in the number of tunable parameters (qubits)~\cite{Heinrich2019}.
Thus, optimizing the search strategy at the algorithmic as well as implementation levels can reap benefits.

\textbf{\emph{Related Work}}:
\label{RW}
~\cite{peterson2012chemical} proposes initializing ansatz parameters such that subsections of the ansatz do not form a `2-design'.
~\cite{mitarai2020quadratic} proposes a perturbative expansion of the cost function from HF initialization. 
This can produce an initialization state very close to the HF, but potentially performs better.
FLIP~\cite{sauvage2021flip} proposes initialization with the help of machine learning.  
MetaVQE~\cite{Cervera2021} encodes the Hamiltonian parameters in the first layers of the quantum circuit.

\section{Conclusion}
\label{Conc}
Advancing NISQ frontiers to real world applicability requires concerted effort on multiple fronts, with support from sophisticated error mitigation, classical computing, and more. In this spirit, CAFQA proposes application-specific classical simulation bootstrapping for VQAs.
CAFQA tackles the problem of finding initial VQA parameters by proposing a ``Clifford Ansatz" --- an ansatz that is a hardware efficient circuit built with only Clifford gates. 
In this ansatz, the initial parameters for the tunable gates are chosen by searching efficiently through the Clifford parameter space via classical simulation, and thereby producing a suitable high accuracy initial state that outperforms state-of-the-art approaches. 
Furthermore, there is considerable potential to extend these findings beyond Cliffords and to other circuit structures.

CAFQA is a promising example of quantum-inspired classical techniques as a supporting methodology for VQAs in the NISQ era and beyond.
It also highlights the potential for a synergistic quantum-classical paradigm to boost NISQ-era quantum computing towards real world applicability.


\begin{acks}
This work is funded in part by EPiQC, an NSF Expedition in Computing, under award CCF-1730449; 
in part by STAQ under award NSF Phy-1818914; in part by NSF award 2110860; 
in part by the US Department of Energy Office  of Advanced Scientific Computing Research, Accelerated Research for Quantum Computing Program; 
and in part by the NSF Quantum Leap Challenge Institute for Hybrid Quantum Architectures and Networks (NSF Award 2016136) 
and in part based upon work supported by the U.S. Department of Energy, Office of Science, National Quantum Information Science Research Centers.  
This research used resources of the Oak Ridge Leadership Computing Facility, which is a DOE Office of Science User Facility supported under Contract DE-AC05-00OR22725.
GSR is supported as a Computing Innovation Fellow at the University of Chicago. This material is based upon work supported by the National Science Foundation under Grant \# 2030859 to the Computing Research Association for the CIFellows Project.
KNS is supported by IBM as a Postdoctoral Scholar at the University of Chicago and the Chicago Quantum Exchange.
YD's work is supported by the National Science Foundation under Grant \# 2030859 to the Computing Research Association for the CIFellows Project.
HH is supported by NSF (grants CCF-2119184, CNS-1956180, CNS-1952050, CCF-1823032, CNS-1764039), ARO (grant W911NF1920321), and a DOE Early Career Award (grant DESC0014195 0003).
WMK acknowledges funding from NSF Grant DGE-1842474.
FTC is Chief Scientist for Quantum Software at ColdQuanta and an advisor to Quantum Circuits, Inc.

\end{acks}

\balance
\bibliographystyle{ACM-Reference-Format}
\bibliography{refs}


\begin{thebibliography}{81}


\ifx \showCODEN    \undefined \def \showCODEN     #1{\unskip}     \fi
\ifx \showDOI      \undefined \def \showDOI       #1{#1}\fi
\ifx \showISBNx    \undefined \def \showISBNx     #1{\unskip}     \fi
\ifx \showISBNxiii \undefined \def \showISBNxiii  #1{\unskip}     \fi
\ifx \showISSN     \undefined \def \showISSN      #1{\unskip}     \fi
\ifx \showLCCN     \undefined \def \showLCCN      #1{\unskip}     \fi
\ifx \shownote     \undefined \def \shownote      #1{#1}          \fi
\ifx \showarticletitle \undefined \def \showarticletitle #1{#1}   \fi
\ifx \showURL      \undefined \def \showURL       {\relax}        \fi
\providecommand\bibfield[2]{#2}
\providecommand\bibinfo[2]{#2}
\providecommand\natexlab[1]{#1}
\providecommand\showeprint[2][]{arXiv:#2}

\bibitem[IBM(2021)]%
        {IBM-SU2}
 \bibinfo{year}{2021}\natexlab{}.
\newblock \bibinfo{title}{IBM Quantum SU2 ansatz}.
\newblock
  \bibinfo{howpublished}{\url{https://qiskit.org/documentation/stubs/qiskit.circuit.library.EfficientSU2.html}}.
\newblock


\bibitem[Mic(2021)]%
        {Microsoft-HF}
 \bibinfo{year}{2021}\natexlab{}.
\newblock \bibinfo{title}{Microsoft Docs Hartree–Fock Theory}.
\newblock
  \bibinfo{howpublished}{\url{https://docs.microsoft.com/en-us/azure/quantum/user-guide/libraries/chemistry/concepts/hartree-fock}}.
\newblock


\bibitem[Orn(2021)]%
        {Ornl-HF}
 \bibinfo{year}{2021}\natexlab{}.
\newblock \bibinfo{title}{Techniques and Applications of Quantum Monte Carlo}.
\newblock
  \bibinfo{howpublished}{\url{https://web.ornl.gov/~kentpr/thesis/pkthnode13.html}}.
\newblock


\bibitem[nis(2022)]%
        {nist}
 \bibinfo{year}{2022}\natexlab{}.
\newblock \bibinfo{title}{NIST Atomic Spectra Database Ionization Energies
  Form}.
\newblock
  \bibinfo{howpublished}{\url{https://physics.nist.gov/PhysRefData/ASD/ionEnergy.html}}.
\newblock
\newblock
\shownote{Accessed: 2022-01-01}.


\bibitem[Abraham et~al\mbox{.}(2019)]%
        {Qiskit}
\bibfield{author}{\bibinfo{person}{H{\'e}ctor Abraham},
  \bibinfo{person}{AduOffei}, \bibinfo{person}{Rochisha Agarwal},
  \bibinfo{person}{Ismail~Yunus Akhalwaya}, \bibinfo{person}{Gadi
  Aleksandrowicz}, \bibinfo{person}{Thomas Alexander}, \bibinfo{person}{Matthew
  Amy}, \bibinfo{person}{Eli Arbel}, \bibinfo{person}{Arijit02},
  \bibinfo{person}{Abraham Asfaw}, \bibinfo{person}{Artur Avkhadiev},
  \bibinfo{person}{Carlos Azaustre}, \bibinfo{person}{AzizNgoueya},
  \bibinfo{person}{Abhik Banerjee}, \bibinfo{person}{Aman Bansal},
  \bibinfo{person}{Panagiotis Barkoutsos}, \bibinfo{person}{George Barron},
  \bibinfo{person}{George~S. Barron}, \bibinfo{person}{Luciano Bello},
  \bibinfo{person}{Yael Ben-Haim}, \bibinfo{person}{Daniel Bevenius},
  \bibinfo{person}{Arjun Bhobe}, \bibinfo{person}{Lev~S. Bishop},
  \bibinfo{person}{Carsten Blank}, \bibinfo{person}{Sorin Bolos},
  \bibinfo{person}{Samuel Bosch}, \bibinfo{person}{Brandon},
  \bibinfo{person}{Sergey Bravyi}, \bibinfo{person}{Bryce-Fuller},
  \bibinfo{person}{David Bucher}, \bibinfo{person}{Artemiy Burov},
  \bibinfo{person}{Fran Cabrera}, \bibinfo{person}{Padraic Calpin},
  \bibinfo{person}{Lauren Capelluto}, \bibinfo{person}{Jorge Carballo},
  \bibinfo{person}{Gin{\'e}s Carrascal}, \bibinfo{person}{Adrian Chen},
  \bibinfo{person}{Chun-Fu Chen}, \bibinfo{person}{Edward Chen},
  \bibinfo{person}{Jielun~(Chris) Chen}, \bibinfo{person}{Richard Chen},
  \bibinfo{person}{Jerry~M. Chow}, \bibinfo{person}{Spencer Churchill},
  \bibinfo{person}{Christian Claus}, \bibinfo{person}{Christian Clauss},
  \bibinfo{person}{Romilly Cocking}, \bibinfo{person}{Filipe Correa},
  \bibinfo{person}{Abigail~J. Cross}, \bibinfo{person}{Andrew~W. Cross},
  \bibinfo{person}{Simon Cross}, \bibinfo{person}{Juan Cruz-Benito},
  \bibinfo{person}{Chris Culver}, \bibinfo{person}{Antonio~D.
  C{\'o}rcoles-Gonzales}, \bibinfo{person}{Sean Dague},
  \bibinfo{person}{Tareq~El Dandachi}, \bibinfo{person}{Marcus Daniels},
  \bibinfo{person}{Matthieu Dartiailh}, \bibinfo{person}{DavideFrr},
  \bibinfo{person}{Abd{\'o}n~Rodr{\'\i}guez Davila}, \bibinfo{person}{Anton
  Dekusar}, \bibinfo{person}{Delton Ding}, \bibinfo{person}{Jun Doi},
  \bibinfo{person}{Eric Drechsler}, \bibinfo{person}{Drew},
  \bibinfo{person}{Eugene Dumitrescu}, \bibinfo{person}{Karel Dumon},
  \bibinfo{person}{Ivan Duran}, \bibinfo{person}{Kareem EL-Safty},
  \bibinfo{person}{Eric Eastman}, \bibinfo{person}{Grant Eberle},
  \bibinfo{person}{Pieter Eendebak}, \bibinfo{person}{Daniel Egger},
  \bibinfo{person}{Mark Everitt}, \bibinfo{person}{Paco~Mart{\'\i}n
  Fern{\'a}ndez}, \bibinfo{person}{Axel~Hern{\'a}ndez Ferrera},
  \bibinfo{person}{Romain Fouilland}, \bibinfo{person}{FranckChevallier},
  \bibinfo{person}{Albert Frisch}, \bibinfo{person}{Andreas Fuhrer},
  \bibinfo{person}{Bryce Fuller}, \bibinfo{person}{MELVIN GEORGE},
  \bibinfo{person}{Julien Gacon}, \bibinfo{person}{Borja~Godoy Gago},
  \bibinfo{person}{Claudio Gambella}, \bibinfo{person}{Jay~M. Gambetta},
  \bibinfo{person}{Adhisha Gammanpila}, \bibinfo{person}{Luis Garcia},
  \bibinfo{person}{Tanya Garg}, \bibinfo{person}{Shelly Garion},
  \bibinfo{person}{Austin Gilliam}, \bibinfo{person}{Aditya Giridharan},
  \bibinfo{person}{Juan Gomez-Mosquera}, \bibinfo{person}{Salvador de~la
  Puente~Gonz{\'a}lez}, \bibinfo{person}{Jesse Gorzinski}, \bibinfo{person}{Ian
  Gould}, \bibinfo{person}{Donny Greenberg}, \bibinfo{person}{Dmitry Grinko},
  \bibinfo{person}{Wen Guan}, \bibinfo{person}{John~A. Gunnels},
  \bibinfo{person}{Mikael Haglund}, \bibinfo{person}{Isabel Haide},
  \bibinfo{person}{Ikko Hamamura}, \bibinfo{person}{Omar~Costa Hamido},
  \bibinfo{person}{Frank Harkins}, \bibinfo{person}{Vojtech Havlicek},
  \bibinfo{person}{Joe Hellmers}, \bibinfo{person}{{\L}ukasz Herok},
  \bibinfo{person}{Stefan Hillmich}, \bibinfo{person}{Hiroshi Horii},
  \bibinfo{person}{Connor Howington}, \bibinfo{person}{Shaohan Hu},
  \bibinfo{person}{Wei Hu}, \bibinfo{person}{Junye Huang},
  \bibinfo{person}{Rolf Huisman}, \bibinfo{person}{Haruki Imai},
  \bibinfo{person}{Takashi Imamichi}, \bibinfo{person}{Kazuaki Ishizaki},
  \bibinfo{person}{Raban Iten}, \bibinfo{person}{Toshinari Itoko},
  \bibinfo{person}{JamesSeaward}, \bibinfo{person}{Ali Javadi},
  \bibinfo{person}{Ali Javadi-Abhari}, \bibinfo{person}{Jessica},
  \bibinfo{person}{Madhav Jivrajani}, \bibinfo{person}{Kiran Johns},
  \bibinfo{person}{Scott Johnstun}, \bibinfo{person}{Jonathan-Shoemaker},
  \bibinfo{person}{Vismai K}, \bibinfo{person}{Tal Kachmann},
  \bibinfo{person}{Naoki Kanazawa}, \bibinfo{person}{Kang-Bae},
  \bibinfo{person}{Anton Karazeev}, \bibinfo{person}{Paul Kassebaum},
  \bibinfo{person}{Josh Kelso}, \bibinfo{person}{Spencer King},
  \bibinfo{person}{Knabberjoe}, \bibinfo{person}{Yuri Kobayashi},
  \bibinfo{person}{Arseny Kovyrshin}, \bibinfo{person}{Rajiv Krishnakumar},
  \bibinfo{person}{Vivek Krishnan}, \bibinfo{person}{Kevin Krsulich},
  \bibinfo{person}{Prasad Kumkar}, \bibinfo{person}{Gawel Kus},
  \bibinfo{person}{Ryan LaRose}, \bibinfo{person}{Enrique Lacal},
  \bibinfo{person}{Rapha{\"e}l Lambert}, \bibinfo{person}{John Lapeyre},
  \bibinfo{person}{Joe Latone}, \bibinfo{person}{Scott Lawrence},
  \bibinfo{person}{Christina Lee}, \bibinfo{person}{Gushu Li},
  \bibinfo{person}{Dennis Liu}, \bibinfo{person}{Peng Liu},
  \bibinfo{person}{Yunho Maeng}, \bibinfo{person}{Kahan Majmudar},
  \bibinfo{person}{Aleksei Malyshev}, \bibinfo{person}{Joshua Manela},
  \bibinfo{person}{Jakub Marecek}, \bibinfo{person}{Manoel Marques},
  \bibinfo{person}{Dmitri Maslov}, \bibinfo{person}{Dolph Mathews},
  \bibinfo{person}{Atsushi Matsuo}, \bibinfo{person}{Douglas~T. McClure},
  \bibinfo{person}{Cameron McGarry}, \bibinfo{person}{David McKay},
  \bibinfo{person}{Dan McPherson}, \bibinfo{person}{Srujan Meesala},
  \bibinfo{person}{Thomas Metcalfe}, \bibinfo{person}{Martin Mevissen},
  \bibinfo{person}{Andrew Meyer}, \bibinfo{person}{Antonio Mezzacapo},
  \bibinfo{person}{Rohit Midha}, \bibinfo{person}{Zlatko Minev},
  \bibinfo{person}{Abby Mitchell}, \bibinfo{person}{Nikolaj Moll},
  \bibinfo{person}{Jhon Montanez}, \bibinfo{person}{Michael~Duane Mooring},
  \bibinfo{person}{Renier Morales}, \bibinfo{person}{Niall Moran},
  \bibinfo{person}{Mario Motta}, \bibinfo{person}{MrF},
  \bibinfo{person}{Prakash Murali}, \bibinfo{person}{Jan M{\"u}ggenburg},
  \bibinfo{person}{David Nadlinger}, \bibinfo{person}{Ken Nakanishi},
  \bibinfo{person}{Giacomo Nannicini}, \bibinfo{person}{Paul Nation},
  \bibinfo{person}{Edwin Navarro}, \bibinfo{person}{Yehuda Naveh},
  \bibinfo{person}{Scott~Wyman Neagle}, \bibinfo{person}{Patrick Neuweiler},
  \bibinfo{person}{Johan Nicander}, \bibinfo{person}{Pradeep Niroula},
  \bibinfo{person}{Hassi Norlen}, \bibinfo{person}{NuoWenLei},
  \bibinfo{person}{Lee~James O'Riordan}, \bibinfo{person}{Oluwatobi Ogunbayo},
  \bibinfo{person}{Pauline Ollitrault}, \bibinfo{person}{Raul Otaolea},
  \bibinfo{person}{Steven Oud}, \bibinfo{person}{Dan Padilha},
  \bibinfo{person}{Hanhee Paik}, \bibinfo{person}{Soham Pal},
  \bibinfo{person}{Yuchen Pang}, \bibinfo{person}{Simone Perriello},
  \bibinfo{person}{Anna Phan}, \bibinfo{person}{Francesco Piro},
  \bibinfo{person}{Marco Pistoia}, \bibinfo{person}{Christophe Piveteau},
  \bibinfo{person}{Pierre Pocreau}, \bibinfo{person}{Alejandro
  Pozas-iKerstjens}, \bibinfo{person}{Viktor Prutyanov},
  \bibinfo{person}{Daniel Puzzuoli}, \bibinfo{person}{Jes{\'u}s P{\'e}rez},
  \bibinfo{person}{Quintiii}, \bibinfo{person}{Rafey~Iqbal Rahman},
  \bibinfo{person}{Arun Raja}, \bibinfo{person}{Nipun Ramagiri},
  \bibinfo{person}{Anirudh Rao}, \bibinfo{person}{Rudy Raymond},
  \bibinfo{person}{Rafael Mart{\'\i}n-Cuevas Redondo}, \bibinfo{person}{Max
  Reuter}, \bibinfo{person}{Julia Rice}, \bibinfo{person}{Marcello~La Rocca},
  \bibinfo{person}{Diego~M. Rodr{\'\i}guez}, \bibinfo{person}{RohithKarur},
  \bibinfo{person}{Max Rossmannek}, \bibinfo{person}{Mingi Ryu},
  \bibinfo{person}{Tharrmashastha SAPV}, \bibinfo{person}{SamFerracin},
  \bibinfo{person}{Martin Sandberg}, \bibinfo{person}{Hirmay Sandesara},
  \bibinfo{person}{Ritvik Sapra}, \bibinfo{person}{Hayk Sargsyan},
  \bibinfo{person}{Aniruddha Sarkar}, \bibinfo{person}{Ninad Sathaye},
  \bibinfo{person}{Bruno Schmitt}, \bibinfo{person}{Chris Schnabel},
  \bibinfo{person}{Zachary Schoenfeld}, \bibinfo{person}{Travis~L. Scholten},
  \bibinfo{person}{Eddie Schoute}, \bibinfo{person}{Joachim Schwarm},
  \bibinfo{person}{Ismael~Faro Sertage}, \bibinfo{person}{Kanav Setia},
  \bibinfo{person}{Nathan Shammah}, \bibinfo{person}{Yunong Shi},
  \bibinfo{person}{Adenilton Silva}, \bibinfo{person}{Andrea Simonetto},
  \bibinfo{person}{Nick Singstock}, \bibinfo{person}{Yukio Siraichi},
  \bibinfo{person}{Iskandar Sitdikov}, \bibinfo{person}{Seyon Sivarajah},
  \bibinfo{person}{Magnus~Berg Sletfjerding}, \bibinfo{person}{John~A. Smolin},
  \bibinfo{person}{Mathias Soeken}, \bibinfo{person}{Igor~Olegovich Sokolov},
  \bibinfo{person}{Igor Sokolov}, \bibinfo{person}{SooluThomas},
  \bibinfo{person}{Starfish}, \bibinfo{person}{Dominik Steenken},
  \bibinfo{person}{Matt Stypulkoski}, \bibinfo{person}{Shaojun Sun},
  \bibinfo{person}{Kevin~J. Sung}, \bibinfo{person}{Hitomi Takahashi},
  \bibinfo{person}{Tanvesh Takawale}, \bibinfo{person}{Ivano Tavernelli},
  \bibinfo{person}{Charles Taylor}, \bibinfo{person}{Pete Taylour},
  \bibinfo{person}{Soolu Thomas}, \bibinfo{person}{Mathieu Tillet},
  \bibinfo{person}{Maddy Tod}, \bibinfo{person}{Miroslav Tomasik},
  \bibinfo{person}{Enrique de~la Torre}, \bibinfo{person}{Kenso Trabing},
  \bibinfo{person}{Matthew Treinish}, \bibinfo{person}{TrishaPe},
  \bibinfo{person}{Davindra Tulsi}, \bibinfo{person}{Wes Turner},
  \bibinfo{person}{Yotam Vaknin}, \bibinfo{person}{Carmen~Recio Valcarce},
  \bibinfo{person}{Francois Varchon}, \bibinfo{person}{Almudena~Carrera
  Vazquez}, \bibinfo{person}{Victor Villar}, \bibinfo{person}{Desiree
  Vogt-Lee}, \bibinfo{person}{Christophe Vuillot}, \bibinfo{person}{James
  Weaver}, \bibinfo{person}{Johannes Weidenfeller}, \bibinfo{person}{Rafal
  Wieczorek}, \bibinfo{person}{Jonathan~A. Wildstrom}, \bibinfo{person}{Erick
  Winston}, \bibinfo{person}{Jack~J. Woehr}, \bibinfo{person}{Stefan Woerner},
  \bibinfo{person}{Ryan Woo}, \bibinfo{person}{Christopher~J. Wood},
  \bibinfo{person}{Ryan Wood}, \bibinfo{person}{Stephen Wood},
  \bibinfo{person}{Steve Wood}, \bibinfo{person}{James Wootton},
  \bibinfo{person}{Daniyar Yeralin}, \bibinfo{person}{David Yonge-Mallo},
  \bibinfo{person}{Richard Young}, \bibinfo{person}{Jessie Yu},
  \bibinfo{person}{Christopher Zachow}, \bibinfo{person}{Laura Zdanski},
  \bibinfo{person}{Helena Zhang}, \bibinfo{person}{Christa Zoufal},
  \bibinfo{person}{Zoufalc}, \bibinfo{person}{a kapila}, \bibinfo{person}{a
  matsuo}, \bibinfo{person}{bcamorrison}, \bibinfo{person}{brandhsn},
  \bibinfo{person}{nick bronn}, \bibinfo{person}{chlorophyll zz},
  \bibinfo{person}{dekel.meirom}, \bibinfo{person}{dekelmeirom},
  \bibinfo{person}{dekool}, \bibinfo{person}{dime10},
  \bibinfo{person}{drholmie}, \bibinfo{person}{dtrenev},
  \bibinfo{person}{ehchen}, \bibinfo{person}{elfrocampeador},
  \bibinfo{person}{faisaldebouni}, \bibinfo{person}{fanizzamarco},
  \bibinfo{person}{gabrieleagl}, \bibinfo{person}{gadial},
  \bibinfo{person}{galeinston}, \bibinfo{person}{georgios ts},
  \bibinfo{person}{gruu}, \bibinfo{person}{hhorii},
  \bibinfo{person}{hykavitha}, \bibinfo{person}{jagunther},
  \bibinfo{person}{jliu45}, \bibinfo{person}{jscott2},
  \bibinfo{person}{kanejess}, \bibinfo{person}{klinvill},
  \bibinfo{person}{krutik2966}, \bibinfo{person}{kurarrr},
  \bibinfo{person}{lerongil}, \bibinfo{person}{ma5x}, \bibinfo{person}{merav
  aharoni}, \bibinfo{person}{michelle4654}, \bibinfo{person}{ordmoj},
  \bibinfo{person}{sagar pahwa}, \bibinfo{person}{rmoyard},
  \bibinfo{person}{saswati qiskit}, \bibinfo{person}{scottkelso},
  \bibinfo{person}{sethmerkel}, \bibinfo{person}{strickroman},
  \bibinfo{person}{sumitpuri}, \bibinfo{person}{tigerjack},
  \bibinfo{person}{toural}, \bibinfo{person}{tsura crisaldo},
  \bibinfo{person}{vvilpas}, \bibinfo{person}{welien},
  \bibinfo{person}{willhbang}, \bibinfo{person}{yang.luh},
  \bibinfo{person}{yotamvakninibm}, {and} \bibinfo{person}{Mantas
  {\v{C}}epulkovskis}.} \bibinfo{year}{2019}\natexlab{}.
\newblock \bibinfo{title}{Qiskit: An Open-source Framework for Quantum
  Computing}.
\newblock
\newblock
\urldef\tempurl%
\url{https://doi.org/10.5281/zenodo.2562110}
\showDOI{\tempurl}


\bibitem[Alipourfard et~al\mbox{.}(2017)]%
        {cherrypick2017}
\bibfield{author}{\bibinfo{person}{Omid Alipourfard},
  \bibinfo{person}{Hongqiang~Harry Liu}, \bibinfo{person}{Jianshu Chen},
  \bibinfo{person}{Shivaram Venkataraman}, \bibinfo{person}{Minlan Yu}, {and}
  \bibinfo{person}{Ming Zhang}.} \bibinfo{year}{2017}\natexlab{}.
\newblock \showarticletitle{Cherrypick: Adaptively Unearthing the Best Cloud
  Configurations for Big Data Analytics}. In
  \bibinfo{booktitle}{\emph{Proceedings of the 14th USENIX Conference on
  Networked Systems Design and Implementation}} (Boston, MA, USA)
  \emph{(\bibinfo{series}{NSDI'17})}. \bibinfo{publisher}{USENIX Association},
  \bibinfo{address}{USA}, \bibinfo{pages}{469–482}.
\newblock
\showISBNx{9781931971379}


\bibitem[Barron and Wood(2020)]%
        {barron2020measurement}
\bibfield{author}{\bibinfo{person}{George~S. Barron} {and}
  \bibinfo{person}{Christopher~J. Wood}.} \bibinfo{year}{2020}\natexlab{}.
\newblock \bibinfo{title}{Measurement Error Mitigation for Variational Quantum
  Algorithms}.
\newblock
\newblock
\showeprint[arxiv]{2010.08520}~[quant-ph]


\bibitem[Biamonte et~al\mbox{.}(2017)]%
        {biamonte2017quantum}
\bibfield{author}{\bibinfo{person}{Jacob Biamonte}, \bibinfo{person}{Peter
  Wittek}, \bibinfo{person}{Nicola Pancotti}, \bibinfo{person}{Patrick
  Rebentrost}, \bibinfo{person}{Nathan Wiebe}, {and} \bibinfo{person}{Seth
  Lloyd}.} \bibinfo{year}{2017}\natexlab{}.
\newblock \showarticletitle{Quantum machine learning}.
\newblock \bibinfo{journal}{\emph{Nature}} \bibinfo{volume}{549},
  \bibinfo{number}{7671} (\bibinfo{year}{2017}), \bibinfo{pages}{195--202}.
\newblock


\bibitem[Boixo et~al\mbox{.}(2018)]%
        {Boixo2018}
\bibfield{author}{\bibinfo{person}{Sergio Boixo}, \bibinfo{person}{Sergei~V.
  Isakov}, \bibinfo{person}{Vadim~N. Smelyanskiy}, \bibinfo{person}{Ryan
  Babbush}, \bibinfo{person}{Nan Ding}, \bibinfo{person}{Zhang Jiang},
  \bibinfo{person}{Michael~J. Bremner}, \bibinfo{person}{John~M. Martinis},
  {and} \bibinfo{person}{Hartmut Neven}.} \bibinfo{year}{2018}\natexlab{}.
\newblock \showarticletitle{Characterizing quantum supremacy in near-term
  devices}.
\newblock \bibinfo{journal}{\emph{Nature Physics}} \bibinfo{volume}{14},
  \bibinfo{number}{6} (\bibinfo{date}{Apr} \bibinfo{year}{2018}),
  \bibinfo{pages}{595–600}.
\newblock
\showISSN{1745-2481}
\urldef\tempurl%
\url{https://doi.org/10.1038/s41567-018-0124-x}
\showDOI{\tempurl}


\bibitem[Botelho et~al\mbox{.}(2021)]%
        {botelho2021error}
\bibfield{author}{\bibinfo{person}{Ludmila Botelho}, \bibinfo{person}{Adam
  Glos}, \bibinfo{person}{Akash Kundu}, \bibinfo{person}{Jarosław~Adam
  Miszczak}, \bibinfo{person}{Özlem Salehi}, {and} \bibinfo{person}{Zoltán
  Zimborás}.} \bibinfo{year}{2021}\natexlab{}.
\newblock \bibinfo{title}{Error mitigation for variational quantum algorithms
  through mid-circuit measurements}.
\newblock
\newblock
\showeprint[arxiv]{2108.10927}~[quant-ph]


\bibitem[Bravyi et~al\mbox{.}(2019)]%
        {Bravyi2019}
\bibfield{author}{\bibinfo{person}{Sergey Bravyi}, \bibinfo{person}{Dan
  Browne}, \bibinfo{person}{Padraic Calpin}, \bibinfo{person}{Earl Campbell},
  \bibinfo{person}{David Gosset}, {and} \bibinfo{person}{Mark Howard}.}
  \bibinfo{year}{2019}\natexlab{}.
\newblock \showarticletitle{Simulation of quantum circuits by low-rank
  stabilizer decompositions}.
\newblock \bibinfo{journal}{\emph{Quantum}}  \bibinfo{volume}{3}
  (\bibinfo{date}{Sep} \bibinfo{year}{2019}), \bibinfo{pages}{181}.
\newblock
\showISSN{2521-327X}
\urldef\tempurl%
\url{https://doi.org/10.22331/q-2019-09-02-181}
\showDOI{\tempurl}


\bibitem[Bravyi and Gosset(2016)]%
        {Bravyi2016}
\bibfield{author}{\bibinfo{person}{Sergey Bravyi} {and} \bibinfo{person}{David
  Gosset}.} \bibinfo{year}{2016}\natexlab{}.
\newblock \showarticletitle{Improved Classical Simulation of Quantum Circuits
  Dominated by Clifford Gates}.
\newblock \bibinfo{journal}{\emph{Physical Review Letters}}
  \bibinfo{volume}{116}, \bibinfo{number}{25} (\bibinfo{date}{Jun}
  \bibinfo{year}{2016}).
\newblock
\showISSN{1079-7114}
\urldef\tempurl%
\url{https://doi.org/10.1103/physrevlett.116.250501}
\showDOI{\tempurl}


\bibitem[Cerezo et~al\mbox{.}(2021)]%
        {Cerezo2021}
\bibfield{author}{\bibinfo{person}{M. Cerezo}, \bibinfo{person}{Akira Sone},
  \bibinfo{person}{Tyler Volkoff}, \bibinfo{person}{Lukasz Cincio}, {and}
  \bibinfo{person}{Patrick~J. Coles}.} \bibinfo{year}{2021}\natexlab{}.
\newblock \showarticletitle{Cost function dependent barren plateaus in shallow
  parametrized quantum circuits}.
\newblock \bibinfo{journal}{\emph{Nature Communications}} \bibinfo{volume}{12},
  \bibinfo{number}{1} (\bibinfo{date}{Mar} \bibinfo{year}{2021}).
\newblock
\showISSN{2041-1723}
\urldef\tempurl%
\url{https://doi.org/10.1038/s41467-021-21728-w}
\showDOI{\tempurl}


\bibitem[Cervera-Lierta et~al\mbox{.}(2021)]%
        {Cervera2021}
\bibfield{author}{\bibinfo{person}{Alba Cervera-Lierta},
  \bibinfo{person}{Jakob~S. Kottmann}, {and} \bibinfo{person}{Alán
  Aspuru-Guzik}.} \bibinfo{year}{2021}\natexlab{}.
\newblock \showarticletitle{Meta-Variational Quantum Eigensolver: Learning
  Energy Profiles of Parameterized Hamiltonians for Quantum Simulation}.
\newblock \bibinfo{journal}{\emph{PRX Quantum}} \bibinfo{volume}{2},
  \bibinfo{number}{2} (\bibinfo{date}{May} \bibinfo{year}{2021}).
\newblock
\showISSN{2691-3399}
\urldef\tempurl%
\url{https://doi.org/10.1103/prxquantum.2.020329}
\showDOI{\tempurl}


\bibitem[Czarnik et~al\mbox{.}(2020)]%
        {czarnik2020error}
\bibfield{author}{\bibinfo{person}{Piotr Czarnik}, \bibinfo{person}{Andrew
  Arrasmith}, \bibinfo{person}{Patrick~J. Coles}, {and} \bibinfo{person}{Lukasz
  Cincio}.} \bibinfo{year}{2020}\natexlab{}.
\newblock \bibinfo{title}{Error mitigation with Clifford quantum-circuit data}.
\newblock
\newblock
\showeprint[arxiv]{2005.10189}~[quant-ph]


\bibitem[De~Raedt et~al\mbox{.}(2019)]%
        {Raedt2019}
\bibfield{author}{\bibinfo{person}{Hans De~Raedt}, \bibinfo{person}{Fengping
  Jin}, \bibinfo{person}{Dennis Willsch}, \bibinfo{person}{Madita Willsch},
  \bibinfo{person}{Naoki Yoshioka}, \bibinfo{person}{Nobuyasu Ito},
  \bibinfo{person}{Shengjun Yuan}, {and} \bibinfo{person}{Kristel Michielsen}.}
  \bibinfo{year}{2019}\natexlab{}.
\newblock \showarticletitle{Massively parallel quantum computer simulator,
  eleven years later}.
\newblock \bibinfo{journal}{\emph{Computer Physics Communications}}
  \bibinfo{volume}{237} (\bibinfo{date}{Apr} \bibinfo{year}{2019}),
  \bibinfo{pages}{47–61}.
\newblock
\showISSN{0010-4655}
\urldef\tempurl%
\url{https://doi.org/10.1016/j.cpc.2018.11.005}
\showDOI{\tempurl}


\bibitem[Ding et~al\mbox{.}(2020a)]%
        {ding2020systematic}
\bibfield{author}{\bibinfo{person}{Yongshan Ding}, \bibinfo{person}{Pranav
  Gokhale}, \bibinfo{person}{Sophia~Fuhui Lin}, \bibinfo{person}{Richard
  Rines}, \bibinfo{person}{Thomas Propson}, {and} \bibinfo{person}{Frederic~T
  Chong}.} \bibinfo{year}{2020}\natexlab{a}.
\newblock \showarticletitle{Systematic Crosstalk Mitigation for Superconducting
  Qubits via Frequency-Aware Compilation}.
\newblock \bibinfo{journal}{\emph{arXiv preprint arXiv:2008.09503}}
  (\bibinfo{year}{2020}).
\newblock


\bibitem[Ding et~al\mbox{.}(2021)]%
        {ding2021}
\bibfield{author}{\bibinfo{person}{Yi Ding}, \bibinfo{person}{Ahsan Pervaiz},
  \bibinfo{person}{Michael Carbin}, {and} \bibinfo{person}{Henry Hoffmann}.}
  \bibinfo{year}{2021}\natexlab{}.
\newblock \showarticletitle{Generalizable and Interpretable Learning for
  Configuration Extrapolation}. In \bibinfo{booktitle}{\emph{Proceedings of the
  29th ACM Joint Meeting on European Software Engineering Conference and
  Symposium on the Foundations of Software Engineering}} (Athens, Greece)
  \emph{(\bibinfo{series}{ESEC/FSE 2021})}. \bibinfo{publisher}{Association for
  Computing Machinery}, \bibinfo{address}{New York, NY, USA},
  \bibinfo{pages}{728–740}.
\newblock
\showISBNx{9781450385626}
\urldef\tempurl%
\url{https://doi.org/10.1145/3468264.3468603}
\showDOI{\tempurl}


\bibitem[Ding et~al\mbox{.}(2020b)]%
        {dingbayesian}
\bibfield{author}{\bibinfo{person}{Yi Ding}, \bibinfo{person}{Ahsan Pervaiz},
  \bibinfo{person}{Sanjay Krishnan}, {and} \bibinfo{person}{Henry Hoffmann}.}
  \bibinfo{year}{2020}\natexlab{b}.
\newblock \showarticletitle{Bayesian Learning for Hardware and Software
  Configuration Co-Optimization}.
\newblock  (\bibinfo{year}{2020}).
\newblock


\bibitem[Du et~al\mbox{.}(2021)]%
        {du2021accelerating}
\bibfield{author}{\bibinfo{person}{Yuxuan Du}, \bibinfo{person}{Yang Qian},
  {and} \bibinfo{person}{Dacheng Tao}.} \bibinfo{year}{2021}\natexlab{}.
\newblock \showarticletitle{Accelerating variational quantum algorithms with
  multiple quantum processors}.
\newblock \bibinfo{journal}{\emph{arXiv preprint arXiv:2106.12819}}
  (\bibinfo{year}{2021}).
\newblock


\bibitem[Elfving et~al\mbox{.}(2020)]%
        {elfving2020quantum}
\bibfield{author}{\bibinfo{person}{V.~E. Elfving}, \bibinfo{person}{B.~W.
  Broer}, \bibinfo{person}{M. Webber}, \bibinfo{person}{J. Gavartin},
  \bibinfo{person}{M.~D. Halls}, \bibinfo{person}{K.~P. Lorton}, {and}
  \bibinfo{person}{A. Bochevarov}.} \bibinfo{year}{2020}\natexlab{}.
\newblock \bibinfo{title}{How will quantum computers provide an industrially
  relevant computational advantage in quantum chemistry?}
\newblock
\newblock
\showeprint[arxiv]{2009.12472}~[quant-ph]


\bibitem[Farhi et~al\mbox{.}(2014)]%
        {farhi2014quantum}
\bibfield{author}{\bibinfo{person}{Edward Farhi}, \bibinfo{person}{Jeffrey
  Goldstone}, {and} \bibinfo{person}{Sam Gutmann}.}
  \bibinfo{year}{2014}\natexlab{}.
\newblock \bibinfo{title}{A Quantum Approximate Optimization Algorithm}.
\newblock
\newblock
\showeprint[arxiv]{1411.4028}~[quant-ph]


\bibitem[Frazier(2018)]%
        {frazier2018tutorial}
\bibfield{author}{\bibinfo{person}{Peter~I Frazier}.}
  \bibinfo{year}{2018}\natexlab{}.
\newblock \showarticletitle{A tutorial on Bayesian optimization}.
\newblock \bibinfo{journal}{\emph{arXiv preprint arXiv:1807.02811}}
  (\bibinfo{year}{2018}).
\newblock


\bibitem[Giurgica-Tiron et~al\mbox{.}(2020)]%
        {giurgica2020digital}
\bibfield{author}{\bibinfo{person}{Tudor Giurgica-Tiron},
  \bibinfo{person}{Yousef Hindy}, \bibinfo{person}{Ryan LaRose},
  \bibinfo{person}{Andrea Mari}, {and} \bibinfo{person}{William~J Zeng}.}
  \bibinfo{year}{2020}\natexlab{}.
\newblock \showarticletitle{Digital zero noise extrapolation for quantum error
  mitigation}. In \bibinfo{booktitle}{\emph{2020 IEEE International Conference
  on Quantum Computing and Engineering (QCE)}}. IEEE,
  \bibinfo{pages}{306--316}.
\newblock


\bibitem[Gokhale et~al\mbox{.}(2019a)]%
        {gokhale2019minimizing}
\bibfield{author}{\bibinfo{person}{Pranav Gokhale}, \bibinfo{person}{Olivia
  Angiuli}, \bibinfo{person}{Yongshan Ding}, \bibinfo{person}{Kaiwen Gui},
  \bibinfo{person}{Teague Tomesh}, \bibinfo{person}{Martin Suchara},
  \bibinfo{person}{Margaret Martonosi}, {and} \bibinfo{person}{Frederic~T
  Chong}.} \bibinfo{year}{2019}\natexlab{a}.
\newblock \showarticletitle{Minimizing state preparations in variational
  quantum eigensolver by partitioning into commuting families}.
\newblock \bibinfo{journal}{\emph{arXiv preprint arXiv:1907.13623}}
  (\bibinfo{year}{2019}).
\newblock


\bibitem[Gokhale et~al\mbox{.}(2019b)]%
        {Gokhale:2019}
\bibfield{author}{\bibinfo{person}{Pranav Gokhale}, \bibinfo{person}{Yongshan
  Ding}, \bibinfo{person}{Thomas Propson}, \bibinfo{person}{Christopher
  Winkler}, \bibinfo{person}{Nelson Leung}, \bibinfo{person}{Yunong Shi},
  \bibinfo{person}{David~I. Schuster}, \bibinfo{person}{Henry Hoffmann}, {and}
  \bibinfo{person}{Frederic~T. Chong}.} \bibinfo{year}{2019}\natexlab{b}.
\newblock \showarticletitle{Partial Compilation of Variational Algorithms for
  Noisy Intermediate-Scale Quantum Machines}.
\newblock \bibinfo{journal}{\emph{Proceedings of the 52nd Annual IEEE/ACM
  International Symposium on Microarchitecture}} (\bibinfo{date}{Oct}
  \bibinfo{year}{2019}).
\newblock
\showISBNx{9781450369381}
\urldef\tempurl%
\url{https://doi.org/10.1145/3352460.3358313}
\showDOI{\tempurl}


\bibitem[Gottesman(1998)]%
        {gottesman1998heisenberg}
\bibfield{author}{\bibinfo{person}{Daniel Gottesman}.}
  \bibinfo{year}{1998}\natexlab{}.
\newblock \showarticletitle{The Heisenberg representation of quantum
  computers}.
\newblock \bibinfo{journal}{\emph{arXiv preprint quant-ph/9807006}}
  (\bibinfo{year}{1998}).
\newblock


\bibitem[Gottesman and Chuang(1999)]%
        {gottesman1999demonstrating}
\bibfield{author}{\bibinfo{person}{Daniel Gottesman} {and}
  \bibinfo{person}{Isaac~L Chuang}.} \bibinfo{year}{1999}\natexlab{}.
\newblock \showarticletitle{Demonstrating the viability of universal quantum
  computation using teleportation and single-qubit operations}.
\newblock \bibinfo{journal}{\emph{Nature}} \bibinfo{volume}{402},
  \bibinfo{number}{6760} (\bibinfo{year}{1999}), \bibinfo{pages}{390--393}.
\newblock


\bibitem[Grimsley et~al\mbox{.}(2019)]%
        {adaptvqe}
\bibfield{author}{\bibinfo{person}{Harper~R. Grimsley},
  \bibinfo{person}{Sophia~E. Economou}, \bibinfo{person}{Edwin Barnes}, {and}
  \bibinfo{person}{Nicholas~J. Mayhall}.} \bibinfo{year}{2019}\natexlab{}.
\newblock \showarticletitle{An adaptive variational algorithm for exact
  molecular simulations on a quantum computer}.
\newblock \bibinfo{journal}{\emph{Nature Communications}} \bibinfo{volume}{10},
  \bibinfo{number}{1} (\bibinfo{date}{Jul} \bibinfo{year}{2019}).
\newblock
\showISSN{2041-1723}
\urldef\tempurl%
\url{https://doi.org/10.1038/s41467-019-10988-2}
\showDOI{\tempurl}


\bibitem[Grover(1996)]%
        {Grover96afast}
\bibfield{author}{\bibinfo{person}{Lov~K. Grover}.}
  \bibinfo{year}{1996}\natexlab{}.
\newblock \showarticletitle{A Fast Quantum Mechanical Algorithm for Database
  Search}. In \bibinfo{booktitle}{\emph{ANNUAL ACM SYMPOSIUM ON THEORY OF
  COMPUTING}}. \bibinfo{publisher}{ACM}, \bibinfo{pages}{212--219}.
\newblock


\bibitem[Gu et~al\mbox{.}(2021)]%
        {gu2021adaptive}
\bibfield{author}{\bibinfo{person}{Andi Gu}, \bibinfo{person}{Angus Lowe},
  \bibinfo{person}{Pavel~A Dub}, \bibinfo{person}{Patrick~J Coles}, {and}
  \bibinfo{person}{Andrew Arrasmith}.} \bibinfo{year}{2021}\natexlab{}.
\newblock \showarticletitle{Adaptive shot allocation for fast convergence in
  variational quantum algorithms}.
\newblock \bibinfo{journal}{\emph{arXiv preprint arXiv:2108.10434}}
  (\bibinfo{year}{2021}).
\newblock


\bibitem[Hartree and Hartree(1935)]%
        {hartree1935self}
\bibfield{author}{\bibinfo{person}{Douglas~Rayner Hartree} {and}
  \bibinfo{person}{William Hartree}.} \bibinfo{year}{1935}\natexlab{}.
\newblock \showarticletitle{Self-consistent field, with exchange, for
  beryllium}.
\newblock \bibinfo{journal}{\emph{Proceedings of the Royal Society of London.
  Series A-Mathematical and Physical Sciences}} \bibinfo{volume}{150},
  \bibinfo{number}{869} (\bibinfo{year}{1935}), \bibinfo{pages}{9--33}.
\newblock


\bibitem[Heinrich and Gross(2019)]%
        {Heinrich2019}
\bibfield{author}{\bibinfo{person}{Markus Heinrich} {and}
  \bibinfo{person}{David Gross}.} \bibinfo{year}{2019}\natexlab{}.
\newblock \showarticletitle{Robustness of Magic and Symmetries of the
  Stabiliser Polytope}.
\newblock \bibinfo{journal}{\emph{Quantum}}  \bibinfo{volume}{3}
  (\bibinfo{date}{Apr} \bibinfo{year}{2019}), \bibinfo{pages}{132}.
\newblock
\showISSN{2521-327X}
\urldef\tempurl%
\url{https://doi.org/10.22331/q-2019-04-08-132}
\showDOI{\tempurl}


\bibitem[Holmes et~al\mbox{.}(2021)]%
        {holmes2021connecting}
\bibfield{author}{\bibinfo{person}{Zoë Holmes}, \bibinfo{person}{Kunal
  Sharma}, \bibinfo{person}{M. Cerezo}, {and} \bibinfo{person}{Patrick~J.
  Coles}.} \bibinfo{year}{2021}\natexlab{}.
\newblock \bibinfo{title}{Connecting ansatz expressibility to gradient
  magnitudes and barren plateaus}.
\newblock
\newblock
\showeprint[arxiv]{2101.02138}~[quant-ph]


\bibitem[Häner and Steiger(2017)]%
        {Haner2017}
\bibfield{author}{\bibinfo{person}{Thomas Häner} {and}
  \bibinfo{person}{Damian~S. Steiger}.} \bibinfo{year}{2017}\natexlab{}.
\newblock \showarticletitle{0.5 petabyte simulation of a 45-qubit quantum
  circuit}.
\newblock \bibinfo{journal}{\emph{Proceedings of the International Conference
  for High Performance Computing, Networking, Storage and Analysis}}
  (\bibinfo{date}{Nov} \bibinfo{year}{2017}).
\newblock
\urldef\tempurl%
\url{https://doi.org/10.1145/3126908.3126947}
\showDOI{\tempurl}


\bibitem[Kandala et~al\mbox{.}(2017)]%
        {kandala2017hardware}
\bibfield{author}{\bibinfo{person}{Abhinav Kandala}, \bibinfo{person}{Antonio
  Mezzacapo}, \bibinfo{person}{Kristan Temme}, \bibinfo{person}{Maika Takita},
  \bibinfo{person}{Markus Brink}, \bibinfo{person}{Jerry~M Chow}, {and}
  \bibinfo{person}{Jay~M Gambetta}.} \bibinfo{year}{2017}\natexlab{}.
\newblock \showarticletitle{Hardware-efficient variational quantum eigensolver
  for small molecules and quantum magnets}.
\newblock \bibinfo{journal}{\emph{Nature}} \bibinfo{volume}{549},
  \bibinfo{number}{7671} (\bibinfo{year}{2017}), \bibinfo{pages}{242--246}.
\newblock


\bibitem[Kirby et~al\mbox{.}(2021)]%
        {kirby2021}
\bibfield{author}{\bibinfo{person}{William~M. Kirby}, \bibinfo{person}{Andrew
  Tranter}, {and} \bibinfo{person}{Peter~J. Love}.}
  \bibinfo{year}{2021}\natexlab{}.
\newblock \showarticletitle{Contextual Subspace Variational Quantum
  Eigensolver}.
\newblock \bibinfo{journal}{\emph{Quantum}}  \bibinfo{volume}{5}
  (\bibinfo{date}{May} \bibinfo{year}{2021}), \bibinfo{pages}{456}.
\newblock
\showISSN{2521-327X}
\urldef\tempurl%
\url{https://doi.org/10.22331/q-2021-05-14-456}
\showDOI{\tempurl}


\bibitem[Larsson et~al\mbox{.}(2022)]%
        {Cr2_2022}
\bibfield{author}{\bibinfo{person}{Henrik~R. Larsson},
  \bibinfo{person}{Huanchen Zhai}, \bibinfo{person}{Cyrus~J. Umrigar}, {and}
  \bibinfo{person}{Garnet Kin-Lic Chan}.} \bibinfo{year}{2022}\natexlab{}.
\newblock \bibinfo{title}{The chromium dimer: closing a chapter of quantum
  chemistry}.
\newblock
\newblock
\urldef\tempurl%
\url{https://doi.org/10.48550/ARXIV.2206.10738}
\showDOI{\tempurl}


\bibitem[Lavrijsen et~al\mbox{.}(2020)]%
        {9259985}
\bibfield{author}{\bibinfo{person}{Wim Lavrijsen}, \bibinfo{person}{Ana Tudor},
  \bibinfo{person}{Juliane Müller}, \bibinfo{person}{Costin Iancu}, {and}
  \bibinfo{person}{Wibe de Jong}.} \bibinfo{year}{2020}\natexlab{}.
\newblock \showarticletitle{Classical Optimizers for Noisy Intermediate-Scale
  Quantum Devices}. In \bibinfo{booktitle}{\emph{2020 IEEE International
  Conference on Quantum Computing and Engineering (QCE)}}.
  \bibinfo{pages}{267--277}.
\newblock
\urldef\tempurl%
\url{https://doi.org/10.1109/QCE49297.2020.00041}
\showDOI{\tempurl}


\bibitem[Li et~al\mbox{.}(2017)]%
        {li2017hyperband}
\bibfield{author}{\bibinfo{person}{Lisha Li}, \bibinfo{person}{Kevin Jamieson},
  \bibinfo{person}{Giulia DeSalvo}, \bibinfo{person}{Afshin Rostamizadeh},
  {and} \bibinfo{person}{Ameet Talwalkar}.} \bibinfo{year}{2017}\natexlab{}.
\newblock \showarticletitle{Hyperband: A novel bandit-based approach to
  hyperparameter optimization}.
\newblock \bibinfo{journal}{\emph{The Journal of Machine Learning Research}}
  \bibinfo{volume}{18}, \bibinfo{number}{1} (\bibinfo{year}{2017}),
  \bibinfo{pages}{6765--6816}.
\newblock


\bibitem[Li and Benjamin(2017)]%
        {li2017efficient}
\bibfield{author}{\bibinfo{person}{Ying Li} {and} \bibinfo{person}{Simon~C.
  Benjamin}.} \bibinfo{year}{2017}\natexlab{}.
\newblock \showarticletitle{Efficient Variational Quantum Simulator
  Incorporating Active Error Minimization}.
\newblock \bibinfo{journal}{\emph{Phys. Rev. X}}  \bibinfo{volume}{7}
  (\bibinfo{date}{Jun} \bibinfo{year}{2017}), \bibinfo{pages}{021050}.
\newblock
Issue 2.
\urldef\tempurl%
\url{https://doi.org/10.1103/PhysRevX.7.021050}
\showDOI{\tempurl}


\bibitem[Marrero et~al\mbox{.}(2021)]%
        {marrero2021entanglement}
\bibfield{author}{\bibinfo{person}{Carlos~Ortiz Marrero},
  \bibinfo{person}{Mária Kieferová}, {and} \bibinfo{person}{Nathan Wiebe}.}
  \bibinfo{year}{2021}\natexlab{}.
\newblock \bibinfo{title}{Entanglement Induced Barren Plateaus}.
\newblock
\newblock
\showeprint[arxiv]{2010.15968}~[quant-ph]


\bibitem[McClean et~al\mbox{.}(2018)]%
        {mcclean2018}
\bibfield{author}{\bibinfo{person}{Jarrod~R. McClean}, \bibinfo{person}{Sergio
  Boixo}, \bibinfo{person}{Vadim~N. Smelyanskiy}, \bibinfo{person}{Ryan
  Babbush}, {and} \bibinfo{person}{Hartmut Neven}.}
  \bibinfo{year}{2018}\natexlab{}.
\newblock \showarticletitle{Barren plateaus in quantum neural network training
  landscapes}.
\newblock \bibinfo{journal}{\emph{Nature Communications}} \bibinfo{volume}{9},
  \bibinfo{number}{1} (\bibinfo{date}{Nov} \bibinfo{year}{2018}).
\newblock
\showISSN{2041-1723}
\urldef\tempurl%
\url{https://doi.org/10.1038/s41467-018-07090-4}
\showDOI{\tempurl}


\bibitem[McClean et~al\mbox{.}(2016)]%
        {mcclean2016theory}
\bibfield{author}{\bibinfo{person}{Jarrod~R McClean}, \bibinfo{person}{Jonathan
  Romero}, \bibinfo{person}{Ryan Babbush}, {and} \bibinfo{person}{Al{\'a}n
  Aspuru-Guzik}.} \bibinfo{year}{2016}\natexlab{}.
\newblock \showarticletitle{The theory of variational hybrid quantum-classical
  algorithms}.
\newblock \bibinfo{journal}{\emph{New Journal of Physics}}
  \bibinfo{volume}{18}, \bibinfo{number}{2} (\bibinfo{year}{2016}),
  \bibinfo{pages}{023023}.
\newblock


\bibitem[Mitarai et~al\mbox{.}(2020)]%
        {mitarai2020quadratic}
\bibfield{author}{\bibinfo{person}{Kosuke Mitarai}, \bibinfo{person}{Yasunari
  Suzuki}, \bibinfo{person}{Wataru Mizukami}, \bibinfo{person}{Yuya~O.
  Nakagawa}, {and} \bibinfo{person}{Keisuke Fujii}.}
  \bibinfo{year}{2020}\natexlab{}.
\newblock \bibinfo{title}{Quadratic Clifford expansion for efficient
  benchmarking and initialization of variational quantum algorithms}.
\newblock
\newblock
\showeprint[arxiv]{2011.09927}~[quant-ph]


\bibitem[Mohseni et~al\mbox{.}(2017)]%
        {45919}
\bibfield{author}{\bibinfo{person}{Masoud Mohseni}, \bibinfo{person}{Peter
  Read}, \bibinfo{person}{Hartmut Neven}, \bibinfo{person}{Sergio Boixo},
  \bibinfo{person}{Vasil Denchev}, \bibinfo{person}{Ryan Babbush},
  \bibinfo{person}{Austin Fowler}, \bibinfo{person}{Vadim Smelyanskiy}, {and}
  \bibinfo{person}{John Martinis}.} \bibinfo{year}{2017}\natexlab{}.
\newblock \showarticletitle{Commercialize Quantum Technologies in Five Years}.
\newblock \bibinfo{journal}{\emph{Nature}}  \bibinfo{volume}{543}
  (\bibinfo{year}{2017}), \bibinfo{pages}{171–174}.
\newblock
\urldef\tempurl%
\url{http://www.nature.com/news/commercialize-quantum-technologies-in-five-years-1.21583}
\showURL{%
\tempurl}


\bibitem[Moll et~al\mbox{.}(2018)]%
        {moll2018quantum}
\bibfield{author}{\bibinfo{person}{Nikolaj Moll}, \bibinfo{person}{Panagiotis
  Barkoutsos}, \bibinfo{person}{Lev~S Bishop}, \bibinfo{person}{Jerry~M Chow},
  \bibinfo{person}{Andrew Cross}, \bibinfo{person}{Daniel~J Egger},
  \bibinfo{person}{Stefan Filipp}, \bibinfo{person}{Andreas Fuhrer},
  \bibinfo{person}{Jay~M Gambetta}, \bibinfo{person}{Marc Ganzhorn},
  {et~al\mbox{.}}} \bibinfo{year}{2018}\natexlab{}.
\newblock \showarticletitle{Quantum optimization using variational algorithms
  on near-term quantum devices}.
\newblock \bibinfo{journal}{\emph{Quantum Science and Technology}}
  \bibinfo{volume}{3}, \bibinfo{number}{3} (\bibinfo{year}{2018}),
  \bibinfo{pages}{030503}.
\newblock


\bibitem[Murali et~al\mbox{.}(2019)]%
        {murali2019noise}
\bibfield{author}{\bibinfo{person}{Prakash Murali}, \bibinfo{person}{Jonathan~M
  Baker}, \bibinfo{person}{Ali Javadi-Abhari}, \bibinfo{person}{Frederic~T
  Chong}, {and} \bibinfo{person}{Margaret Martonosi}.}
  \bibinfo{year}{2019}\natexlab{}.
\newblock \showarticletitle{Noise-adaptive compiler mappings for noisy
  intermediate-scale quantum computers}. In
  \bibinfo{booktitle}{\emph{Proceedings of the Twenty-Fourth International
  Conference on Architectural Support for Programming Languages and Operating
  Systems}}. \bibinfo{pages}{1015--1029}.
\newblock


\bibitem[Murali et~al\mbox{.}(2020)]%
        {murali2020software}
\bibfield{author}{\bibinfo{person}{Prakash Murali}, \bibinfo{person}{David~C
  McKay}, \bibinfo{person}{Margaret Martonosi}, {and} \bibinfo{person}{Ali
  Javadi-Abhari}.} \bibinfo{year}{2020}\natexlab{}.
\newblock \showarticletitle{Software mitigation of crosstalk on noisy
  intermediate-scale quantum computers}. In
  \bibinfo{booktitle}{\emph{Proceedings of the Twenty-Fifth International
  Conference on Architectural Support for Programming Languages and Operating
  Systems}}. \bibinfo{pages}{1001--1016}.
\newblock


\bibitem[Nardi et~al\mbox{.}(2019)]%
        {nardi2019practical}
\bibfield{author}{\bibinfo{person}{Luigi Nardi}, \bibinfo{person}{David
  Koeplinger}, {and} \bibinfo{person}{Kunle Olukotun}.}
  \bibinfo{year}{2019}\natexlab{}.
\newblock \bibinfo{title}{Practical Design Space Exploration}.
\newblock
\newblock
\showeprint[arxiv]{1810.05236}~[cs.LG]


\bibitem[Nielsen and Chuang(2002)]%
        {nielsen2002quantum}
\bibfield{author}{\bibinfo{person}{Michael~A Nielsen} {and}
  \bibinfo{person}{Isaac Chuang}.} \bibinfo{year}{2002}\natexlab{}.
\newblock \bibinfo{title}{Quantum computation and quantum information}.
\newblock
\newblock


\bibitem[O’Gorman and Campbell(2017)]%
        {O_Gorman_2017}
\bibfield{author}{\bibinfo{person}{Joe O’Gorman} {and}
  \bibinfo{person}{Earl~T. Campbell}.} \bibinfo{year}{2017}\natexlab{}.
\newblock \showarticletitle{Quantum computation with realistic magic-state
  factories}.
\newblock \bibinfo{journal}{\emph{Physical Review A}} \bibinfo{volume}{95},
  \bibinfo{number}{3} (\bibinfo{date}{Mar} \bibinfo{year}{2017}).
\newblock
\showISSN{2469-9934}
\urldef\tempurl%
\url{https://doi.org/10.1103/physreva.95.032338}
\showDOI{\tempurl}


\bibitem[Patel and Tiwari(2020)]%
        {Clite2020}
\bibfield{author}{\bibinfo{person}{Tirthak Patel} {and} \bibinfo{person}{Devesh
  Tiwari}.} \bibinfo{year}{2020}\natexlab{}.
\newblock \showarticletitle{CLITE: Efficient and QoS-Aware Co-Location of
  Multiple Latency-Critical Jobs for Warehouse Scale Computers}. In
  \bibinfo{booktitle}{\emph{2020 IEEE International Symposium on High
  Performance Computer Architecture (HPCA)}}. \bibinfo{pages}{193--206}.
\newblock
\urldef\tempurl%
\url{https://doi.org/10.1109/HPCA47549.2020.00025}
\showDOI{\tempurl}


\bibitem[Patti et~al\mbox{.}(2021)]%
        {PhysRevResearch.3.033090}
\bibfield{author}{\bibinfo{person}{Taylor~L. Patti}, \bibinfo{person}{Khadijeh
  Najafi}, \bibinfo{person}{Xun Gao}, {and} \bibinfo{person}{Susanne~F.
  Yelin}.} \bibinfo{year}{2021}\natexlab{}.
\newblock \showarticletitle{Entanglement devised barren plateau mitigation}.
\newblock \bibinfo{journal}{\emph{Phys. Rev. Research}}  \bibinfo{volume}{3}
  (\bibinfo{date}{Jul} \bibinfo{year}{2021}), \bibinfo{pages}{033090}.
\newblock
Issue 3.
\urldef\tempurl%
\url{https://doi.org/10.1103/PhysRevResearch.3.033090}
\showDOI{\tempurl}


\bibitem[Peruzzo et~al\mbox{.}(2014)]%
        {peruzzo2014variational}
\bibfield{author}{\bibinfo{person}{Alberto Peruzzo}, \bibinfo{person}{Jarrod
  McClean}, \bibinfo{person}{Peter Shadbolt}, \bibinfo{person}{Man-Hong Yung},
  \bibinfo{person}{Xiao-Qi Zhou}, \bibinfo{person}{Peter~J Love},
  \bibinfo{person}{Al{\'a}n Aspuru-Guzik}, {and} \bibinfo{person}{Jeremy~L
  O’brien}.} \bibinfo{year}{2014}\natexlab{}.
\newblock \showarticletitle{A variational eigenvalue solver on a photonic
  quantum processor}.
\newblock \bibinfo{journal}{\emph{Nature communications}}  \bibinfo{volume}{5}
  (\bibinfo{year}{2014}), \bibinfo{pages}{4213}.
\newblock


\bibitem[Peterson et~al\mbox{.}(2012)]%
        {peterson2012chemical}
\bibfield{author}{\bibinfo{person}{Kirk~A Peterson}, \bibinfo{person}{David
  Feller}, {and} \bibinfo{person}{David~A Dixon}.}
  \bibinfo{year}{2012}\natexlab{}.
\newblock \showarticletitle{Chemical accuracy in ab initio thermochemistry and
  spectroscopy: current strategies and future challenges}.
\newblock \bibinfo{journal}{\emph{Theoretical Chemistry Accounts}}
  \bibinfo{volume}{131}, \bibinfo{number}{1} (\bibinfo{year}{2012}),
  \bibinfo{pages}{1--20}.
\newblock


\bibitem[Preskill(2018)]%
        {preskill2018quantum}
\bibfield{author}{\bibinfo{person}{John Preskill}.}
  \bibinfo{year}{2018}\natexlab{}.
\newblock \showarticletitle{Quantum Computing in the NISQ era and beyond}.
\newblock \bibinfo{journal}{\emph{Quantum}}  \bibinfo{volume}{2}
  (\bibinfo{year}{2018}), \bibinfo{pages}{79}.
\newblock


\bibitem[Ravi et~al\mbox{.}(2021)]%
        {ravi2021vaqem}
\bibfield{author}{\bibinfo{person}{Gokul~Subramanian Ravi},
  \bibinfo{person}{Kaitlin~N. Smith}, \bibinfo{person}{Pranav Gokhale},
  \bibinfo{person}{Andrea Mari}, \bibinfo{person}{Nathan Earnest},
  \bibinfo{person}{Ali Javadi-Abhari}, {and} \bibinfo{person}{Frederic~T.
  Chong}.} \bibinfo{year}{2021}\natexlab{}.
\newblock \bibinfo{title}{VAQEM: A Variational Approach to Quantum Error
  Mitigation}.
\newblock
\newblock
\showeprint[arxiv]{2112.05821}~[quant-ph]


\bibitem[Roffe(2019)]%
        {QECIntro}
\bibfield{author}{\bibinfo{person}{Joschka Roffe}.}
  \bibinfo{year}{2019}\natexlab{}.
\newblock \showarticletitle{Quantum error correction: an introductory guide}.
\newblock \bibinfo{journal}{\emph{Contemporary Physics}} \bibinfo{volume}{60},
  \bibinfo{number}{3} (\bibinfo{date}{Jul} \bibinfo{year}{2019}),
  \bibinfo{pages}{226–245}.
\newblock
\showISSN{1366-5812}
\urldef\tempurl%
\url{https://doi.org/10.1080/00107514.2019.1667078}
\showDOI{\tempurl}


\bibitem[Romero et~al\mbox{.}(2018)]%
        {romero2018strategies}
\bibfield{author}{\bibinfo{person}{Jonathan Romero}, \bibinfo{person}{Ryan
  Babbush}, \bibinfo{person}{Jarrod~R. McClean}, \bibinfo{person}{Cornelius
  Hempel}, \bibinfo{person}{Peter Love}, {and} \bibinfo{person}{Alán
  Aspuru-Guzik}.} \bibinfo{year}{2018}\natexlab{}.
\newblock \bibinfo{title}{Strategies for quantum computing molecular energies
  using the unitary coupled cluster ansatz}.
\newblock
\newblock
\showeprint[arxiv]{1701.02691}~[quant-ph]


\bibitem[Rosenberg et~al\mbox{.}(2021)]%
        {Rosenberg2021}
\bibfield{author}{\bibinfo{person}{Eliott Rosenberg}, \bibinfo{person}{Paul
  Ginsparg}, {and} \bibinfo{person}{Peter~L. McMahon}.}
  \bibinfo{year}{2021}\natexlab{}.
\newblock \showarticletitle{Experimental error mitigation using linear
  rescaling for variational quantum eigensolving with up to 20 qubits}.
\newblock \bibinfo{journal}{\emph{Quantum Science and Technology}}
  (\bibinfo{date}{Nov} \bibinfo{year}{2021}).
\newblock
\showISSN{2058-9565}
\urldef\tempurl%
\url{https://doi.org/10.1088/2058-9565/ac3b37}
\showDOI{\tempurl}


\bibitem[Roy et~al\mbox{.}(2021)]%
        {Bliss2021}
\bibfield{author}{\bibinfo{person}{Rohan~Basu Roy}, \bibinfo{person}{Tirthak
  Patel}, \bibinfo{person}{Vijay Gadepally}, {and} \bibinfo{person}{Devesh
  Tiwari}.} \bibinfo{year}{2021}\natexlab{}.
\newblock \bibinfo{booktitle}{\emph{Bliss: Auto-Tuning Complex Applications
  Using a Pool of Diverse Lightweight Learning Models}}.
\newblock \bibinfo{publisher}{Association for Computing Machinery},
  \bibinfo{address}{New York, NY, USA}, \bibinfo{pages}{1280–1295}.
\newblock
\showISBNx{9781450383912}
\urldef\tempurl%
\url{https://doi.org/10.1145/3453483.3454109}
\showURL{%
\tempurl}


\bibitem[Ryabinkin et~al\mbox{.}(2018)]%
        {ryabinkin2018constrained}
\bibfield{author}{\bibinfo{person}{Ilya~G. Ryabinkin},
  \bibinfo{person}{Scott~N. Genin}, {and} \bibinfo{person}{Artur~F. Izmaylov}.}
  \bibinfo{year}{2018}\natexlab{}.
\newblock \bibinfo{title}{Constrained variational quantum eigensolver: Quantum
  computer search engine in the Fock space}.
\newblock
\newblock
\showeprint[arxiv]{1806.00461}~[physics.chem-ph]


\bibitem[Sauvage et~al\mbox{.}(2021)]%
        {sauvage2021flip}
\bibfield{author}{\bibinfo{person}{Frederic Sauvage}, \bibinfo{person}{Sukin
  Sim}, \bibinfo{person}{Alexander~A. Kunitsa}, \bibinfo{person}{William~A.
  Simon}, \bibinfo{person}{Marta Mauri}, {and} \bibinfo{person}{Alejandro
  Perdomo-Ortiz}.} \bibinfo{year}{2021}\natexlab{}.
\newblock \bibinfo{title}{FLIP: A flexible initializer for arbitrarily-sized
  parametrized quantum circuits}.
\newblock
\newblock
\showeprint[arxiv]{2103.08572}~[quant-ph]


\bibitem[Shor(1997)]%
        {Shor_1997}
\bibfield{author}{\bibinfo{person}{Peter~W. Shor}.}
  \bibinfo{year}{1997}\natexlab{}.
\newblock \showarticletitle{Polynomial-Time Algorithms for Prime Factorization
  and Discrete Logarithms on a Quantum Computer}.
\newblock \bibinfo{journal}{\emph{SIAM J. Comput.}} \bibinfo{volume}{26},
  \bibinfo{number}{5} (\bibinfo{date}{Oct} \bibinfo{year}{1997}),
  \bibinfo{pages}{1484–1509}.
\newblock
\showISSN{1095-7111}
\urldef\tempurl%
\url{https://doi.org/10.1137/s0097539795293172}
\showDOI{\tempurl}


\bibitem[Smith et~al\mbox{.}(2021)]%
        {smith2021error}
\bibfield{author}{\bibinfo{person}{Kaitlin~N Smith},
  \bibinfo{person}{Gokul~Subramanian Ravi}, \bibinfo{person}{Prakash Murali},
  \bibinfo{person}{Jonathan~M Baker}, \bibinfo{person}{Nathan Earnest},
  \bibinfo{person}{Ali Javadi-Abhari}, {and} \bibinfo{person}{Frederic~T
  Chong}.} \bibinfo{year}{2021}\natexlab{}.
\newblock \showarticletitle{Error Mitigation in Quantum Computers through
  Instruction Scheduling}.
\newblock \bibinfo{journal}{\emph{arXiv preprint arXiv:2105.01760}}
  (\bibinfo{year}{2021}).
\newblock


\bibitem[Strikis et~al\mbox{.}(2021)]%
        {strikis2021learningbased}
\bibfield{author}{\bibinfo{person}{Armands Strikis}, \bibinfo{person}{Dayue
  Qin}, \bibinfo{person}{Yanzhu Chen}, \bibinfo{person}{Simon~C. Benjamin},
  {and} \bibinfo{person}{Ying Li}.} \bibinfo{year}{2021}\natexlab{}.
\newblock \bibinfo{title}{Learning-based quantum error mitigation}.
\newblock
\newblock
\showeprint[arxiv]{2005.07601}~[quant-ph]


\bibitem[Sun et~al\mbox{.}(2017)]%
        {sun2017pythonbased}
\bibfield{author}{\bibinfo{person}{Qiming Sun}, \bibinfo{person}{Timothy~C.
  Berkelbach}, \bibinfo{person}{Nick~S. Blunt}, \bibinfo{person}{George~H.
  Booth}, \bibinfo{person}{Sheng Guo}, \bibinfo{person}{Zhendong Li},
  \bibinfo{person}{Junzi Liu}, \bibinfo{person}{James McClain},
  \bibinfo{person}{Elvira~R. Sayfutyarova}, \bibinfo{person}{Sandeep Sharma},
  \bibinfo{person}{Sebastian Wouters}, {and} \bibinfo{person}{Garnet Kin-Lic
  Chan}.} \bibinfo{year}{2017}\natexlab{}.
\newblock \bibinfo{title}{The Python-based Simulations of Chemistry Framework
  (PySCF)}.
\newblock
\newblock
\showeprint[arxiv]{1701.08223}~[physics.chem-ph]


\bibitem[Takagi et~al\mbox{.}(2021)]%
        {takagi2021fundamental}
\bibfield{author}{\bibinfo{person}{Ryuji Takagi}, \bibinfo{person}{Suguru
  Endo}, \bibinfo{person}{Shintaro Minagawa}, {and} \bibinfo{person}{Mile Gu}.}
  \bibinfo{year}{2021}\natexlab{}.
\newblock \bibinfo{title}{Fundamental limits of quantum error mitigation}.
\newblock
\newblock
\showeprint[arxiv]{2109.04457}~[quant-ph]


\bibitem[Tang et~al\mbox{.}(2021)]%
        {CutQC}
\bibfield{author}{\bibinfo{person}{Wei Tang}, \bibinfo{person}{Teague Tomesh},
  \bibinfo{person}{Martin Suchara}, \bibinfo{person}{Jeffrey Larson}, {and}
  \bibinfo{person}{Margaret Martonosi}.} \bibinfo{year}{2021}\natexlab{}.
\newblock \showarticletitle{CutQC: using small Quantum computers for large
  Quantum circuit evaluations}.
\newblock \bibinfo{journal}{\emph{Proceedings of the 26th ACM International
  Conference on Architectural Support for Programming Languages and Operating
  Systems}} (\bibinfo{date}{Apr} \bibinfo{year}{2021}).
\newblock
\urldef\tempurl%
\url{https://doi.org/10.1145/3445814.3446758}
\showDOI{\tempurl}


\bibitem[Tannu and Qureshi(2019)]%
        {tannu2019not}
\bibfield{author}{\bibinfo{person}{Swamit~S Tannu} {and}
  \bibinfo{person}{Moinuddin~K Qureshi}.} \bibinfo{year}{2019}\natexlab{}.
\newblock \showarticletitle{Not all qubits are created equal: a case for
  variability-aware policies for NISQ-era quantum computers}. In
  \bibinfo{booktitle}{\emph{Proceedings of the Twenty-Fourth International
  Conference on Architectural Support for Programming Languages and Operating
  Systems}}. \bibinfo{pages}{987--999}.
\newblock


\bibitem[Temme et~al\mbox{.}(2017)]%
        {temme2017error}
\bibfield{author}{\bibinfo{person}{Kristan Temme}, \bibinfo{person}{Sergey
  Bravyi}, {and} \bibinfo{person}{Jay~M Gambetta}.}
  \bibinfo{year}{2017}\natexlab{}.
\newblock \showarticletitle{Error mitigation for short-depth quantum circuits}.
\newblock \bibinfo{journal}{\emph{Physical review letters}}
  \bibinfo{volume}{119}, \bibinfo{number}{18} (\bibinfo{year}{2017}),
  \bibinfo{pages}{180509}.
\newblock


\bibitem[Tilly et~al\mbox{.}(2021)]%
        {tilly2021variational}
\bibfield{author}{\bibinfo{person}{Jules Tilly}, \bibinfo{person}{Hongxiang
  Chen}, \bibinfo{person}{Shuxiang Cao}, \bibinfo{person}{Dario Picozzi},
  \bibinfo{person}{Kanav Setia}, \bibinfo{person}{Ying Li},
  \bibinfo{person}{Edward Grant}, \bibinfo{person}{Leonard Wossnig},
  \bibinfo{person}{Ivan Rungger}, \bibinfo{person}{George~H. Booth}, {and}
  \bibinfo{person}{Jonathan Tennyson}.} \bibinfo{year}{2021}\natexlab{}.
\newblock \bibinfo{title}{The Variational Quantum Eigensolver: a review of
  methods and best practices}.
\newblock
\newblock
\showeprint[arxiv]{2111.05176}~[quant-ph]


\bibitem[Turney et~al\mbox{.}(2012)]%
        {turney2012psi4}
\bibfield{author}{\bibinfo{person}{Justin~M Turney}, \bibinfo{person}{Andrew~C
  Simmonett}, \bibinfo{person}{Robert~M Parrish}, \bibinfo{person}{Edward~G
  Hohenstein}, \bibinfo{person}{Francesco~A Evangelista},
  \bibinfo{person}{Justin~T Fermann}, \bibinfo{person}{Benjamin~J Mintz},
  \bibinfo{person}{Lori~A Burns}, \bibinfo{person}{Jeremiah~J Wilke},
  \bibinfo{person}{Micah~L Abrams}, {et~al\mbox{.}}}
  \bibinfo{year}{2012}\natexlab{}.
\newblock \showarticletitle{Psi4: an open-source ab initio electronic structure
  program}.
\newblock \bibinfo{journal}{\emph{Wiley Interdisciplinary Reviews:
  Computational Molecular Science}} \bibinfo{volume}{2}, \bibinfo{number}{4}
  (\bibinfo{year}{2012}), \bibinfo{pages}{556--565}.
\newblock


\bibitem[Uvarov and Biamonte(2021)]%
        {Uvarov2021}
\bibfield{author}{\bibinfo{person}{A~V Uvarov} {and} \bibinfo{person}{J~D
  Biamonte}.} \bibinfo{year}{2021}\natexlab{}.
\newblock \showarticletitle{On barren plateaus and cost function locality in
  variational quantum algorithms}.
\newblock \bibinfo{journal}{\emph{Journal of Physics A: Mathematical and
  Theoretical}} \bibinfo{volume}{54}, \bibinfo{number}{24} (\bibinfo{date}{May}
  \bibinfo{year}{2021}).
\newblock
\showISSN{1751-8121}
\urldef\tempurl%
\url{https://doi.org/10.1088/1751-8121/abfac7}
\showDOI{\tempurl}


\bibitem[Vancoillie et~al\mbox{.}(2016)]%
        {vancoillie2016potential}
\bibfield{author}{\bibinfo{person}{Steven Vancoillie},
  \bibinfo{person}{Per~{\AA}ke Malmqvist}, {and} \bibinfo{person}{Valera
  Veryazov}.} \bibinfo{year}{2016}\natexlab{}.
\newblock \showarticletitle{Potential energy surface of the chromium dimer
  re-re-revisited with multiconfigurational perturbation theory}.
\newblock \bibinfo{journal}{\emph{Journal of chemical theory and computation}}
  \bibinfo{volume}{12}, \bibinfo{number}{4} (\bibinfo{year}{2016}),
  \bibinfo{pages}{1647--1655}.
\newblock


\bibitem[Veitch et~al\mbox{.}(2014)]%
        {Veitch2014}
\bibfield{author}{\bibinfo{person}{Victor Veitch}, \bibinfo{person}{S~A
  Hamed~Mousavian}, \bibinfo{person}{Daniel Gottesman}, {and}
  \bibinfo{person}{Joseph Emerson}.} \bibinfo{year}{2014}\natexlab{}.
\newblock \showarticletitle{The resource theory of stabilizer quantum
  computation}.
\newblock \bibinfo{journal}{\emph{New Journal of Physics}}
  \bibinfo{volume}{16}, \bibinfo{number}{1} (\bibinfo{date}{Jan}
  \bibinfo{year}{2014}), \bibinfo{pages}{013009}.
\newblock
\showISSN{1367-2630}
\urldef\tempurl%
\url{https://doi.org/10.1088/1367-2630/16/1/013009}
\showDOI{\tempurl}


\bibitem[Wang et~al\mbox{.}(2021)]%
        {wang2021error}
\bibfield{author}{\bibinfo{person}{Samson Wang}, \bibinfo{person}{Piotr
  Czarnik}, \bibinfo{person}{Andrew Arrasmith}, \bibinfo{person}{M. Cerezo},
  \bibinfo{person}{Lukasz Cincio}, {and} \bibinfo{person}{Patrick~J. Coles}.}
  \bibinfo{year}{2021}\natexlab{}.
\newblock \bibinfo{title}{Can Error Mitigation Improve Trainability of Noisy
  Variational Quantum Algorithms?}
\newblock
\newblock
\showeprint[arxiv]{2109.01051}~[quant-ph]


\bibitem[Wang et~al\mbox{.}(2020)]%
        {wang2020noise}
\bibfield{author}{\bibinfo{person}{Samson Wang}, \bibinfo{person}{Enrico
  Fontana}, \bibinfo{person}{Marco Cerezo}, \bibinfo{person}{Kunal Sharma},
  \bibinfo{person}{Akira Sone}, \bibinfo{person}{Lukasz Cincio}, {and}
  \bibinfo{person}{Patrick~J Coles}.} \bibinfo{year}{2020}\natexlab{}.
\newblock \showarticletitle{Noise-induced barren plateaus in variational
  quantum algorithms}.
\newblock \bibinfo{journal}{\emph{arXiv preprint arXiv:2007.14384}}
  (\bibinfo{year}{2020}).
\newblock


\bibitem[Zhang et~al\mbox{.}(2021)]%
        {zhang2021variational}
\bibfield{author}{\bibinfo{person}{Yu Zhang}, \bibinfo{person}{Lukasz Cincio},
  \bibinfo{person}{Christian~FA Negre}, \bibinfo{person}{Piotr Czarnik},
  \bibinfo{person}{Patrick Coles}, \bibinfo{person}{Petr~M Anisimov},
  \bibinfo{person}{Susan~M Mniszewski}, \bibinfo{person}{Sergei Tretiak}, {and}
  \bibinfo{person}{Pavel~A Dub}.} \bibinfo{year}{2021}\natexlab{}.
\newblock \showarticletitle{Variational quantum eigensolver with reduced
  circuit complexity}.
\newblock \bibinfo{journal}{\emph{arXiv preprint arXiv:2106.07619}}
  (\bibinfo{year}{2021}).
\newblock


\bibitem[Zhou et~al\mbox{.}(2020)]%
        {PhysRevX.10.041038}
\bibfield{author}{\bibinfo{person}{Yiqing Zhou}, \bibinfo{person}{E.~Miles
  Stoudenmire}, {and} \bibinfo{person}{Xavier Waintal}.}
  \bibinfo{year}{2020}\natexlab{}.
\newblock \showarticletitle{What Limits the Simulation of Quantum Computers?}
\newblock \bibinfo{journal}{\emph{Phys. Rev. X}}  \bibinfo{volume}{10}
  (\bibinfo{date}{Nov} \bibinfo{year}{2020}), \bibinfo{pages}{041038}.
\newblock
Issue 4.
\urldef\tempurl%
\url{https://doi.org/10.1103/PhysRevX.10.041038}
\showDOI{\tempurl}


\end{thebibliography}

\end{document}